\documentclass[12pt]{article}
\pdfoutput=1
\usepackage{putex}
\usepackage{amsmath,hyperref,amssymb,amsfonts,bm,amscd,slashed}
\usepackage{cite}
\usepackage{graphicx}
\usepackage{multirow}
\usepackage{verbatim}
\usepackage{appendix}
\usepackage{fancybox}
\usepackage{array}
\usepackage{url}
\usepackage{float}
\usepackage{mathtools}
\usepackage{mathrsfs}
\usepackage{bbm}

\DeclareMathAlphabet{\mathpzc}{OT1}{pzc}{m}{it}

\newcommand{\bea}{\begin{eqnarray}}
\newcommand{\eea}{\end{eqnarray}}
\newcommand{\be}{\begin{equation}}
\newcommand{\ee}{\end{equation}}

\def \beaa {\begin{equation}\begin{aligned}}
\def \eeaa {\end{aligned}\end{equation}}

\newcommand{\Z}{{\mathbb Z}}
\newcommand{\R}{{\mathbb R}}
\newcommand{\C}{{\mathbb C}}

\newcommand{\cZ}{{\mathcal{Z}}}
\newcommand{\cF}{{\mathcal{F}}}
\newcommand{\cN}{{\mathcal{N}}}

\newcommand{\dd}{{\rm d}}

\def\Tr{{\rm Tr \,}}

\def\m{\mu}
\def\n{\nu}

\def\tilde{\widetilde}
\def\hat{\widehat}
\def\bar{\overline}

\def\cA{{\mathcal A}}

\def\cC{{\mathcal C}}
\def\cD{{\mathcal D}}

\def\cF{{\mathcal F}}
\def\cG{{\mathcal G}}
\def\cH{{\mathcal H}}
\def\cI{{\mathcal I}}
\def\cJ{{\mathcal J}}

\def\cL{{\mathcal L}}
\def\cM{{\mathcal M}}
\def\cN{{\mathcal N}}
\def\cO{{\mathcal O}}

\def\cQ{{\mathcal Q}}
\def\cR{{\mathcal R}}
\def\cS{{\mathcal S}}

\def\cV{{\mathcal V}}
\def\cW{{\mathcal W}}

\def\cZ{{\mathcal Z}}

\def\rR{{\mathbf R}}
\def\rQ{{\mathbbmtt{Q}\,}}
\def\sZ{{\mathpzc{Z}}}

\renewcommand{\bar}{\overline}
\renewcommand{\hat}{\widehat}

\numberwithin{equation}{section}
\newcommand{\ii}{\mathrm{i}}

\begin{document}
\preprint{CALT-TH 2019-011, IPMU19-0045}

\institution{Caltech}{Walter Burke Institute for Theoretical Physics, California Institute of Technology, \cr Pasadena, CA 91125, USA}
\institution{IPMU}{Kavli Institute for the Physics and Mathematics of the Universe (WPI), University of Tokyo, \cr Kashiwa, 277-8583, Japan}

\title{Chiral Algebra, Localization, Modularity, Surface defects, And All That}
\authors{Mykola Dedushenko\worksat{\Caltech} and Martin Fluder\worksat{\Caltech,\IPMU}}

\abstract{
We study the 2D vertex operator algebra (VOA) construction in 4D $\mathcal{N}=2$ superconformal field theories (SCFT) on $S^3 \times S^1$, focusing both on old puzzles as well as new observations. The VOA lives on a two-torus $\mathbb{T}^2\subset S^3\times S^1$, it is $\frac12\mathbb{Z}$-graded, and this torus is equipped with the natural choice of spin structure (1,0) for the $\mathbb{Z} +\frac12$-graded operators, corresponding to the NS sector vacuum character. By analyzing the possible refinements of the Schur index that preserve the VOA, we find that it admits discrete deformations, which allow access to the remaining spin structures (1,1), (0,1) and (0,0), of which the latter two involve the inclusion of a particular surface defect. For Lagrangian theories, we perform the detailed analysis: we describe the natural supersymmetric background, perform localization, and derive the gauged symplectic boson action on a torus in any spin structure. In the absence of flavor fugacities, the 2D and 4D path integrals precisely match, including the Casimir factors. We further analyze the 2D theory: we identify its integration cycle, the two-point functions, and interpret flavor holonomies as screening charges in the VOA. Next, we make some observations about modularity; the $T$-transformation acts on our four partition functions and lifts to a large diffeomorphism on $S^3\times S^1$. More interestingly, we generalize the four partition functions on the torus to an infinite family labeled both by the spin structure and the integration cycle inside the complexified maximal torus of the gauge group. Members of this family transform into one another under the full modular group, and we confirm the recent observation that the $S$-transform of the Schur index in Lagrangian theories exhibits logarithmic behavior. Finally, we comment on how locally our background reproduces the $\Omega$-background. 

}

\date{}

\maketitle

\tableofcontents
\setlength{\unitlength}{1mm}

\newpage

\section{Introduction}

Accessing the strongly coupled and non-perturbative dynamics of a quantum field theory is hard, and even upon incorporating simplifying assumptions, such as supersymmetry or conformal symmetry, techniques allowing for control in such regimes are rare. Certain supersymmetric or protected quantities in supersymmetric field theories nevertheless prove amenable to analytic study, and much of the focus in the last decades has been on such examples. One prominent class of such theories are four-dimensional $\cN=2$ quantum field theories. The infrared (IR) dynamics on the Coulomb branch of such theories has long been understood since the release of the seminal set of works~\cite{Seiberg:1994aj,Seiberg:1994rs,Nekrasov:2002qd,Nekrasov:2003rj}. Incorporating techniques of supersymmetric localization was another major achievement in getting analytical control of such theories~\cite{Pestun:2007rz}. The advent of a whole new set of structures in these theories also resulted from the invention of the $\Omega$-background \cite{Nekrasov:2002qd,Nekrasov:2003rj,Losev:2003py,Nekrasov:2009rc,Nekrasov:2010ka}.

Recent years have seen another surge of research activity on four-dimensional $\cN=2$ theories. More specifically, the discovery and precise formulation of a connection between vertex operator algebras (VOA) and four-dimensional superconformal field theories (SCFTs) in~\cite{Beem:2013sza}, has sparked a rekindled research interest -- see \emph{e.g.}~\cite{Beem:2014rza,Lemos:2014lua,Cordova:2015nma,Bobev:2015kza,Liendo:2015ofa,Lemos:2015orc,Cecotti:2015lab,Buican:2016arp,Cordova:2016uwk,Arakawa:2016hkg,Bonetti:2016nma,Song:2016yfd,Fredrickson:2017yka,Cordova:2017mhb,Song:2017oew,Buican:2017fiq,Beem:2017ooy,Neitzke:2017cxz,Pan:2017zie,Fluder:2017oxm,Buican:2017rya,Choi:2017nur,Arakawa:2017fdq,Niarchos:2018mvl,Feigin:2018bkf,Creutzig:2018lbc,Bonetti:2018fqz,Nishinaka:2018zwq,Agarwal:2018zqi,Beem:2018duj,Costello:2018zrm,Xie:2019yds,Pan:2019bor,Beem:2019tfp,Oh:2019bgz,Jeong:2019pzg} for subsequent developments. Such VOAs are rich yet rigid objects which, quite surprisingly, allow to gain a non-perturbative access to an algebraically closed sector in the operator product expansion (OPE) data of the superconformal field theory. The latter fact is completely independent on whether the theory admits a weakly-coupled Lagrangian description or not. This has grown into a subfield by itself, sometimes referred to as the ``SCFT/VOA correspondence".

In the current work we are set to contribute to this subfield by combining the latest developments for the (older) Lagrangian technique of supersymmetric localization with the newer ideas on four-dimensional $\cN=2$ superconformal field theories. The original construction of~\cite{Beem:2013sza} was formulated in flat Euclidean space $\R^4$, though the need to also understand it in other backgrounds was repeatedly raised over the years. Here, we study the chiral algebra construction\footnote{Note that we interchangeably use terms ``the chiral algebra'' and ``the VOA'' in this paper, \emph{not} following the terminology where these are different mathematical objects.} on the $S^3\times S^1$ geometry, which, incidentally, is also the geometry relevant for studying the superconformal index, including its Schur limit~\cite{Gadde:2011ik,Gadde:2011uv}. Given that the VOA precisely captures the so-called Schur sector of the theory (\emph{i.e.} states counted by the Schur index), we take it as no coincidence, but rather as an indication that the $S^3\times S^1$ background is in many ways an inherently natural playground for the study of SCFT/VOA correspondence.

As a curtesy to busy readers, and to avoid getting lost in technicalities, we now provide a somewhat more detailed overview of the main results presented in this work.

\paragraph{Chiral algebra on $S^3\times S^1$.}

The original construction~\cite{Beem:2013sza} of the two-dimensional vertex operator algebra as a subsector of the full four-dimensional $\cN=2$ superconformal field theory, relies on a powerful ``abstract" operator approach to the four-dimensional theory, which does not require a Lagrangian description. From such a point of view, any conformal theory can be placed on $S^3\times \R$ via a Weyl transformation (on Weyl invariance of conformal theories, see \emph{e.g.}~\cite{Farnsworth:2017tbz}). If we further close $\R$ to a circle, which we label as $S^1_y$ throughout this paper, we find that the theory breaks supersymmetry, unless we define it in a twisted sector with respect to the R-symmetry. In that case, one manages to preserve half of the flat space supercharges, namely those commuting with $E-R$, where $E$ is the dilatation generator (that generates translations along $\R$ in the $S^3\times \R$ geometry), and $R$ is a Cartan generator of the $SU(2)_R$ symmetry. The surviving supercharges, together with isometries of $S^3$ and the $U(1)_R\times U(1)_r\subset SU(2)_R\times U(1)_r$ R-symmetry subgroup, form a supersymmetry algebra,
\begin{equation}
\label{main_SUSY}
\mathfrak{su}(2|1)_\ell\oplus \mathfrak{su}(2|1)_r\,,
\end{equation}
which is centrally extended by $E-R$. This should be regarded as an $\cN=2$ superconformal symmetry on $S^3\times S^1$. Conformal invariance is crucial as this algebra contains $U(1)_r$ which is only known to be unbroken in conformal theories. A similar algebra without central extension has previously appeared as three-dimensional $\cN=4$ supersymmetry on $S^3$ in~\cite{Dedushenko:2016jxl}. In that case, the algebra admitted central extensions due to the addition of masses and Fayet-Iliopoulos (FI) parameters in Lagrangian theories. In the present context, this $E-R$ should be regarded as a mass-like central charge. One can also introduce additional mass-like central charges by turning on flavor symmetry holonomies around the $S^1_y$ circle (corresponding to flavor fugacities in the index), and they can be regarded as lifts of the three-dimensional mass terms. However, there is no interesting lift of three-dimensional FI parameters, because we study four-dimensional superconformal field theories, and even if they admit Lagrangian descriptions, they do not have any abelian factors in their gauge groups.

The algebra~\eqref{main_SUSY} contains everything required to define the two-dimensional VOA sector. All ingredients of the original construction from~\cite{Beem:2013sza} can be easily translated to this context, and one finds that the VOA is now supported on the torus $S^1_\varphi \times S^1_y$, where $S^1_\varphi\subset S^3$ is a great circle. The labeling of circles stems from the coordinates we use; $S^1_y$ is parametrized by $y \in [0,\beta\ell]$, while $S^3$ is parametrized, just like in~\cite{Dedushenko:2016jxl}, by the fibration coordinates $(\theta,\varphi, \tau)$, in which $S^1_\varphi$ is a circle parametrized by $\varphi$ that sits at $\theta=\pi/2$, where the $\tau$-circle shrinks to a point (see Section~\ref{sec:TheBG} for details). Since we essentially view $S^3$ as $D^2\times S^1_\tau$ with $S^1_\tau$ shrinking at $\partial D^2= S^1_\varphi$, we sometimes refer to $S^1_\varphi$ as the ``boundary'', even though $S^3$ is of course closed.

\paragraph{Localization on $S^{3}\times S^{1}$.}

With the exception of the above generalities, in this paper we focus on Lagrangian $\cN=2$ superconformal field theories in four dimensions. In this case, by employing supersymmetric localization on a rigid background of the form $S^3\times S^1_y$, we explicitly localize a given Lagrangian superconformal field theory, and find that it indeed reproduces the expected two-dimensional VOA on the torus $S^{1}_{\varphi}\times S^{1}_{y}\subset S^{3}\times S^{1}_{y}$, described as a gauged symplectic boson. The symplectic boson VOA is also known as a $\beta-\gamma$ system of weight $\frac12$ and we interchangeably use the two terms.

To derive the two-dimensional VOA on the torus, we first define the appropriate rigid supersymmetric $S^{3}\times S^{1}_y$ background, reproducing the superconformal index. In doing so, we analyze the supersymmetry algebra and classify possible fugacities and their preserved subalgebras. In order to retain the VOA construction, the minimal amount of supersymmetry we ought to preserve is $\mathfrak{su}(1|1)_\ell \oplus \mathfrak{su}(1|1)_r$. We find that we may turn on fugacities preserving an $\mathfrak{su}(1|1)_\ell \oplus \mathfrak{su}(2|1)_r$ subalgebra (which can be further broken to the minimal one by defects). Surprisingly, this goes beyond the fugacity in the Schur limit, which is well-known to be relevant for the chiral algebra construction. Namely, we are allowed to turn on \emph{discrete} fugacities $M,N\in \mathbb{Z}$, where $N$ corresponds to an insertion of $e^{2\pi\ii N (R+r)}$ in the Schur index, while non-zero $M$ modifies the geometry. For non-zero $M$, the $S^3\times S^1_y$ is no longer equipped with a product metric, but rather one rotates $S^3$ by $\Delta\varphi=\Delta\tau=2\pi M$ as we go around the $S^1_y$. As we shall argue, these deformations \emph{do not} affect the VOA construction, but change the complex structure of the torus and affect the boundary conditions (spin structure) upon going around one of the cycles, $S^{1}_{y}$ (see below). The inclusion of $M$ actually does not affect the partition function, but throughout we keep both $M$ and $N$ as generic integers.

We perform localization for the $\cN=2$ vector multiplets; in this case, the two-dimensional theory is determined \emph{indirectly}. As we show, the Yang-Mills action is $Q$-exact, and thus the four-dimensional theory solely localizes to a one-loop determinant piece. Nonetheless, the two-dimensional action can be ``bootstrapped" from the knowledge of the partition function and the four-dimensional propagators, and is given by the \emph{small} $bc$ ghost system on the torus.

Localization for the four-dimensional $\cN=2$ hypermultiplets more straightforwardly reproduces the two-dimensional theory. Namely, the remnant ``classical" piece in the localization precisely reduces to the two-dimensional symplectic boson theory on the boundary torus $S^{1}_{\varphi}\times S^{1}_{y}\subset S^{3}\times S^{1}_{y}$. As alluded above, the discrete label $M$ changes the complex structure of the torus (essentially, $M\neq 0$ differs from $M=0$ by precisely $M$ Dehn twists, or a $T^M$ modular transformation), while $N$ changes the spin structure (periodicity) of symplectic bosons along $S^1_y$.

The localization from four to two dimensions produces almost no overall one-loop determinant, except for a simple Casimir energy factor. The part of it that depends on the gauge holonomy around the $S^1_y$ cancels between the hypermultiplets and vector multiplets as a consequence of conformal invariance. If we turn on flavor holonomies, however, there is a simple leftover term that survives, and describes the mismatch of the four- and two-dimensional path integrals. It is given by
\begin{equation}
q^{-\frac14 \sum_{w_f\in\mathbf{R}_f}\langle w_f,a_f\rangle^2}\,,
\end{equation}
where the sum goes over all weights in a flavor symmetry representation $\mathbf{R}_f$ in which the matter transforms, $a_f$ is the background holonomy in the Cartan of flavor group, and $q=e^{2\pi\ii\tau}$, where $\tau$ is a complex structure of the torus.

\paragraph{Spin structures.}
All operators in the VOA have either integral or half-integral conformal dimensions, as a consequence of the SCFT/VOA correspondence. Thus in order to place the VOA on a torus, one has to pick a particular spin structure. The canonical choice that follows from how we put a four-dimensional theory on $S^3\times S^1_y$ is $(1,0)$, meaning that spinors are anti-periodic (in the NS sector) along $S^1_\varphi$ and periodic (in the R sector) along $S^1_y$.

We mentioned above that the nontrivial $N \mod 2$ amounts to flipping the spin structure of symplectic bosons along $S^{1}_{y}$. This is in fact more general and holds in an arbitrary VOA (arising from a superconformal field theory): starting with the ``standard" NS sector vacuum character, corresponding to $(1,0)$ spin structure, and turning on $N\mod2$ in the four-dimensional background, we arrive at the torus partition function with $(1,1)$ spin structure. Given our general localization, we immediately get the corresponding four-dimensional result, $Z^{(1,1)}$, by setting $N=1$. Of course, it precisely agrees with the NS-NS character.

In two dimensions, we can also change the periodicity of the fields along the other, here $S^{1}_{\varphi}$, cycle to obtain the remaining $(0,1)$ and $(0,0)$ spin structures. However, because the $S^{1}_{\varphi}$-cycle is contractible in $S^{3}$, we cannot continuously change the periodicity of the four-dimensional fields along it; thus, we have to make $S^1_\varphi$ non-contractible. To do so in a physically sensible way, we have to introduce a surface defect at $\theta=0$, \emph{i.e.} describing $S^{3}$ as a circle fibration over the disk $D^{2}$, $\theta=0$ is at the origin of $D^{2}$, and the defect extends in the $\tau$ and $y$ directions.

\paragraph{The R-symmetry defect.}
The surface defect that can switch the $S^1_\varphi$ spin structure should act similarly to the $e^{2\pi\ii(R+r)}$ monodromy that was able to alter the $S^1_y$ spin structure. Namely, we first define the symmetry interface that does the transformation $e^{2\pi\ii(R+r)}$ on a theory -- we call it the canonical R-symmetry interface. Then, we claim that the relevant surface defect should sit at the boundary of such an interface. If we place the surface defect at $\theta=0$, with the canonical R-symmetry interface ending on it, we get what we want: operators that have half-integral dimension in the VOA will obtain the opposite periodicity around $S^1_\varphi$. In this way we can achieve the $(0,1)$ and $(0,0)$, or NS-R and R-R spin structures from four dimensions.

Such surface defects should exist in general superconformal field theories, but they are by no means unique. In fact, they correspond to the Ramond sector modules of the VOA, and there might be many of them. For Lagrangian theories, we define in the main text and in Appendix~\ref{app:defect} the simplest possible pair of such defects. Because $e^{2\pi\ii(R+r)}$ acts trivially on the vector multiplets, we define defects that only directly couple to hypermultiplets by changing their asymptotic behavior in the vicinity of the defect. This is closely related to the monodromy defect considered in~\cite{Cordova:2017mhb}, and we elaborate more on their relation in Section~\ref{sec:R_defect}. We further generalize the localization answer to include this defect at $\theta=0$. In this way, we end up with four partition functions, $Z^{(1,0)}$, $Z^{(1,1)}$, $Z^{(0,1)}$, $Z^{(0,0)}$, labeled by the spin structure, where the first one is the usual Schur index and the second one is the ``modified Schur index" considered by Razamat~\cite{Razamat:2012uv}.

\paragraph{The two-dimensional theory.}

Once we have the two-dimensional action of gauged symplectic bosons, it is straightforward to study various of its properties, such as the proper integration cycle, the two-point functions in all spin structures, \emph{etc}. One interesting observation that we make is that in the presence of flavor holonomies, -- which appear as mass-like central charges in the supersymmetry algebra, -- vertex operators charged under the flavor symmetries fail to remain holomorphic. The sector that remains holomorphic is formed by flavor-neutral operators. At the level of the VOA it corresponds to the well-known operation of screening. Somewhat formally, it expands the class of vertex algebras we might study. One simple example we describe is a free hypermultiplet: turning on its $U(1)_F\subset SU(2)_F$ flavor holonomy screens the free symplectic boson VOA to a slightly simpler VOA described by a pair chiral bosons.

\paragraph{Modularity of the Schur index.} 

It has long been suggested that the four-dimensional Schur index observes modular properties (see for example~\cite{Spiridonov:2012ww,Razamat:2012uv}).\footnote{The study of modular properties in dimensions three and four have recently attracted attention, see for instance~\cite{Cheng:2018vpl,Shaghoulian:2016gol}.} Naively, one expects that modular properties are related to an exchange of the ``thermal cycle" with the Hopf fiber of the index. Of course, with the advent of the work~\cite{Beem:2013sza}, the identification of the Schur index with some vacuum character of a two-dimensional chiral algebra immediately suggest that the Schur index ought to be part of a modular vector. Indeed, in the work~\cite{Beem:2017ooy}, differential equations observing modular properties were derived, and it was suggested, that the vacuum character -- or Schur index -- is a solution of them. 

Nevertheless, a four-dimensional understanding of the modular properties of the two-dimensional character remains an unresolved problem. For Lagrangian theories, our discussion of the spin structures of the torus partition function and its relation to the four-dimensional Schur index, gives us immediately the $T$-modular transformation of the latter. In particular, we explicitly find the following general result
\beaa
Z^{(1,0)} (\tau + 1, a_f) & \ \propto \ Z^{(1,1)} (\tau , a_f) \,,\qquad & Z^{(1,1)} (\tau + 1, a_f) & \ \propto \ Z^{(1,0)} (\tau , a_f) \,, \\
Z^{(0,0)} (\tau + 1,a_f) & \ \propto \ Z^{(0,0)} (\tau,a_f) \,,\qquad & Z^{(0,1)} (\tau + 1,a_f) & \ \propto \ Z^{(0,1)} (\tau,a_f) \,.
\eeaa

In order to shed light on the action of the modular $S$-transformation, we introduce novel (formal) partition functions labeled by two additional indices, $m,n\in\mathbb{Z}$, $Z_{(m,n)}^{(\nu_1,\nu_2)}$. They are defined as the partition function in given spin structure $(\nu_1, \nu_2)$ but now with modified contour $\mathbb{T}_{(m,n)}$ of the holonomy integral in the localization formula, labeled by the integers $m$ and $n$. The upshot of introducing this extended, infinite set of partition functions is that they exhibit a simple behavior under modular transformations.\footnote{We refer to Section~\ref{sec:taste} for more details, including some comments on the convergence of these objects upon turning off flavor fugacities.} For instance, we find that
\beaa
Z_{(m,n)}^{(\nu_1,\nu_2)} \left( -\frac{1}{\tau}, \frac{a_f}{\tau} \right) & \ \propto \  Z_{(-n,m)}^{(\nu_2,\nu_1)} \left( \tau, a_f \right) \,,\\
Z_{(m,n)}^{(\nu_1,\nu_2)} \left(\tau+1, a_f  \right) & \ \propto \  Z_{(m+n,n)}^{(\nu_1,\nu_2+\nu_1)} \left( \tau, a_f \right) \,.\\
\eeaa
Thus, we suggest that the objects $Z_{(m,n)}^{(\nu_1,\nu_2)} $ furnish an (infinite-dimensional projective) representation of $SL(2,\Z)$, and given the relation to two-dimensional chiral algebras, we expect it to truncate, \emph{i.e.} there are relations among them, and a -- possibly finite -- set of them transform as some modular vector~\cite{Beem:2017ooy}. Physically, the remanent independent objects $Z_{(m,n)}^{(\nu_1,\nu_2)}$ are expected to correspond to partition functions in the presence of general defects (corresponding to non-trivial modules of the two-dimensional chiral algebra~\cite{Beem:2013sza,Cordova:2016uwk,Cordova:2017ohl,Cordova:2017mhb}).

Finally, we discuss two simple examples, and explicitly observe that the $S$-transformation of the Schur index exhibits logarithmic behavior, which is in accordance with the expected solutions to the modular differential equations derived in~\cite{Beem:2017ooy}.

\paragraph{Relation to the (flat) $\Omega$-background $\R^{2}_{\epsilon}\oplus \R^{2}$.}

For a while, it has been suggested~\cite{OldIdea} (see~also~\cite{Costello:2018fnz,Costello:2018zrm}) that the two-dimensional chiral algebra can be obtained by an $\Omega$-deformation \cite{Nekrasov:2002qd,Nekrasov:2003rj} of the holomorphic-topological twist, introduced by Kapustin in~\cite{Kapustin:2006hi} (see~\cite{Oh:2019bgz,Jeong:2019pzg} for recent results to this end). Indeed, our analysis of the $S^{3}\times S^{1}$ background, with the two-dimensional theory arising on the torus, $\mathbb{T}^{2}\subset S^{3}\times S^{1}$, suggests that the (flat space) $\Omega$-background is hidden as some ``local" (and decompactified) version near the torus.\footnote{Indeed, in the three-dimensional $\cN=4$ analogue of our setup, there are alternative approaches, one relying on the $S^{3}$ localization~\cite{Dedushenko:2016jxl,Dedushenko:2017avn,Dedushenko:2018icp}, another introducing an $\Omega$-background $\R_{\epsilon}^{2}\oplus \R$~\cite{Bullimore:2016hdc}, and for superconformal $\cN=4$ theories, there is a third approach~\cite{Chester:2014mea,Beem:2016cbd}, which is more closely related to the operator approach of~\cite{Beem:2013sza} in four-dimensions. All of these determine a one-dimensional theory quantizing the Coulomb and Higgs-branch chiral rings.} In Section~\ref{sec:omega}, we comment on this connection in four dimensions, determining an expansion of the $S^{3}\times S^{1}$ background in the vicinity of the torus and thereby explicitly obtain the flat $\Omega$-background $\R^{2}_{\epsilon}\oplus\R^{2}$. The theory then effectively localizes to the tip of the $\Omega$-background, $\R^{2}_{\epsilon}$, giving the symplectic boson action in the remaining (flat) two-dimensional space. We remark here that our background seems to be in accordance with the recent results in~\cite{Oh:2019bgz,Jeong:2019pzg}.


\vspace{.2 in}

The remainder of this paper is organized as follows. We start in Section~\ref{sec:bg} by setting the stage, and define the supersymmetric background on $S^{3}\times S^{1}$. We further discuss the possible fugacities including their preserved subalgebra, and argue that we may slightly deform away from the Schur limit while still preserving the VOA construction. Furthermore, we introduce the notion of spin structures on the torus to the game and discuss how they are realized from a four-dimensional point of view, which marks the inception of the canonical R-symmetry interface and surface defect. In Section~\ref{sec:loc}, we explicitly localize the $S^{3}\times S^{1}$ partition function onto the gauged symplectic bosons in two-dimensions. This includes the results for the (four) different spin structures, and the R-symmetry surface defect. We point out various subtleties along the way. In Section~\ref{sec:2dthy}, we consequently discuss various aspects of the resulting two-dimensional theory. This includes the determination of its integration cycle, the propagators, as well as interpreting flavor fugacities as screening charges in two dimensions. In Section~\ref{sec:taste}, we then discuss some implications of our results towards understanding the modular properties of the Schur index. Finally, in Section~\ref{sec:omega}, we mention the connection to the flat $\Omega$-background. We close in Section~\ref{sec:disc} with a brief discussion and mention some work for the future. We collect some more technical details in Appendices~\ref{sec:Notation}--\ref{app:omega}.


\vspace{.2 in}


\emph{\textbf{Note added:} During the final stages of this project, the paper~\cite{Pan:2019bor} appeared, which has partial overlap with some results in our work. Additionally, the recent papers~\cite{Oh:2019bgz,Jeong:2019pzg} have some overlap with our Section~\ref{sec:omega}}.


\section{The Index, the Background, and the Algebra}
\label{sec:bg}

In this section, we provide the necessary details on the constructions explored in the remainder of this paper. We start by recalling the definition of the superconformal index of four-dimensional $\cN=2$ superconformal field theories, and choose a particular representation of it convenient for our purposes. We then proceed to describe the rigid background in four-dimensional $\cN=2$ conformal supergravity that is appropriate for studying the index. Such a background of course ought to have the topology of $S^3\times S^1_y$ (we denote the Euclidean time direction by $y$). However, we may also replace $S^1_y$ by an interval or $\R$. In the latter case, it can be obtained by a Weyl transformation from flat space. 

We further explicitly describe the most general fugacities in the superconformal index compatible with the supercharge $\cQ^H$, which is required to be preserved for the vertex operator algebra (VOA) construction, and explain in detail how the VOA construction works on $S^3\times S^1_y$. In particular, we obtain a VOA on the torus $\mathbb{T}^2\subset S^3\times S^1_y$. Interestingly, such a specialization of fugacities goes slightly beyond the well-known Schur limit of the superconformal index. Namely, we add a discrete parameter that switches the spin structure of the torus (along $S^1_y$) at the level of the VOA. Furthermore, we introduce a surface defect which switches the spin structure along the other cycle of the torus -- such defects correspond to Ramond sector modules for the VOA.

\subsection{Different representations of the superconformal index}
The superconformal index in four-dimensional $\cN=2$ superconformal field theories was introduced in~\cite{Kinney:2005ej,Romelsberger:2005eg,Romelsberger:2007ec}. In its most basic form, and given a choice of the supercharge $\cQ$, it is defined as
\be
\label{index}
\cI(\mu_i) \ = \  {\rm Tr}_{\cH_{S^3}} (-1)^F e^{-\mu_i T_i} e^{-L\delta}\,,
\ee
where the trace is taken over the Hilbert space of radial quantization $\cH_{S^3}$, $F$ is the fermion number, and we introduced the definition
\be
 \delta \ \equiv \ 2\left\{\cQ, \cQ^\dagger\right\} \,.
\ee
Furthermore, $T_i$ is a maximal set of mutually commuting generators that necessarily also commute with $\cQ$. Mirroring the choice in~\cite{Gadde:2011uv}, we pick the particular supercharge $\cQ=\tilde\cQ_{1\dot-}$ to define the index. Thus, we fix 
\begin{equation}
\delta \ = \ \tilde\delta_{1\dot -} \ = \ E-2j_2-2R+r\,,
\end{equation}
and the maximal set of commuting generators consists of the following anti-commutators
\begin{align}
\delta_{1-} 		& \ = \ E-2j_1-2R-r\,,\\
\delta_{1+} 		& \ = \ E+2j_1-2R-r\,,\\
\tilde\delta_{2\dot+} 	& \ = \ E + 2j_2+2R+r \,.
\end{align}
A key property of the superconformal index is its independence of $L$. This is due to the pairwise cancellations of non-zero modes of~$\tilde\delta_{1\dot -}$. Thus, we may shift $L$, as long as we are preserving convergence of~\eqref{index}. In particular, shifting $L$ by various linear combinations of chemical potentials $\mu_i$ is equivalent to redefining $T_i \to T_i + c \delta$, where $c$ is some number. Because such redefinitions do not affect the answer, we may write the following equivalent formulae for the index,
\begin{align}
\label{ind1}\cI(\rho,\sigma,\tau)& \ = \ {\rm Tr}(-1)^F \rho^{\frac12 (E-2j_1-2R-r)}\sigma^{\frac12 (E+2j_1-2R-r)}\tau^{\frac12 \tilde\delta_{2\dot+}}e^{-L\tilde\delta_{1\dot-}}\,,\\
\label{ind2}\cI(p,q,t)& \ = \ {\rm Tr}(-1)^F p^{\frac12 (E+2j_1-2R-r)}q^{\frac12 (E-2j_1-2R-r)}t^{R+r}e^{-L\tilde\delta_{1\dot-}}\,,\\
\label{ind3}\cI(p,q,t)& \ = \ {\rm Tr}(-1)^F p^{j_2+j_1-r}q^{j_2-j_1-r}t^{R+r}e^{-L\tilde\delta_{1\dot-}} \,.
\end{align}
Passing from~\eqref{ind1} to~\eqref{ind2} involves the change of variables $p=\tau\sigma$, $q=\tau \rho$, $t=\tau^2$, and a shift of $L$, while passing from~\eqref{ind2} to~\eqref{ind3} is accomplished solely by shifting $L$.

Notice that even though the answer is $L$-independent, $L$ also plays the role of a regulator, regularizing the trace over an infinite-dimensional Hilbert space. The importance of this factor differs in the alternative representations of the index. For example, since the factor $(\rho\sigma\tau)^{\frac12 E}$ ensures convergence,  as long as $|\rho\sigma\tau|<1$ (with some extra assumptions), we may safely put $L=0$ in~\eqref{ind1}. Similarly, we can set $L=0$ in~\eqref{ind2} as long as $|pq|<1$. However, we \emph{cannot} put $L=0$ in~\eqref{ind3}, as none of the other factors can obviously serve as a regulator.

In the following we will set $L=0$ and work with the representation of the superconformal index given in~\eqref{ind2}. We further perform the following change of variables
\begin{align}
\label{pqtbetazetagamma1}
(pq)^{1/2}& \ = \ e^{-\beta}\,,\\
\label{pqtbetazetagamma2}
\frac{p}{q}& \ = \ e^{2\ii\beta\zeta}\,,\\
\label{pqtbetazetagamma3}
\frac{t}{(pq)^{1/2}}& \ = \ e^{\ii\beta\gamma}\,,
\end{align}
upon which the index takes the following form
\begin{equation}
\label{index_PI}
\cI(\beta, \zeta, \gamma) \ = \ {\rm Tr}_{\cH_{S^3}} (-1)^F e^{-\beta E} e^{\beta R} e^{2\ii\beta\zeta j_1} e^{\ii\beta\gamma(R+r)}\,.
\end{equation}
Indeed, this representation is the most convenient for the path integral interpretation; We put the theory on a space $S^3$, and the factor $e^{-\beta E}$ suggests that we have to evolve it for the Euclidean time $\beta$. The additional factors $e^{\beta R} e^{2\ii\beta\zeta j_1} e^{\ii\beta\gamma(R+r)}$ further imply that afterwards we have to perform various symmetry transformations on the system, and the trace means that we close the time direction into a circle $S^{1}_y$. Lastly, $(-1)^F$ implies that fermions ought to have periodic boundary conditions (up to the twists introduced by $e^{\beta R} e^{2\ii\beta\zeta j_1} e^{\ii\beta\gamma(R+r)}$). 

Thus, we have to put the theory on a Euclidean space given by $S^3\times S^1_y$, defined in a twisted sector, where upon going once around $S^1_y$, we perform the R-symmetry transformations $e^{\beta R} e^{\ii\beta\gamma(R+r)}$, and rotate $S^3$ by $e^{2\ii\beta\zeta j_1}$. The latter means that the geometry can be constructed by first taking $S^3\times I$ with the product metric, where $I$ is an interval, and then identifying the two boundary three-spheres by a rotation generated by $e^{2\ii\beta\zeta j_1}$. It is straightforward to describe a background satisfying these criteria, and we do so in the next subsection. Quite importantly, this background should preserve both $U(1)_R\subset SU(2)_R$ and $U(1)_r$ R-symmetries, simply because we work in a twisted sector with respect to both $R$ and $R+r$. 

On the contrary, the background does not have to preserve the supercharge $\tilde\cQ_{1\dot -}$ used in the original definition of the index~\eqref{index}. Once we have arrived at the expression~\eqref{index_PI} that concerns a particular way to count states in the radial quantization Hilbert space $\cH_{S^3}$, it no longer matters what supercharge we started with in~\eqref{index}. In what follows, we will investigate what supersymmetry is preserved by such a background for various values of the fugacities in~\eqref{index_PI}.

\subsection{The Background}\label{sec:TheBG}

Using off-shell supergravity to construct rigid supersymmetric backgrounds has become a standard technique in supersymmetry literature~\cite{Festuccia:2011ws}.\footnote{This technique goes back to as early as~\cite{Karlhede:1988ax,Galperin:1990nm,Johansen:1994aw}, in the context of topologically twisted theories. Of course, it is also possible to define (rigid) supersymmetry on the $S^3\times \R$ background using other methods~\cite{Sen:1989bg}. Nevertheless, we are going to follow the general approach of~\cite{Festuccia:2011ws}.} The relevant supergravity theory for our problem is given by the four-dimensional $\cN=2$ off-shell conformal supergravity of~\cite{deWit:1979dzm,deWit:1980lyi,deWit:1984rvr} based on the standard Weyl multiplet.\footnote{General rigid supersymmetric backgrounds have been analyzed in~\cite{Klare:2013dka} within the setting of four-dimensional conformal supergravity.} It has been successfully used to put general four-dimensional $\cN=2$ gauge theories on the ellipsoid in~\cite{Hama:2012bg,Hosomichi:2016flq}, in which case the theory does not have to be conformal. We follow the conventions of reference~\cite{Hosomichi:2016flq} (see also Appendix~\ref{sec:Notation}). As we will see, theories that are not conformal quantum-mechanically appear to break supersymmetry in our $S^3 \times S^1$ background, and therefore, for the scope of this paper, we focus solely on conformal theories.\footnote{An unconventional $S^{3} \times S^{1}$ background that would allow to define the Schur index for non-conformal $\cN=2$ theories is being studied in a long-awaited work~\cite{AreYouEverGonnaPublishThis}.}

We parametrize the space $S^3 \times S^1_y$ by variables $(\theta, \varphi, \tau, y)$, where $\theta\in[0,\pi/2]$, $\varphi\in[-\pi,\pi]$ and $\tau\in[0,2\pi]$ are ``fibration'' coordinates for the three-sphere $S^3$ of radius $\ell$, and $y\in[0,\beta\ell]$ parametrizes the circle $S^1_y$ of circumference $\beta \ell$. The corresponding metric then reads
\be\label{metric}
\dd s^2 \ = \ \dd y^2 + \ell^2 \left[\dd \theta^2 + \sin^2\theta \left(\dd\varphi+\frac{\zeta}{\ell} \dd y\right)^2 + \cos^2\theta \left(\dd\tau + \frac{\zeta}{\ell}\dd y\right)^2\right] \,,
\ee
where we further introduced the additional deformation $\zeta$, which is related to the ``standard" fugacities $p$, $q$ and $t$ via equations~\eqref{pqtbetazetagamma1}~--~\eqref{pqtbetazetagamma3}. Notice that for $\zeta=0$, this is just a product of a round $S^3$ with $S^1_y$, while non-zero $\zeta$ introduces a twist, \emph{i.e.} as we go around the $S^1_y$, the  three-sphere is rotated by $\Delta\varphi=\Delta\tau=\beta\zeta$. Alternatively, we could write our metric in terms of the coordinates $\tilde\varphi=\varphi+\zeta y/\ell$ and $\tilde\tau=\tau + \zeta y/\ell$, in which it would simply become a product metric, where gluing a patch $y\in (0,\beta\ell)$ into a circle, \emph{i.e.}, identifying $y=0$ with $y=\beta\ell$, would involve a rotation (``twist") by $\Delta\varphi=\Delta\tau=\beta\zeta$. This rotation is precisely generated by $j_1$, which appears in the definition of the superconformal index in equations~\eqref{ind1}~--~\eqref{ind3}. We prefer to use the coordinates $(\theta, \varphi, \tau, y)$, in which the metric takes the form~\eqref{metric}, and up to the R-symmetry twists introduced in the previous subsection all the variables of the theory are periodic in the $y$ direction. We conveniently work with the following vierbein
\beaa
\label{fframe}
e^1 & \ = \ \ell \dd\theta\,, \qquad 	& e^2 &  \ = \ \ell \sin\theta \left(\dd\varphi+\frac{\zeta}{\ell} \dd y\right)\,,\cr
e^4 & \ = \ \dd y\,, \qquad 			& e^3 &  \ = \ \ell \cos\theta \left(\dd\tau + \frac{\zeta}{\ell}\dd y\right) \,,
\eeaa
in which the spin connection has the following non-vanishing components:
\beaa
\Omega_\varphi^{21} & \ = \ -\Omega_\varphi^{12} \ = \ \cos\theta\,, \qquad & \Omega_\tau^{13} & \ = \ -\Omega_\tau^{31} \ = \ \sin\theta\,,\cr
\Omega_y^{13} & \ = \ -\Omega_y^{31} \ = \ \frac{\zeta}{\ell}\sin\theta\,, \qquad & \Omega_y^{21} & \ = \ -\Omega_y^{12} \ = \ \frac{\zeta}{\ell}\cos\theta\,.
\eeaa

We define rigid supersymmetry for four-dimensional $\cN=2$ superconformal field theories by coupling to the $\cN=2$ standard Weyl supergravity multiplet (in the rigid limit). Its component fields are 
\be
\left( g_{\mu\nu}, \ (V_{\mu})^{A}{}_{B} , \ \tilde{V}_\mu, \  \bar{T}_{\mu\nu}, \ T_{\mu\nu}, \ M , \  \psi_{\mu A}, \ \bar{\psi}_{\mu A} , \ \eta_A, \ \bar\eta_A \right) \,,
\ee
where $g_{\mu\nu}$ is the metric, $(V_\mu)^A{}_B$ is a gauge field for the $SU(2)_R$ symmetry, $\tilde{V}_\mu$ a gauge field for the $U(1)_r$ symmetry, $\bar{T}_{\mu\nu}$ ($T_{\mu\nu}$) is an (anti-)self-dual tensor, $M$ a scalar, $\psi_{\mu A}$ ($\bar{\psi}_{\mu A}$) is an (anti-)chiral gravitino, and finally $\eta_A$ ($\bar\eta_A$) is an (anti-)chiral spinor. Then, supersymmetry is determined by nontrivial chiral spinors $\xi_{\alpha A}$ and anti-chiral spinors $\bar\xi_A^{\dot \alpha}$, where $A=1,2$ is the $SU(2)_R$ index, which together with appropriate background supergravity fields allow for the vanishing of the supersymmetry variations of the fermions $\bar{\psi}_{\mu A}$, $\psi_{\mu A}$, $\bar\eta_A$, and $\eta_A$. This implies that we ought to solve the following system of conformal Killing spinor equations (see Appendix~\ref{sec:Notation} for relevant notation)
\begin{align}
\label{KSE1}
\cD_\mu \xi_A + T^{\nu\rho}\sigma_{\nu\rho}\sigma_\mu\bar\xi_A& \ = \ -\ii\sigma_\mu\bar{\xi}'_A\,,\\
\label{KSE2}
\cD_\mu \bar\xi_A + \bar{T}^{\nu\rho}\bar\sigma_{\nu\rho}\bar\sigma_\mu\xi_A& \ = \ -\ii\bar\sigma_\mu\xi'_A\,,\\
\label{KSE3}
\sigma^\mu\bar\sigma^\nu\cD_\mu \cD_\nu\xi_A + 4\cD_\rho T_{\mu\nu}\sigma^{\mu\nu}\sigma^\rho \bar\xi_A & \ = \  M\xi_A\,,\\
\label{KSE4}
\bar\sigma^\mu \sigma^\nu\cD_\mu \cD_\nu\bar\xi_A + 4\cD_\rho \bar{T}_{\mu\nu}\bar\sigma^{\mu\nu}\bar\sigma^\rho\xi_A & \ = \  M\bar\xi_A\,,
\end{align}
where the covariant derivatives are defined as follows
\begin{align}
\label{covxi1}
\cD_\mu \xi_A & \ = \ \left( \partial_\mu + \frac14 \Omega_\mu^{ab}\sigma_{ab} \right)\xi_A +i\xi_B (V_\mu)^B{}_A -\ii \tilde{V}_\mu \xi_A\,,\\
\label{covxi2}
\cD_\mu \bar\xi_A & \ = \ \left( \partial_\mu + \frac14 \Omega_\mu^{ab}\bar\sigma_{ab} \right)\bar\xi_A +\ii \bar\xi_B (V_\mu)^B{}_A +\ii \tilde{V}_\mu \bar\xi_A\,.
\end{align}
For the background in consideration, we require the $U(1)_r$-symmetry to be preserved, and thus $T_{\mu\nu}$ and $\bar{T}_{\mu\nu}$ necessarily have to vanish, for they have nontrivial $U(1)_r$-charges given by $\pm2$. Thus, we have to set
\begin{equation}
T_{\mu\nu}\ = \ \bar{T}_{\mu\nu} \ = \ 0\,.
\end{equation}
Then, in this case, using~\eqref{KSE1} and~\eqref{KSE2}, equations~\eqref{KSE3} and~\eqref{KSE4} can equivalently be written as\footnote{Actually, these equations follow more straightforwardly directly from the supersymmetry conditions of $\eta_A$ and $\bar{\eta}_A$.}
\begin{align}
\label{simpleM1}
(3M+R)\xi_A 		& \ = \  2\ii\sigma^{\mu\nu}(V_{\mu\nu})^B{}_A\xi_B+4\ii\sigma^{\mu\nu}\tilde{V}_{\mu\nu}\xi_A\,,\\
\label{simpleM2}
(3M+R)\bar{\xi}_A	& \ = \ 2\ii\bar\sigma^{\mu\nu}(V_{\mu\nu})^B{}_A\bar\xi_B-4\ii\bar\sigma^{\mu\nu}\tilde{V}_{\mu\nu}\bar\xi_A\,.
\end{align}
Here, $(V_{\mu\nu})^B{}_A$ and $\tilde{V}_{\mu\nu}$ denote the field strengths of the respective R-symmetry gauge fields, and $R=\frac{6}{\ell^2}$ is the scalar curvature of $S^3\times S^1$. We pick a solution for which both the left and the right hand sides are zero. Namely, we have
\begin{equation}
M \ = \ -\frac{R}{3}\,,
\end{equation}
while the R-symmetry gauge fields have vanishing field strengths,
\be
(V_{\mu\nu})^B{}_A \ = \ \tilde{V}_{\mu\nu} \ = \ 0 \,.
\ee
However, we may of course turn on holonomies for $(V_\mu)^B{}_A$ and $\tilde{V}_\mu$. For reasons that will become clear soon, we only consider a holonomy for $R+r$; in particular, we set
\begin{align}
\label{VN2}
\tilde{V}_{\mu} \dd x^{\mu} & \ = \  -\frac{\alpha}{2\ell}\dd y\,,\\
\label{tildeVN2}
(V_{\mu})^{B}{}_{A} \dd x^{\mu} & \ = \  \frac{\alpha}{2\ell}\dd y \, (\tau_3)^{B}{}_{A} \,,
\end{align}
where $\tau_3$ is the third Pauli matrix. Notice that by performing non-single-valued $(R+r)$ gauge transformations with parameter $e^{\ii c y}$, for some $c\in\R$, we can arbitrarily shift the holonomy $\alpha$, and in particular make it zero. The latter is possible at a cost of introducing an extra $(R+r)$ monodromy (``twisting'' in the sense of twisted sectors) around $S^1_y$. Recall that $\gamma$ in~\eqref{index_PI} parametrizes such a monodromy, \emph{i.e.} the gluing cocycle for the $(R+r)$-symmetry flat bundle: as we go once around $S^1_y$, we identify fibers by an $e^{\ii\gamma(R+r)}$ rotation. We can reverse the logic, and completely gauge away $\gamma$, and instead keep the holonomy $\alpha$ generic. We find the latter more convenient, and adhere to such a convention below. We could try to similarly gauge away the $e^{\beta R}$ twisted sector in~\eqref{index_PI}, however this would introduce an imaginary background holonomy (manifesting that our background is non-unitary), which we find slightly inconvenient. Therefore, we work in a twisted sector with respect to the $R$ symmetry only. For any variable $X$ in our theory -- be it a field or a Killing spinor -- the periodicity is determined by its R charge, \emph{i.e.}
\begin{equation}
\label{twisted_sector}
X(y + \beta\ell)  \ = \  e^{-\beta R}X(y)\,.
\end{equation}

Now, solving the remaining equations~\eqref{KSE1} and~\eqref{KSE2}, and throwing away half of the solutions that cannot satisfy the twisted sector periodicity (due to having the wrong sign in front of $\frac{y}{2\ell}$ in the exponent), we find the following set of Killing spinors
\beaa
\label{KS_sols}
\xi_1 & \ = \  e^{-\frac{y}{2\ell}-\ii\alpha\frac{y}{\ell}}e^{\frac{\ii}2 \tau_1 \theta} e^{\frac{\ii}2 \tau_3 (\varphi+\tau+ 2\zeta \frac{y}{\ell})}\epsilon\,,\cr
\bar\xi_1& \ = \ e^{-\frac{y}{2\ell}}e^{-\frac{\ii}2 \tau_1 \theta} e^{\frac{\ii}2 \tau_3 (\varphi-\tau)}\bar\epsilon\,,\cr
\xi_2& \ = \ e^{\frac{y}{2\ell}}e^{-\frac{\ii}2 \tau_1 \theta} e^{\frac{\ii}2 \tau_3 (\varphi-\tau)}\eta\,,\cr
\bar\xi_2& \ = \  e^{\frac{y}{2\ell}+\ii\alpha\frac{y}{\ell}}e^{\frac{\ii}2 \tau_1 \theta} e^{\frac{\ii}2 \tau_3 (\varphi+\tau+2\zeta \frac{y}{\ell})}\bar\eta \,,
\eeaa
where $\epsilon, \bar\epsilon, \eta, \bar\eta$ are constant two-component spinors. Part of these can still be broken due to the incorrect periodicity at generic $\alpha$ and $\zeta$, which will be discussed in more detail below.

Setting the deformations $\alpha=\zeta=0$, and if we replace $S^1$ by $\R$, we can think of these solutions as the flat space solutions conformally mapped to the cylinder $S^3\times \R$. This perspective allows us to easily identify the map between the conformal supercharges and the parameters in $\epsilon, \bar\epsilon, \eta, \bar\eta$ as follows
\beaa
\cQ^1_\alpha  & \ \ \longleftrightarrow \ \ \epsilon_\alpha \,,\\
\tilde{\cQ}_{2\dot{\alpha}}& \  \ \longleftrightarrow \ \ \bar\epsilon^{\dot{\alpha}} \,,\\
\tilde{\cS}^{2\dot{\alpha}} & \ \ \longleftrightarrow \ \ \eta^{\dot{\alpha}} \,,\\
\cS_1^\alpha & \ \ \longleftrightarrow \ \ \bar\eta_\alpha \,.
\eeaa
These are the eight out of sixteen four-dimensional $\cN=2$ conformal supercharges that commute with the combination $E-R$, and thus their Killing spinors obey the twisted sector periodicity. The remaining supercharges, which are not commuting with $E-R$, acquire the wrong periodicity in the $y$ direction and are broken on $S^3 \times S^1$. 

One finds that the remaining eight preserved supercharges form an $\mathfrak{su}(2|1)_\ell\oplus \mathfrak{su}(2|1)_r$ superalgebra, which is centrally extended by the element $E-R$, with nontrivial anti-commutation relations given by
\beaa
\{\cQ_\alpha^1, \cS_1^\beta\} & \ = \ \frac12 \delta_\alpha^\beta (E-R) + \cM_\alpha{}^\beta -\delta_\alpha^\beta\,\frac{r+R}2\,,\\
\{\tilde{\cS}^{2\dot{\alpha}},\tilde{\cQ}_{2\dot{\beta}}\} & \ = \ \frac12 \delta^{\dot{\alpha}}_{\dot{\beta}}(E-R) +\cM^{\dot \alpha}{}_{\dot \beta} + \delta^{\dot \alpha}_{\dot{\beta}}\,\frac{r-R}2\,.
\eeaa
Here, $\cM_\alpha{}^\beta$ and $\cM^{\dot \alpha}{}_{\dot{\beta}}$ are the generators of left- and right-rotations. In our conventions, $\cQ^1_\alpha$ and $\cS_1^\alpha$ generate the left factor, $\mathfrak{su}(2|1)_{\ell}$, and $\tilde\cS^{2\dot{\alpha}}$ together with $\tilde\cQ_{2\dot{\beta}}$ generate the right factor, $\mathfrak{su}(2|1)_{r}$. Notice that if we turn on non-vanishing deformations $\alpha\neq 0$ or $\zeta\neq 0$, we further break part of the $\mathfrak{su}(2|1)_\ell\oplus \mathfrak{su}(2|1)_r$ supersymmetry. We will investigate this below.

Now, we proceed to present the four-dimensional $\cN=2$ Yang-Mills and matter actions in our background.\footnote{Here, we mostly follow reference~\cite{Hosomichi:2016flq}, except that we redefine their $\bar\phi$ to $-\bar\phi$ to have a more natural-looking reality condition for $\phi$.} The four-dimensional $\cN=2$ vector multiplet consists of a gauge field $A_\mu$ with field strength $F_{\m\n}$, a complex scalar $\phi$, (anti)chiral gaugini ($\bar\lambda_A^{\dot{\alpha}}$) $\lambda_{A \alpha}$, and the auxiliary $SU(2)_R$-triplet $D_{AB}$. The $\cN=2$ Yang-Mills action coupled to the rigid standard Weyl multiplet is then given by~\cite{deWit:1979dzm,deWit:1980lyi,deWit:1984rvr}
\begin{align}\label{YM_act}
\cL_{\rm YM}& \ = \ 
\frac{1}{g_{\rm YM}^2}{\rm Tr}\bigg(\frac12 F_{\mu\nu}F^{\mu\nu} + 4 \cD_\mu \bar\phi \cD^\mu\phi +\frac{2R}{3} \bar\phi \phi -2\ii \lambda^A \sigma^\mu \cD_\mu \bar\lambda_A+ 2\lambda^A [\bar\phi,\lambda_A] + 2\bar\lambda^A[\phi,\bar\lambda_A]\nonumber\\ 
&\hspace{2.5 cm}+4[\phi,\bar\phi]^2-\frac12 D^{AB}D_{AB} \bigg) 
+ \frac{\ii\theta}{32\pi^2} {\rm Tr}\Big(\varepsilon^{\mu\nu\lambda\rho}F_{\mu\nu}F_{\lambda\rho} \Big) \,.
\end{align}
Here, $g_{\rm YM}$ is the four-dimensional Yang-Mills coupling, $R$ the curvature of the background, $\theta$ the $\theta$-angle, and the covariant derivatives are understood (see \emph{e.g.}~\eqref{covxi1} and~\eqref{covxi2}).

The four-dimensional $\cN=2$ hypermultiplet $(q_{AI}, \psi_{I\alpha}, \bar\psi_I{}^{\dot \alpha}, F_{\check{A}I})$ contains the scalars $q_{AI}$, the fermions $\psi_{I\alpha}$, $\bar\psi_{I\alpha}$ as well as auxiliary fields $F_{\check{A}I}$. Here $I, J,\dots$ are flavor indices of the hypermultiplets, and $\varepsilon^{IJ}$ is their pseudoreal structure. For instance, for a single free hypermultiplet $I, J,\dots=1,2$, and the flavor symmetry is $SU(2)_F$. More generally, $2N_f$ half-hypermultiplets transform in the fundamental representation of the flavor symmetry group $USp(2N_{f})_F$, part of which can be gauged by the dynamical vector multiplets. After gauging $G\subset USp(2N_f)_F$, the fundamental representation of $USp(2N_f)_F$ becomes some representation $\rR$ of $G$, which is pseudoreal in general, and for full hypermultiplets is of the form $\rR\cong \cR\oplus \bar\cR$, in which case we say that hypermultiplets are valued in the (complex) representation $\cR$. The action of (half-)hypermultiplets (coupled to vector multiplets) including auxiliary fields is then given by\footnote{The auxiliary fields are necessary to close part of the hypermultiplet supersymmetry off-shell, which is required for supersymmetric localization. Of course, it is not possible to close the full $\cN=2$ supersymmetry for hypermultiplets with finitely many auxiliary fields.}
\beaa
\label{hmaction}
\cL_{\rm mat}& \ = \ \frac{1}{2} \cD_\mu q^A \cD^\mu q_A +q^A \{\phi,\bar\phi\}q_A +\frac{\ii}2 q^A D_{AB} q^B + \frac23 R q^A q_A -\frac{\ii}{2}\bar\psi \bar\sigma^\mu\cD_\mu \psi\cr 
&\hspace{.85 cm}-\frac12 \psi\phi\psi +\frac12 \bar\psi\bar\phi\bar\psi - q^A\lambda_A\psi +\bar\psi\bar\lambda_A q^A - \frac12 F^{\check{A}} F_{\check A} \,, 
\eeaa
where we suppress $I,J,\dots$ indices, and the covariant derivative for the scalars is given by
\be
\cD_{\mu} q_{AI} \ = \ \partial_{\mu} q_{AI} - \ii (A_{\mu})_{I}{}^{J} q_{AJ} + \ii V^{B}{}_{A} q_{BI} \,.
\ee

The vector multiplet and hypermultiplet actions in~\eqref{YM_act} and~\eqref{hmaction} are invariant under $\cN=2$ supersymmetry. The vector multiplet components transform as follows
\beaa
QA_\mu& \ = \ \ii\xi^A\sigma_\mu\bar\lambda_A - \ii\bar\xi^A\bar\sigma_\mu\lambda_A\,,\\
Q\phi& \ = \ -\ii\xi^A\lambda_A\,,\\
Q\bar\phi& \ = \ -\ii\bar\xi^A\bar\lambda_A\,,\\
Q\lambda_A& \ = \ \frac{1}{2} \sigma^{\mu\nu}\xi_A F_{\mu\nu} +2\sigma^\mu \bar\xi_A\cD_\mu\phi + \sigma^\mu\cD_\mu\bar\xi_A \phi -2\ii\xi_A[\phi,\bar\phi]+D_{AB}\xi^B\,,\\
Q\bar\lambda_A& \ = \ \frac12\bar\sigma^{\mu\nu}\bar\xi_A F_{\mu\nu} - 2\bar\sigma^\mu\xi_A\cD_\mu\bar\phi -\bar\sigma^\mu\cD_\mu\xi_A\bar\phi +2\ii\bar\xi_A[\phi,\bar\phi]+D_{AB}\bar\xi^B\,,\\
QD_{AB}& \ = \ -\ii\bar\xi_A\bar\sigma^\mu\cD_\mu\lambda_B -\ii\bar\xi_B\bar\sigma^\mu\cD_\mu\lambda_A+\ii\xi_A\sigma^\mu\cD_\mu\bar\lambda_B+\ii\xi_B\sigma^\mu\cD_\mu\bar\lambda_A\\
&\hspace{.75 cm}-2[\phi,\bar\xi_A\bar\lambda_B+\bar\xi_B\bar\lambda_A]-2[\bar\phi,\xi_A\lambda_B+\xi_B\lambda_A]\,.
\eeaa
The gauge field $A_\mu$ is real, the reality condition of the scalar reads $\phi^*=\bar\phi$, and the auxiliary fields satisfy
\begin{equation}
(D_{AB})^* \ = \ -e^{-\frac{y}{\ell}(s_A + s_B)}D^{AB} \ = \ -e^{-\frac{y}{\ell}(s_A + s_B)}\varepsilon^{A\tilde{A}}\varepsilon^{B\tilde{B}}D_{\tilde{A}\tilde{B}} \,,
\end{equation}
where $s_1=1$ and $s_2=-1$. Here, the factor $e^{-\frac{y}{\ell}(s_A + s_B)}$ has to be added because we work in a twisted sector~\eqref{twisted_sector}, where the $y$-periodicity of fields is dependent on their R-charge.

Similarly, the hypermultiplet supersymmetry transformations are given by
\beaa
\label{Qhm1}
Qq_A& \ = \ -\ii\xi_A\psi+\ii\bar\xi_A\bar\psi\,,\\
Q\psi& \ = \ 2\sigma^\mu\bar\xi_A\cD_\mu q^A + \sigma^\mu\cD_\mu\bar\xi_A q^A+4\ii\xi_A \bar\phi q^A +2\check{\xi}_{\check A}F^{\check A}\,,\\
Q\bar\psi& \ = \ 2\bar\sigma^\mu\xi_A\cD_\mu q^A + \bar\sigma^\mu\cD_\mu\xi_A q^A - 4\ii\bar\xi_A \phi q^A + 2\bar{\check \xi}_{\check A}F^{\check A}\,,\\
QF_{\check A}& \ = \ \ii\check{\xi}_{\check A}\sigma^\mu\cD_\mu\bar\psi - 2\check{\xi}_{\check A}\phi\psi-2\check{\xi}_{\check A}\lambda_B q^B -\ii\bar{\check \xi}_{\check A}\bar\sigma^\mu\cD_\mu\psi \\ 
&\hspace{.75 cm} -2\bar{\check \xi}_{\check A}\bar\phi \bar\psi + 2\bar{\check \xi}_{\check A}\bar\lambda_B q^B\,.
\eeaa
We impose the following reality conditions for the bosonic fields of the hypermultiplets
\beaa
\label{hyp_real}
(q_{AI})^*& \ = \ e^{-\frac{y}{\ell}s_A}q^{AI} \ \equiv \ e^{-\frac{y}{\ell}s_A} \varepsilon^{AB}\varepsilon^{IJ}q_{BJ}\,,\cr
(F_{{\check A}I})^*& \ = \ -e^{-\frac{y}{\ell}s_{\check A}}F^{{\check A}I} \ \equiv \ -e^{-\frac{y}{\ell}s_{\check A}}\varepsilon^{\check A\check B}\varepsilon^{IJ}F_{{\check B}J}\,,
\eeaa
where again, $s_{1,2}=\pm 1$ accounts for the twisted sector. Notice that we have introduced the same twisted sector for the auxiliary $SU(2)$ index $\check{A}$ as for the $SU(2)_R$ index $A$ -- this is because in the rest of this paper, it will be convenient to identify $\check{A}$ with $A$.

Finally, we stress that we assume our matter content to always satisfies the $U(1)_r$-anomaly cancellation condition, which, incidentally, also ensures that the theory is conformal and free of global anomalies. For each simple factor $G_i$ of the gauge group, this condition reads~\cite{Bhardwaj:2013qia}
\begin{equation}
\label{noanom}
2h^\vee(G_i)  \ = \  \sum_x C(\rR_{i,x})\prod_{j\neq i}\dim \rR_{j,x} + \sum_y 2 C(\cR_{i,y})\prod_{j\neq i}\dim \cR_{j,y} \,,
\end{equation}
where $h^\vee$ is the dual Coxeter number, $C(\cR)$ is an index of the representation $\cR$, the sum over $y$ goes over all \emph{full} hypermultiplets $\rR_y=\cR_y\oplus\bar\cR_y$, whereas the sum over $x$ goes over the remaining half-hypermultiplets $\rR_x$. Furthermore, by $\rR_{i,x}$ we denote the representation of the $x$-th half-hypermultiplet under the simple gauge group factor $G_i$, and similarly by $\rR_{i,y}=\cR_{i,y}\oplus\bar\cR_{i,y}$ we denote the representation of the $y$-th full hypermultiplet under $G_i$. For abelian groups $h^\vee=0$, and thus~\eqref{noanom} implies that abelian factors are decoupled and free. Hence, they are not interesting and excluded from our discussion.

\subsection{Fugacities and the supersymmetry they preserve}\label{sec:fug_susy}

Let us now establish the correspondence between fugacities of the superconformal index and the supersymmetry algebra preserved by our background in their presence. To make statements less cumbersome, we drop the characterization ``centrally extended'', as all supersymmetry algebras we deal with are centrally extended by $E-R$.

In the presence of non-vanishing $\alpha$ and $\zeta$ in~\eqref{KS_sols}, part of the solutions in~\eqref{KS_sols} are broken on $S^3\times S^1_y$ due to the spinors having incorrect periodicity upon going around $S^1_y$. Indeed, in the twisted sector~\eqref{twisted_sector}, our spinors ought to satisfy the following conditions,
\beaa
\label{R_twisted_periodicity}
\xi_1(y+\beta\ell)  \ = \ & e^{-\frac{\beta}{2}}\xi_1(y)\,,\qquad & \bar\xi_1(y+\beta\ell)  \ = \ & e^{-\frac{\beta}{2}}\bar\xi_1(y)\,,\cr
\xi_2(y+\beta\ell)  \ = \ & e^{\frac{\beta}{2}}\xi_2(y)\,,\qquad & \bar\xi_2(y+\beta\ell)  \ = \ &e^{\frac{\beta}{2}}\bar\xi_2(y)\,.
\eeaa
Thus, $\xi_1$ and $\bar\xi_2$ are generically broken in~\eqref{KS_sols}, as they do not obey this periodicity, whereas, $\bar\xi_1$ and $\xi_2$ are preserved. Therefore, only the $\mathfrak{su}(2|1)_r$ inside $\mathfrak{su}(2|1)_\ell\oplus \mathfrak{su}(2|1)_r$ is unbroken, and we conclude:

\begin{itemize}
\item\emph{Keeping all three fugacities $(\beta, \zeta, \alpha)$ -- or equivalently $(p,q,t)$ of~\eqref{ind2} -- in the index~\eqref{index_PI} only preserves $\mathfrak{su}(2|1)_r$.}
\end{itemize}

If we additionally want to preserve $\mathfrak{su}(2|1)_\ell$, we have to ensure that all conformal Killing spinors in~\eqref{KS_sols} satisfy the periodicity conditions~\eqref{R_twisted_periodicity}. This is attained if and only if
\be\label{alphazetaconds}
\alpha \ = \ \frac{2\pi N}{\beta} \,, \quad \text{and} \quad \zeta \ = \ \frac{2\pi M}{\beta} \,, \quad M,N \in \mathbb{Z} \,.
\ee
Recall that, by an $(R+r)$-symmetry gauge transformation, we could equivalently consider $\gamma=2\pi N/\beta$ instead. The value $M=N=0$ corresponds to specializing~\eqref{index_PI} to 
\be\label{Schurind}
\cI\left( \beta \right)  \ = \  {\rm Tr}_{\cH_{S^3}}(-1)^F e^{-\beta (E-R)}\,,
\ee
which is the standard ``Schur" specialization of the superconformal index~\cite{Gadde:2011ik,Gadde:2011uv}. Thus, we draw yet another general conclusion:
\begin{itemize}
\item\emph{The Schur index specialization preserves $\mathfrak{su}(2|1)_\ell\oplus\mathfrak{su}(2|1)_r$.}
\end{itemize}

Allowing in~\eqref{alphazetaconds} for non-zero integer $M \neq 0$, corresponds to an insertion of $e^{4\pi \ii M j_1}$ into the Schur index~\eqref{Schurind}. Since $j_1$ is half-integral, such an insertion is immaterial to the index. However, a non-zero integer $N\neq 0$ in~\eqref{alphazetaconds} corresponds to an insertion of $e^{2\pi \ii N (R+r)}$ into the Schur index, which can be non-trivial: because in a Lagrangian theory $R+r$ is half-integral, the insertion of $e^{2\pi \ii N (R+r)}$ depends on $N \mod 2$, and thus becomes nontrivial for $N \equiv 1\mod 2$. (As explained in the footnote~\ref{foot:Rr}, this is also true in general superconformal field theory.) This is a tiny quirk of the index, to which we will return in Section~\ref{sec:spin_str}. In fact, in our localization computations, we will keep $M$ and $N$ generic.

Interestingly, there is an additional, intermediate case, where we tune $\zeta = \alpha$.\footnote{Shifting $\zeta$ by $2\pi M/\beta$ in this case again introduces a trivial factor of $e^{4\pi \ii M j_1}=1$ in the Schur index.} In this case, the two competing phases in~\eqref{KS_sols} cancel each other for the solutions corresponding to $\cQ_{-}^1$ and $\cS_1^{-}$. These two generate the $\mathfrak{su}(1|1)_\ell$ subalgebra of $\mathfrak{su}(2|1)_\ell$. It is easy to check that $\alpha=\zeta$ corresponds to turning on the fugacity for $2j_1 + R + r$, which -- upon shifting by $\tilde\delta_{1\dot-}=E-2j_2 -2R + r$ and redefinition of the fugacity for $E-R$ -- is equivalent to introducing an additional fugacity ``$s$" for the symmetry
\begin{equation}
\cZ \ = \ j_1 - j_2 + r \,.
\end{equation}
That is, such a specialization of the background computes 
\be\label{Schurs}
{\rm Tr}(-1)^F q^{E-R} s^\cZ \,.
\ee
We will argue soon that $s$ actually cancels, and this again reduces to the Schur index. Thus, we conclude:
\begin{itemize}
\item \emph{The $\zeta=\alpha$ specialization preserves $\mathfrak{su}(1|1)_\ell \oplus \mathfrak{su}(2|1)_r$, but still computes the Schur index.}
\end{itemize}
We can also contemplate the possibility of setting $\zeta=-\alpha$, in which case the surviving left-handed supercharges are $\cQ_+^1$ and $\cS_1^+$. However, they are not of interest for the purpose of this paper.

Lastly, as we briefly touch upon in Section~\ref{sec:R_defect}, one can further refine the Schur index by the insertion of an $\cN=(2,2)$-supersymmetric surface defect at $\theta=0$, extended along $\tau$ and $y$. Such a defect must preserve $\mathfrak{su}(1|1)_\ell\oplus \mathfrak{su}(1|1)_r$, which is still enough for all the constructions discussed in the current paper to work.

\subsection{The chiral algebra}

In this section, let us assume for simplicity that $M=N=0$, that is $\zeta=\alpha=0$, so we deal with the pure Schur index background. We will comment on non-zero $M$ and $N$ towards the end of this section, and will work in such a more general setting in later sections. 

We notice that the two supercharges relevant for the chiral algebra construction of~\cite{Beem:2013sza}, 
\beaa
\rQ_1& \ = \  \cQ^1_- + \frac1{\ell}\tilde{\cS}^{2\dot -}\,,\cr
\rQ_2& \ = \ \frac1{\ell}\cS_1^- - \tilde{\cQ}_{2\dot -}\,,
\eeaa
belong to the algebra $\mathfrak{su}(2|1)_\ell\oplus\mathfrak{su}(2|1)_r$ described above. Furthermore, they also belong to the smaller subalgebra $\mathfrak{su}(1|1)_\ell\oplus\mathfrak{su}(2|1)_r$ (or even $\mathfrak{su}(1|1)_\ell\oplus\mathfrak{su}(1|1)_r$) mentioned at the end of the previous subsection. Because of relation
\begin{equation}
\{\rQ_1,\rQ_2\} \ = \ -\frac1{\ell}\cZ  \ = \  \frac1{\ell}\left(j_2 - j_1 -r\right) \,,
\end{equation}
any operator in the cohomology of $\rQ_i$ -- or in the equivariant cohomology of $\cQ^H = \rQ_1+\rQ_2$ -- necessarily satisfies $\cZ=0$. This immediately shows that the additional fugacity $s$ for the $\cZ$ symmetry in~\eqref{Schurs} indeed cancels: only observables in the cohomology of $\cQ^H$ contribute to the index,\footnote{This is true because the index~\eqref{Schurs} contains only fugacities for $E-R$ and $\cZ$, and both of these symmetries commute with the supercharges $\rQ_i$ and $\cQ^H$.} and they have $\cZ=0$.

Let us proceed to explore the structure of $\rQ_i$- or $\cQ^H$-cohomology. We note that the supercharge $\cQ^H$ is, of course, the dimensional uplift of an analogous supercharge in three-dimensional $\cN=4$ theories studied in~\cite{Dedushenko:2016jxl}. It is therefore suggestive to view the two-dimensional chiral algebra construction as the dimensional uplift of the one-dimensional topological quantum field theory construction studied in that reference.

We introduce the following rotation generators:
\beaa
\cM^\perp & \ = \  \cM_+{}^+ - \cM^{\dot{+}}{}_+  \ = \  P_\tau\,,\\
\cM& \ = \  \cM_+{}^+ + \cM^{\dot{+}}{}_{\dot+}  \ = \  P_\varphi \,,
\eeaa
where, akin to the notation in~\cite{Dedushenko:2016jxl,Dedushenko:2017avn}, by writing $P_\tau$ and $P_\varphi$ we emphasize that these generators simply act by rotating the $\tau$- and $\varphi$-circles in our fibration coordinates on $S^3$, respectively. Next, we introduce the twisted $\varphi$-rotation,
\begin{equation}
\widehat{P}_\varphi  \ = \  P_\varphi + R \,,
\end{equation}
which, again in accordance with the three-dimensional case, generates a $\cQ^H$-closed operation. However, unlike in three dimensions, this operation is not $\cQ^H$-exact. Additionally, in our four-dimensional case we also have a ``twisted time translation'', given by the central element of the algebra
\begin{equation}
H  \ = \  E-R \,,
\end{equation}
which generates twisted translations in the $S^1_y$ direction of the $S^3\times S^1_y$ geometry. Being a central charge, it is of course $\cQ^H$-closed, but it is not $\cQ^H$-exact. Following~\cite{Beem:2013sza}, we consider the holomorphic and anti-holomorphic linear combinations of these and find the latter to be $\rQ_i$ -- and thus $\cQ^H$ --  exact:
\begin{align}
2L_{-1} & \ = \  H + \widehat{P}_\varphi = E + P_\varphi\,,\\
2\bar{L}_{-1}& \ = \  H-\widehat{P}_\varphi =  E - P_\varphi - 2R \cr
& \ = \ \{\rQ_1, \cS_1^- + \ell\tilde\cQ_{2\dot -}\}  \ = \ \{\rQ_2, \ell\cQ^1_- - \tilde\cS^{2\dot -}\} \ = \ \{\cQ^H, \cS_1^- + \ell\tilde\cQ_{2\dot -}\} \,.
\end{align}
This readily implies, as in~\cite{Beem:2013sza}, that the $\cQ^H$-cohomology has the structure of a chiral algebra, or vertex operator algebra (VOA), as we call them interchangeably in this paper.

More precisely, because $(\cQ^H)^2=\{\rQ_1,\rQ_2\}=\frac1\ell (\cM^\perp + r)$, the chiral algebra is going to live at the fixed point locus of $\cM^\perp$. This is a torus located at $\theta=\pi/2$ and parametrized by our coordinates $\varphi$ and $y$ -- we often refer to it as $S^1_\varphi\times S^1_y$. Recall that $E$ is a dilatation generator, which acts as $E=\Delta + r\partial_r$ in flat space on observables of conformal dimension $\Delta$. Then, upon moving to $S^3\times \R$ and $S^3\times S^1_y$, it becomes the operator $\ell\partial_y$ (the additional numeric term $\Delta$ disappears because $\cO^{\rm cylinder}=(r/\ell)^\Delta \cO^{\rm flat}$). Similarly, on $S^3\times S^1_y$ we have that $P_\varphi=-\ii\partial_\varphi$. Note that
\begin{align}
\ell\frac{\partial}{\partial y} -\ii\frac{\partial}{\partial\varphi}& \ = \ 2\partial_w \,,\cr
\ell\frac{\partial}{\partial y}+ \ii\frac{\partial}{\partial\varphi} & \ = \ 2\partial_{\bar w}\,,
\end{align}
where we have introduced the coordinates $w$ and $\bar w$ as
\begin{equation}
w \ = \ \frac{y}{\ell} + \ii\varphi \,, \qquad \bar{w} \ = \ \frac{y}{\ell} - \ii\varphi \,.
\end{equation}
In the case when $S^1_y$ is replaced by $\R$, these are precisely coordinates on the cylinder related to the flat space ones via the usual map
\begin{equation}
z \ = \ e^w \,, \qquad \bar{z} \ = \ e^{\bar w} \,.
\end{equation}

Recall that the chiral algebra is formed by the twisted-translated Schur operators~\cite{Beem:2013sza}. These are defined as the ordinary Schur operators translated along the ``chiral algebra surface" by employing the operators $L_{-1}$ and $\bar{L}_{-1}$ as translation generators. On $S^1_\varphi \times S^1_y\subset S^3\times S^1_y$, the twisted-translated Schur operators are constructed as follows
\begin{equation}
\label{tw_tr_def}
\cO(w,\bar{w}) \ = \ \cO_{A_1 \dots A_n} u^{A_1}\dots u^{A_n}\,, \quad \text{where} \quad  u^A \ = \ (e^{-\bar{w}/2}, e^{\bar{w}/2})\,.
\end{equation}
Here, the $SU(2)_R$ highest weight component $\cO_{+\dots +}$ is the usual Schur operator, while $\cO(w,\bar{w})$ is its twisted-translated cousin. With this definition, it is straightforward to check that
\beaa
\partial_w \cO(w,\bar{w}) & \ = \  \frac12 [E+P_\varphi,\cO(w,\bar{w})] & \ = \ [L_{-1}, \cO(w,\bar{w})]\,,\cr
\partial_{\bar w} \cO(w,\bar{w}) & \ = \  \frac12 [E-P_\varphi-2R,\cO(w,\bar{w})] & \ = \ [\bar{L}_{-1}, \cO(w,\bar{w})]\,,
\eeaa
where the latter is $\rQ_i$- and thus $\cQ^H$-exact because $\bar{L}_{-1}$ is exact and $\cO(w,\bar{w})$ is closed. As usual, it implies that in the $\rQ_i$ or $\cQ^H$ cohomology, $\partial_{\bar w}\cO(w,\bar{w})$ is trivial, so the cohomology classes
\begin{equation}
\cO(w) \ = \ \left[\cO(w,\bar{w}) \right]
\end{equation}
are holomorphic in $w$, and furthermore, have the VOA structure induced by the four-dimensional OPE.

In fact, by a Weyl transformation one can straightforwardly connect our definitions to the flat space description of~\cite{Beem:2013sza}, and see that at the level of an abstract VOA, one has:
\begin{equation}
\label{cyl_flat_relation}
\cO^{\rm cyl}(w)  \ = \  z^h \cO^{\rm flat}(z) \,,
\end{equation}
where the superscripts ``cyl'' and ``flat'' are there to distinguish the construction on a cylinder $S^3\times \R$ from the flat space one of~\cite{Beem:2013sza}. Furthermore, $h$ is the (holomorphic) conformal dimension of a VOA operator $\cO$, which is related to quantum numbers of the Schur operator $\cO_{+\dots +}$ by~\cite{Beem:2013sza}
\begin{equation}
\label{hdim}
h \ = \ R + j_1 + j_2 \,.
\end{equation}

Finally, notice that the above construction admits a generalization to non-zero $\zeta=2\pi M/\beta$ and $\alpha=2\pi N/\beta$. The parameter $\zeta$, being a geometric twist of the background, modifies the ``time translation'' generator $E$ to $E-\ii\zeta P_\varphi - \ii\zeta P_\tau$. This amounts to shifting the complex structure of the torus -- which we still refer to as $S^1_\varphi\times S^1_y$ throughout the paper\footnote{At general $\zeta=2\pi M/\beta\neq0$, which is our main case of interest, the induced metric on $S^1_\varphi\times S^1_y$ is \emph{not} a product metric, see Section~\ref{sec:betagamma}.} -- in such a way that in new complex coordinates
\begin{equation}\label{w_with_zeta}
w \ = \ \frac{y}{\ell} + \ii\varphi + \ii \zeta\frac{y}{\ell} \,, \qquad \bar{w} \ = \ \frac{y}{\ell} - \ii\varphi - \ii \zeta\frac{y}{\ell} \,,
\end{equation}
the two-dimensional theory looks the same. Namely, the $\partial_{\bar w}$ translation is $\cQ^H$-exact, and the cohomology has the emergent VOA structure. This shift, however, might alter the periodicity in the $y$ direction. The latter is also true for non-zero $\alpha=2\pi N/\beta$, which, despite not being a geometric parameter of the four-dimensional background, acquires an emergent geometric interpretation in two dimensions. In the following we will perform the localization with general non-zero $M$ and $N$, but before doing so, we first give a more unified view on the discrete refinements of the construction.

\subsection{Spin structures on $S^1_\varphi \times S^1_y$}\label{sec:spin_str}

Since $R$, $j_1$ and $j_2$ are either integers or half-integers, the relation~\eqref{hdim} implies that all vertex operators in the chiral algebra have either integral or half-integral spins. Whenever half-integral spins are present, one has to choose a spin structure. In the case at hand, we ought to consider the spin structure on the torus $S^1_\varphi \times S^1_y$ on which our chiral algebra lives. As a matter of fact, the construction from the previous subsection provides a natural choice of spin structure: On the one hand, the chiral algebra is periodic in the $y$ direction, that is spinors have Ramond boundary conditions along $S^1_y$. This is simply the reflection of the fact that our theory is defined in a twisted sector, where the R-charge dependent non-periodicity of $\cO_{A_1\dots A_n}$ is canceled by explicit non-periodic factors $u^{A_1}\dots u^{A_n}$. This can alternatively be understood as a consequence of the fact that twisted $y$-translations are generated by $E-R$, which is a symmetry of our system. On the other hand, the $S^1_\varphi$ direction has NS sector boundary conditions. It is most straightforward to see this from equation~\eqref{cyl_flat_relation}: since vertex operators are single-valued in flat space, they are in the NS sector on the cylinder. Indeed, equation~\eqref{cyl_flat_relation} implies that on the cylinder, the $\varphi$-periodicity is determined by
\begin{equation}
z^h  \ = \  e^{\frac{y}{\ell}h + \ii\varphi h} \,,
\end{equation}
and while operators with integral $h$ are periodic, those with half-integral $h$ are anti-periodic. One can alternatively check this using the explicit definition~\eqref{tw_tr_def}. Thus, in conclusion, the chiral algebra construction provides us with a natural choice of spin structure on $S^1_\varphi \times S^1_y$, which is in the NS sector along $S^1_\varphi$ and in the R sector along $S^1_y$; we will refer to this choice as $(1,0)$.\footnote{We use the notation $(\nu_1, \nu_2)$, indicating that spinors pick up a phase $e^{\pi \ii\nu_1}$, $e^{\pi\ii\nu_2}$  upon going once around $S^1_\varphi$, $S^1_y$, respectively.}

However, we realize that there exists a simple modification of the original construction which flips the spin structure to $(1,1)$. Recall that in the subsection~\ref{sec:fug_susy}, we noticed that the Schur index could be modified by an insertion of $e^{2\pi i N (R+r)}$, which did not break any supersymmetry and only depended on $N\mod 2$. In particular, it is nontrivial for $N=1$, corresponding to an insertion of
\begin{equation}
\label{spin_str_flip}
(-1)^{2(R+r)}\,.
\end{equation}
Such an insertion can be realized by turning on an $(R+r)$-holonomy in our background, and it flips the $y$-periodicity for those operators that have half-integral $R+r$. The operators that contribute to this slightly modified index are still the Schur operators, because such a refinement preserves $\cQ^H$. In particular, they still obey $Z= j_1 - j_2 + r=0$, and using~\eqref{hdim}, we can find another expression for the chiral algebra conformal dimension,
\begin{equation}
\label{hdim_alt}
h \ = \ R+r+2j_1 \,.
\end{equation}
This equation clearly shows that $R+r$ is (half-)integral if and only if $h$ is (half-)integral.\footnote{\label{foot:Rr}This shows that the $R+r$ of Schur operators are always (half-)integral, and therefore the refinement of the Schur index by $e^{2\pi \ii N (R+r)}$ depends only on $N\mod 2$ for \emph{any} superconformal field theory, not only the Lagrangian ones.} Hence the insertion of~\eqref{spin_str_flip} changes the periodicity along $y$ for all operators that have half-integral $h$. In other words, the modification~\eqref{spin_str_flip} flips the spin structure along $S^1_y$.

Finally, there also exists a way to flip the $S^1_\varphi$ spin structure, though it does not follow from any of our earlier discussions yet. Similar to how we were able to flip the $S^1_y$ spin structure, it must be possible to achieve this by turning on the monodromy $(-1)^{2(R+r)}$ for the $S^1_\varphi$ circle. Since this circle is contractible, this ought to be done by an insertion of a codimension-two defect with non-zero vorticity for the $R+r$ symmetry.\footnote{Codimension-two defects characterized by a vortex-type singularity have recently been a subject of increasing attention, see \emph{e.g.} ~\cite{Gukov:2006jk,Gukov:2008sn,Gaiotto:2012xa,Kapustin:2012iw,Drukker:2012sr,Gadde:2013dda,Hosomichi:2017dbc}.} Since this surface defect is expected to change the NS boundary condition along $S^1_\varphi$ into the R one, at the level of $\cQ^H$ cohomology, it is expected to give the Ramond sector module(s) of the VOA (that is, twisted modules). We will discuss it in the next subsection.

To summarize, in this subsection we have explained that:
\begin{enumerate}
	\item The chiral algebra on $S^1_\varphi\times S^1_y$ comes with the natural choice of the $(1,0)$ spin structure, where spinorial vertex operators are anti-periodic along $S^1_\varphi$ and periodic along $S^1_y$.
	\item We can flip the $S^1_y$ spin structure by turning on the holonomy $(-1)^{2(R+r)}$ along $S^1_y$.
	\item We can flip the $S^1_\varphi$ spin structure by the insertion of an $(R+r)$ monodromy-creating surface defect at $\theta=0$.
\end{enumerate}

In fact, the choice of spin structure is intimately related to modular properties of the chiral algebra characters. We will discuss some aspects of it later in this work, while leaving a detailed investigation of this connection for the future. Now, let us look closer at the defect mentioned in the third item above.

\subsection{The canonical R-symmetry interface and surface defects}\label{sec:R_defect}
The possibility to refine the Schur index by an extra factor of $e^{2\pi\ii (R+r)}=(-1)^{2(R+r)}$, as discussed in the previous subsection and in Section~\ref{sec:fug_susy}, was attributed to the possibility of turning on the background holonomy $\alpha=2\pi/\beta$ for $R+r$, without breaking any supersymmetry. In Section~\ref{sec:fug_susy} it was equivalently interpreted as a twisted sector for the $U(1)_{R+r}$ symmetry, with twisting parameter $\gamma=2\pi/\beta$. In the latter case, we can think of it as placing a codimension-one symmetry defect at any fixed point ${\rm pt}\in S^1_y$ (and extending it in the $S^3$ directions) that implements the $(-1)^{2(R+r)}$ transformation. We are going to call this symmetry defect a canonical R-symmetry codimension-one defect (or interface), because it plays an important role in various constructions.

For arbitrary operators in a general superconformal field theory, the value of $R+r$ does not necessarily possess any nice properties. However, we have seen in~\eqref{hdim_alt} that for Schur operators, $R+r\in\frac12 \Z$. So $(-1)^{2(R+r)}$ becomes a $\Z_2$ defect in the Schur sector. In Lagrangian theories (which are the case of primary interest in this paper), $R+r\in\frac12 \Z$ for \emph{all} operators, and $(-1)^{2(R+r)}$ is a $\Z_2$ symmetry defect in the full theory. 

As is well-known, the Schur sector, and in particular the chiral algebra construction, admits a refinement by $\cN=(2,2)$ surface defects orthogonal to the chiral algebra plane and intersecting it at the origin~\cite{Beem:2013sza} (see also~\cite{Cordova:2016uwk,Cordova:2017ohl,Cordova:2017mhb,Nishinaka:2018zwq}). In our $S^3\times S^1_y$ geometry, such defects no longer intersect the chiral algebra torus: they are extended along the torus located at
\begin{equation}
\theta \ = \ 0: \quad \text{surface defect extended along } S^1_\tau\times S^1_y\,.
\end{equation}

With the canonical R-symmetry interface at hand, we can consider more general configurations with the R-symmetry interface ending at the surface defect. Geometrically, we place the R-symmetry interface at any fixed value of $\varphi$ and extend it in the $\theta, \tau$ and $y$ directions, ending it at the surface defect sitting at $\theta=0$. Following the nomenclature of~\cite{Cordova:2017mhb}, the latter type of defects living at the boundary of the R-symmetry interface will be referred to as ``twisted'' or ``monodromy'' surface defects. They have the distinguishing property that operators with half-integral $R+r$ have monodromy $(-1)$ around such defects.

Recall from~\eqref{hdim_alt} that $R+r$ is (half-)integral whenever $h$ is (half-)integral. Therefore, \emph{only} operators that have half-integral $h$ receive the monodromy $(-1)$ around the twisted defect. Thus, at the level of VOA, such defects will change the spin structure of $S^1_\varphi$ from NS to R, and they correspond to twisted, or ``Ramond sector", modules.

The above discussion is completely general and holds for arbitrary four-dimensional ${\cN=2}$ superconformal field theories: in such theories, all our twisted surface defects of interest would sit at the end of the topological R-symmetry interface defined above. To further specify the surface defect, one ought to provide more details on what exactly happens at the boundary of the R-symmetry interface. Since our focus here is on Lagrangian theories, we are going to provide more details for this case.

First, we note that for all fields in the vector multiplet, $(-1)^{2(R+r)} \equiv 1$. Hence, the R-symmetry interface is completely transparent and can be ignored for the vector multiplets.

However, for all fields in the hypermultiplet, the opposite is true, \emph{i.e.} $(-1)^{2(R+r)} \equiv -1$. Notice that this is the same as the flavor symmetry interface for the hypermultiplets, which performs the $(-1) \in U(1)_F\subset SU(2)_F$ flavor symmetry transformation across the defect. The ``flavor monodromy" defects for free hypermultiplets were considered in~\cite{Cordova:2017mhb}, and indeed were used there to define the spectral flow connecting the NS and R sectors in the $\beta-\gamma$ system. We thus see that for Lagrangian theories, our R-symmetry interface coincides with the flavor interface of~\cite{Cordova:2017mhb} for the transformation $(-1)\in U(1)_F$. However, the two types of defects are members of different families: while the defect of~\cite{Cordova:2017mhb} admits a continuous generalization to $e^{\ii\alpha}\in U(1)_F$ in the hypermultiplet theory, our R-symmetry defect is always \emph{discrete} but can be defined in an \emph{arbitrary} $\cN=2$ superconformal field theory.

We proceed to define the simplest monodromy surface defects in a Lagrangian theory. Because the R-symmetry interface ending on them is not visible to vector multiplets, we define the simplest surface defects to be invisible for vector multiplets as well. As for the hypermultiplets, the detailed definition involving specifying the asymptotic behavior of all fields at the defect location is given in Appendix~\ref{app:defect}. Such a choice of asymptotic behavior breaks half of the supersymmetry, and in Appendix~\ref{app:defect} we define the defects in a way that preserves the following four supercharges
\begin{equation}
\label{defect_SUSY}
\text{preserved supercharges:} \qquad \cQ^1_-,\ \cS_1^-,\ \tilde\cQ_{2\dot -},\ \tilde\cS^{2\dot -} \,,
\end{equation}
forming the $\mathfrak{su}(1|1)_\ell\oplus \mathfrak{su}(1|1)_r$ subalgebra of $\mathfrak{su}(2|1)_\ell\oplus \mathfrak{su}(2|1)_r$. It can be interpreted as the $\cN=(2,2)$ supersymmetry on $S^1_\varphi\times S^1_y$.

Notice that the supercharges~\eqref{defect_SUSY} are exactly the ones required for the chiral algebra construction. In particular, this means that the $\cQ^H$ is still preserved, and its cohomology classes are still represented by twisted-translated Schur operators. The index in the presence of a defect still counts Schur operators, albeit again in a slightly modified way. In Appendix~\ref{app:defect}, we actually define two such simplest defects that correspond to the two Ramond modules of the symplectic boson.\footnote{Recall that the symplectic boson, \emph{i.e.} the weight-$\frac12$ $\beta-\gamma$ system, has two distinct Ramond modules that satisfy either $\beta_0|0\rangle=0$ or $\gamma_0|0\rangle=0$.}

\section{Localization}
\label{sec:loc}

In this section, we move towards applying supersymmetric localization to (Lagrangian) four-dimensional $\cN=2$ superconformal field theories, in order to derive the two-dimensional chiral algebra action. We will explain certain crucial points along the way. Akin to~\cite{Dedushenko:2016jxl}, we are localizing with respect to $\cQ^H = \rQ_1 + \rQ_2$. This corresponds to choosing the conformal Killing spinor in~\eqref{KS_sols} with constant spinors $\epsilon$, $\bar \epsilon$, $\eta$ and $\bar \eta$ explicitly given by
\beaa
 \epsilon \ = \ & \left(\begin{matrix}
-1\cr 0
\end{matrix} \right)\,,\qquad 
& \bar\epsilon \ = \ & \left(\begin{matrix}
1\cr 0
\end{matrix} \right)\,,\cr
\eta \ = \ & \left(\begin{matrix}
0\cr 1
\end{matrix} \right)\,,\qquad 
&\bar\eta \ = \ & \left(\begin{matrix}
0\cr 1
\end{matrix} \right)\,.
\eeaa
Then, the conformal Killing spinors are written in terms of components as follows\footnote{Recall that we work in the frame of equation~\eqref{fframe}, and with the gamma matrices in equation~\eqref{gammas}.}
\beaa\label{KillingSpins}
 \xi_{1}  \ = \ &   
\left( 
\begin{array}{c} 
	- e^{\frac{\ii (\tau+\varphi) }{2}-\frac{y}{2\ell}+\ii (\zeta-\alpha)\frac{y}{\ell}} \cos\frac{\theta }{2}\\
	-\ii e^{\frac{\ii (\tau+\varphi) }{2}-\frac{y}{2\ell}+\ii (\zeta -\alpha)\frac{y}{\ell}} \sin\frac{\theta }{2} 
\end{array}
\right) \,, \quad 
& \xi_2   \ = \ &  \left( 
\begin{array}{c} 
	-\ii e^{\frac{\ii (\tau-\varphi) }{2}+\frac{y}{2 \ell }}\sin \frac{\theta }{2} \\
	e^{\frac{\ii (\tau- \varphi) }{2}+\frac{y}{2 \ell }} \cos \frac{\theta }{2}
\end{array}
\right) \,,\\
\bar\xi_{1}  \ = \ &   
\left( 
\begin{array}{c} 
	e^{-\frac{\ii \tau }{2}+\frac{\ii \varphi }{2}-\frac{y}{2 \ell }} \cos \frac{\theta }{2} \\
	-\ii e^{-\frac{\ii \tau }{2}-\frac{y}{2 \ell }+\frac{\ii \varphi }{2}} \sin \frac{\theta }{2}
\end{array}
\right)\,, \quad \quad\quad
& \bar\xi_2   \ = \ &  \left( 
\begin{array}{c} 
	\ii e^{-\frac{\ii (\tau+\varphi) }{2}+\frac{y}{2\ell}-\ii (\zeta -\alpha)\frac{y}{\ell}}\sin \frac{\theta }{2}\\
	e^{-\frac{\ii (\tau+\varphi) }{2}+\frac{y}{2\ell}-\ii (\zeta -\alpha)\frac{y}{\ell}} \cos \frac{\theta }{2}
\end{array}
\right) \,,
\eeaa
with their conformal cousins $\xi^{\prime}$ and $\bar\xi^{\prime}$ given by
\beaa
\xi_{1}^{\prime}  \ = \ &
-\frac{1}{2 \ell } \left( 
\begin{array}{c} 
	\ii e^{\frac{\ii (\varphi-\tau) }{2}-\frac{y}{2 \ell }} \cos \frac{\theta }{2} \\
	e^{\frac{\ii (\varphi-\tau) }{2}-\frac{y}{2 \ell }} \sin \frac{\theta }{2} 
\end{array}
\right) \,, \quad
& \xi_2^{\prime}  \ = \ &
\frac{1}{2 \ell } \left( 
\begin{array}{c} 
	-e^{-\frac{\ii (\tau+\varphi) }{2}+\frac{y}{2\ell}-\ii (\zeta -\alpha)\frac{y}{\ell} }\sin \frac{\theta }{2}  \\
	\ii e^{-\frac{\ii (\tau+\varphi) }{2}+\frac{y}{2\ell}-\ii (\zeta -\alpha)\frac{y}{\ell} }\cos \frac{\theta }{2} 
\end{array}
\right) \,,\\
\bar\xi_{1}^{\prime}  \ = \ &
\frac{1}{2 \ell }\left( 
\begin{array}{c} 
	\ii e^{\frac{\ii (\tau+\varphi) }{2}-\frac{y}{2\ell}+\ii (\zeta-\alpha)\frac{y}{\ell} }\cos \frac{\theta }{2} \\
	- e^{\frac{\ii (\tau+\varphi) }{2}-\frac{y}{2\ell}+\ii (\zeta-\alpha)\frac{y}{\ell} }\sin \frac{\theta }{2}
\end{array}
\right)\,,\quad
& \bar\xi_2^{\prime}  \ = \ & \frac{1}{2 \ell } \left( 
\begin{array}{c} 
	e^{\frac{\ii (\tau-\varphi) }{2}+\frac{y}{2 \ell }}\sin \frac{\theta }{2} \\
	\ii  e^{\frac{\ii (\tau-\varphi) }{2}+\frac{y}{2 \ell }}\cos \frac{\theta }{2}
\end{array}
\right) \,.
\eeaa
As previously mentioned, in the following we will set 
\be
\gamma \ = \ 0\,, \qquad \zeta \ = \ \frac{ 2\pi M}{\beta}\,,\quad \text{and} \quad \alpha \ = \ \frac{2\pi N}{\beta}\,,
\ee
with generic $M, N \in \mathbb{Z}$. This is the most general situation allowing for the full $\mathfrak{su}(2|1)_\ell\oplus \mathfrak{su}(2|1)_r$ to be preserved.

Finally, we will be required to introduce auxiliary spinors $\check{\xi}_{\check{A}}$ and $\bar{\check{\xi}}_{\check{A}}$ such that the supercharge $\cQ^H$ closes off-shell for the hypermultiplets.\footnote{Notice that here we identify the indices of the auxiliary fields and spinors with the $SU(2)_{R}$ ones. This is consistent with the choice of twisted periodicity for the spinors given in equation~\eqref{checkxifix1} and~\eqref{checkxifix2}.} Those auxiliary spinors are to satisfy
\beaa
\xi_A\check\xi_B-\bar\xi_A\bar{\check\xi}_B &  \ = \  0 \,, \qquad 
& \xi^A\xi_A+\bar{\check\xi}^A\bar{\check\xi}_A & \ = \ 0 \,, \\
\label{auxspinors}
\bar\xi^A\bar\xi_A+\check\xi^A\check\xi_A &  \ = \  0 \,, \qquad
& \xi^A\sigma^m\bar\xi_A+\check\xi^A\sigma^m\bar{\check\xi}_A & \ = \ 0 \,,
\eeaa
which fixes them up to an $SL(2,\mathbb{C})$-rotation. Indeed, we may use the remaining freedom to rotate them into the simple form\footnote{It might be tempting to follow the approach that works well for ellipsoids in~\cite{Hama:2012bg,Hosomichi:2016flq}, and choose $\check{\xi}_A=c \xi_A$, $\bar{\check{\xi}}_A=-c^{-1} \bar\xi_A$ on $S^3\times S^1$ as well, as was done \emph{e.g.} in~\cite{Pan:2019bor}. On $S^3\times S^1$ the constraints~\eqref{auxspinors} imply $c\propto e^{-\ii\tau}$, which is \emph{not} a smooth function on $S^3$. Such $\check{\xi}_A$ and $\bar{\check{\xi}}_A$ would simply not be smooth sections of the appropriate spinor bundles, even though the formalism requires them to be smooth. In fact, one can check that choosing such auxiliary spinors would result in \emph{wrong} BPS equations (meaning they would not have naturally expected properties). On the other hand, the choice in~\eqref{checkxifix1}-\eqref{checkxifix2} gives perfectly smooth auxiliary spinors that produces correct BPS equations later in this section.}
\bea\label{checkxifix1}
\check{\xi}_A & = & 2\ell \, \xi^{\prime}_{A} \,, \\
\label{checkxifix2}
\bar{\check{\xi}}_A & = &  2\ell \, \bar{\xi}^{\prime}_{A} \,.
\eea

\subsection{Vector multiplets}

The first step is to localize vector multiplets. Here, akin to~\cite{Dedushenko:2016jxl}, a crucial and non-trivial observation is that the Yang-Mills action~\eqref{YM_act} is $\cQ^H$-exact on $S^3\times S^1_y$ -- we refer to Appendix~\ref{sec:YM_exact} for details. It thus immediately follows that the classical Yang-Mills action vanishes on the localization locus (LL), \emph{i.e.}
\begin{equation}
S_{\rm YM}\big|_{\rm LL}  \ = \  0 \,,
\end{equation}
and furthermore the LL itself is simply given by
\be
F_{\mu\nu} \ = \ \phi \ = \ D_{AB} \ = \ 0 \,, \quad \text{and} \quad \bar\lambda_A \ = \ \lambda_A \ = \ 0\,.
\ee
The latter follows from the fact that $S^3\times S^1_y$ has positive scalar curvature, and thus the Dirac operator has no zero modes, hence there are no fermionic directions on the LL. We conclude that the LL is simply given by solutions to $F_{\mu\nu}=0$, that is flat connections on $S^3\times S^1_y$. On $S^3\times S^1_y$, modulo gauge transformations, flat connections are parametrized by the holonomy around $S^1_y$, valued in the maximal torus of the gauge group $\mathbb{T}\subset G$. Therefore, the path integral over vector multiplets localizes to an integral over $\mathbb{T}$.

Some readers might find this concerning; the vector multiplets are known to produce a small $bc$ ghost system in the cohomology~\cite{Beem:2013sza}, so one naively expects to find the fermionic zero modes on the localization locus. However, we just argued that there are none. This has a simple explanation, which (due to its importance) we will discuss in the separate subsection~\ref{sec:no_bc}.

Let us proceed with the vector multiplet localization. Having dealt with the classical piece, we now turn towards computing the one-loop determinant in the background of a flat connection parametrized by the holonomy
\begin{equation}
\label{flat_conn}
u  \ = \  e^{\ii a} \ \in \ \mathbb{T}
\end{equation}
along the circle. Using the classical action $S_{\rm YM}$ as a localizing deformation corresponds to taking the weak-coupling limit $g_{\rm YM}\to 0$. In this limit, the vector multiplet becomes a free multiplet coupled to the background flat connection $u$. The corresponding one-loop determinant is therefore the same as the one appearing in the localization computations of superconformal indices for $\cN=2$ Lagrangian theories (in the Schur limit). More precisely -- and to keep track of the more general fugacity structure -- we can employ localization results for $\cN=1$ theories on $S^3 \times S^1_y$~\cite{Closset:2013sxa,Assel:2014paa}. In order to be allowed to use these results, we embed our background into the one used in~\cite{Assel:2014paa}, which further requires turning on an additional background holonomy to obtain the $\cN=2$ fugacities from the $\cN=1$ ones. We refer to Appendix~\ref{App:N1intoN2} for details on this procedure. The vector multiplet localization results in
\begin{align}
\frac1{|\cW|}\int_{\mathbb{T}} \frac{\dd u}{2\pi\ii u}\, \Delta_1(u)\,Z_{\rm vec}(u) \int \cD\cH\,  e^{-\cS_{\rm mat}[\cH, u]}\,,
\end{align}
where $|\cW|$ is the order of the Weyl group, $\cD\cH$ denotes the path integral over the hypermultiplet fields in the background of a flat connection $u\in\mathbb{T}$, the factor $\Delta_1(u)$ combines the Vandermonde determinant with part of the Faddeev-Popov determinant, following~\cite{Assel:2014paa},
\begin{equation}
\Delta_1(u) \ = \ \prod_{\alpha\in\Delta\setminus \{0\}}(1-u^\alpha) \ = \ \prod_{\alpha\in\Delta_+}(1-u^\alpha)(1-u^{-\alpha})\,,
\end{equation}
and the one-loop determinant $Z_{\rm vec}(u)$ is given in~\eqref{ZvmSchur} (it can of course also be read off from the Schur index in the existing literature~\cite{Gadde:2011ik,Gadde:2011uv,Benini:2011nc,Lemos:2012ph,Aharony:2013dha,Closset:2013sxa,Assel:2014paa,Rastelli:2014jja,Martelli:2015kuk,Rastelli:2016tbz})\footnote{The notation here is standard, and we refer to Appendix~\ref{App:N1intoN2} for more details.}
\begin{align}
Z_{\rm vec}(u) \ = \ q^{\ell E_0} \cI_v \,,
\end{align}
where
\begin{align}
\cI_v  & \ = \  \prod_{\alpha\in\Delta} (q u^\alpha; q)^2  \ = \  (q;q)^{2r_G}\prod_{\alpha\in\Delta_+} (q u^\alpha; q)^2 (q u^{-\alpha}; q)^2\,,
\end{align}
the nome is $q=e^{-\beta(1+\ii\zeta)}$, and $E_0$ is the Casimir energy given by~\cite{Assel:2014paa,Lorenzen:2014pna,Assel:2015nca,Bobev:2015kza}
\begin{equation}
\label{Cas_vm}
\ell E_0 \ = \ \sum_{\alpha\in\Delta^+}\langle \alpha,a\rangle^2 + \frac{\dim G}{12} \,.
\end{equation}

As we will see, the first (holonomy-dependent) term in~\eqref{Cas_vm} will cancel in conformal theories against the similar one in the Casimir energy for the hypermultiplets. One can check that the surviving part of $E_0$, together with the index $\cI_v$, is reproduced by the path integral of a small $bc$ ghost system (of weights $(h_b,h_c)=(1,0)$) on a torus $S^1_\varphi \times S^1_y$
\beaa
\label{bc_PI}
&q^{\frac{1}{12} \dim G}\prod_{\alpha\in\Delta} (q u^\alpha; q)^2\cr
&\hspace{.75 cm} \ = \  \int \cD b\cD c\, \delta\left(\int b(\varphi,y)\dd\varphi\right) \delta\left(\int c(\varphi,y)\dd\varphi\right) e^{-\Tr\int_{S^1_\varphi\times S^1_y}\dd\varphi\dd y \, b\left(\bar\partial_w - \ii \frac{a}{2\beta}\right) c} \,.
\eeaa
Here, the fields $b$ and $c$ are periodic fermions on $S^1_\varphi \times S^1_y$ valued in the adjoint of the gauge group $G$ and coupled to the holonomy $a$ (\emph{c.f.}~\eqref{flat_conn}). The delta-functions in~\eqref{bc_PI} kill modes that ought to be eliminated in a \emph{small} $bc$ ghost system, known to correspond to the four-dimensional $\cN=2$ vector multiplets. This is quite different from the ordinary $bc$ system (\emph{e.g.} encountered in string theory), where one simply saturates zero modes by local insertions of $b$ and $c$ (one for each zero mode). On the contrary, for the small $bc$ ghost system, we have to perform a non-local operation of killing all $\varphi$-independent modes. To understand the origin of this prescription, recall that the Hilbert space $\cH_0$ of the small $bc$ system is defined as a subspace of $\cH_{bc}$ without the $c_0$ zero mode,
\begin{equation}
\cH_0  \ = \  \{\psi\in\cH_{bc}| b_0 \psi=0\} \,.
\end{equation}
Correlation functions on the torus (including the vacuum character) are then defined through
\beaa
\langle \cO_1\dots \cO_n\rangle  \ = \  & \Tr_{\cH_0}\left( (-1)^F q^{L_0} \widehat{\cO}_n(\varphi_n, y_n)\dots \widehat{\cO}_1(\varphi_1, y_1)\right)\cr
 \ = \ & \Tr_{\cH_0}\left( (-1)^F e^{-(\beta\ell-y_{n})\widehat{H}} \widehat{\cO}_n(\varphi_n) e^{-(y_n-y_{n-1}) \widehat{H}}\dots e^{-(y_2-y_1) \widehat{H}}\widehat{\cO}_1(\varphi_1) e^{-y_1 \widehat{H} }\right)\,,
\eeaa
where we have assumed the ordering $0<y_1<y_2<\dots < y_n<\beta\ell$, and $\widehat{H}=\frac{E}{\ell}$ generates translations in the $y$ direction. In this equation, we treat the $y$ coordinate as time, $\widehat{\cO}_k(\varphi_k)$ are the Schr\"odinger picture local operators, and $\widehat{\cO}_k(\varphi_k, y_k)=e^{y_k \widehat{H}}\widehat{\cO}_k(\varphi_k) e^{-y_k \widehat{H}}$ are the Heisenberg picture local operators acting on the Hilbert space $\cH_0$. The operators $\widehat{\cO}_k(\varphi_k)$ are constructed from the modes of
\beaa
\label{schrod_modes}
\widehat{b}(\varphi) \ = \ & \sum_{m\neq 0} \widehat{b}_m e^{-\ii m\varphi}\,,\cr
\widehat{c}(\varphi) \ = \ & \sum_{m\neq 0} \widehat{c}_m e^{-\ii m\varphi}\,,
\eeaa
which are fields of the small $bc$ system quantized on the $y={\rm const}$ slice. Using standard arguments to obtain the path integral representation of correlators, these $\widehat{b}(\varphi)$ and $\widehat{c}(\varphi)$ correspond to fermionic fields $b(\varphi,y)$ and $c(\varphi,y)$ that have no $m=0$ (\emph{i.e.} $\varphi$-independent) modes in their Fourier expansions. These are precisely the modes eliminated by the delta-functions in~\eqref{bc_PI}. This derivation also shows that correlation functions in~\eqref{bc_PI} ought to be inserted at separate ``times'' $y_i$ in order for the path integral representation to make sense. The final answer is of course going to be meromorphic in $w_k=\frac{y_k}{\ell}+\ii\varphi_k$, which allows extending correlators outside the regime where the path integral~\eqref{bc_PI} makes sense (\emph{i.e.} some $y_k$ and $y_n$ may coincide in the final formula as long as $\varphi_k$ and $\varphi_n$ are distinct).

Before moving to other computations, we should add a word of caution. Notice that the determinant $\Delta_1(u)$ can be written as
\begin{equation}
\Delta_1(u) \ \sim \ \prod_{\alpha\in\Delta\setminus\{0\}} \langle\alpha, a\rangle \prod_{\alpha\in\Delta}\frac{\sin\frac{\langle\alpha,a\rangle}{2}}{\frac{\langle\alpha,a\rangle}{2}}
 \ = \ \cJ(a) \Delta_0(a)\,,
\end{equation}
where $\cJ(a)$ is the usual Vandermonde and $\Delta_0(a)$ is a part of the Faddeev-Popov determinant that is also equal to the standard Haar measure on the Lie group. This $\Delta_0(a)$ is equal to the determinant of the one-dimensional analog of the $bc$ system, and can provide the ``missing'' zero modes in the two-dimensional small $bc$ system. Therefore, we find an alternative path integral representation, at least for the partition function,
\begin{align}
\label{alternative_bc}
Z  \ = \  \frac1{|\cW|} \int_{\mathbb{T}} \frac{\dd u}{2\pi\ii u}\cJ(a)\int \cD' b\cD' c\, e^{-\Tr\int_{S^1_\varphi\times S^1_y}\dd\varphi\dd y \, b\left(\bar\partial_w - \ii \frac{a}{2\beta}\right) c} \int \cD\cH\,  e^{-\cS_{\rm mat}[\cH, u]}\,,
\end{align}
where primes on the $bc$ measure mean that we simply drop the zero modes having $m=n=0$. This version of the $bc$ action can be interpreted as arising from the gauge fixing in two dimensions (after we localize hypermultiplets in the next section), with the absence of zero modes meaning that constant gauge transformations are not summed over. However, if we start computing correlation functions with this formula, we will find that it produces the wrong Green's function for the small $bc$ system, because this formula contains modes that do not act on the \emph{small} $bc$ system Hilbert space $\cH_0$. On the other hand,~\eqref{bc_PI} gives the right propagator, as we will see.

Yet, if we use the path integral in~\eqref{alternative_bc} to compute correlators of operators containing only derivatives of the $c$ ghost and not $c$ itself (as appropriate in the \emph{small} $bc$ system), and furthermore only focus on \emph{gauge-invariant} operators, not minding what (not gauge-invariant) Green's function $\langle b(0) \nabla_w c(w)\rangle$ we use in the process, we will find that~\eqref{alternative_bc} reproduces the correct correlation functions as well, with all the unwanted pieces canceling. Despite this, and to avoid possible confusion, we will stick to the version~\eqref{bc_PI} in this work, as it manifestly describes the small $bc$ system by eliminating all the unnecessary modes.

The computation in~\eqref{bc_PI} is merely a check of the statement about the small $bc$ system; essentially we \emph{guess} the action that reproduces the correct partition function and check that it indeed corresponds to the small $bc$ system. Since our localization locus does not involve gaugini zero modes, we cannot obtain the $bc$ action directly from the vector multiplet localization. Nevertheless, the two-dimensional action in~\eqref{bc_PI} is the correct one to compute correlators of Schur operators containing gaugini. We now pause to elaborate on this point in more detail.

\subsection{More on the $bc$ action}\label{sec:no_bc}

It is well-known from~\cite{Beem:2013sza}, that certain components of the four-dimensional gaugini become fields of the small $bc$ ghost system in the chiral algebra. In fact, the computation we performed in~\eqref{bc_PI} confirms this statement. In particular, for free $\cN=2$ vector multiplets, one exactly obtains the small $bc$ ghost system. However, if we look closer at the vector multiplet -- either in flat space or on $S^3\times S^1$ -- and check how $\cQ^H$ or $\rQ_i$ act on the corresponding fermions, we discover the following curious fact. For example, in flat space, using notations similar to those in~\cite{Beem:2013sza}, we find:
\beaa
\left\{\rQ_1, u^{A}(\bar z) \lambda_{A +}(z,\bar{z}) \right\}& \ = \ u^{A}(\bar z) u^{B}(\bar z) D_{AB}\,,\qquad \left\{\rQ_2, u^{A}(\bar z) \lambda_{A +}(z,\bar{z}) \right\} \ = \ 0\,,\cr
\left\{\rQ_2, u^{A}(\bar z) \bar\lambda_{A +}(z,\bar{z}) \right\}& \ = \ u^{A}(\bar z) u^{B}(\bar z) D_{AB}\,,\qquad \left\{\rQ_1, u^{A}(\bar z) \bar\lambda_{A +}(z,\bar{z}) \right\} \ = \ 0\,,
\eeaa
where $u^{A}=(1,\bar{z})$. Thus, the right-hand sides are not all vanishing, and in particular, the cohomology of $\rQ_1$ only contains $u^{A}(\bar z) \bar\lambda_{A +}(z,\bar{z})$, the cohomology of $\rQ_2$ only contains $u^{A}(\bar z) \lambda_{A +}(z,\bar{z})$, the cohomology of $\cQ^H=\rQ_1 + \rQ_2$ only contains the combination $u^{A}(\bar z) \lambda_{A +}(z,\bar{z})-u^{A}(\bar z) \bar\lambda_{A +}(z,\bar{z})$, while the simultaneous cohomology of $\rQ_1$ and $\rQ_2$ appears to be empty. All these statements might look surprising, because from~\cite{Beem:2013sza} we could expect that either of these four cohomologies would give the small two-dimensional $bc$ ghost system. 

Of course, we know the resolution: \emph{on-shell}, auxiliary fields vanish for free vector multiplets, \emph{i.e.} $D_{AB}=0$, so the four cohomology theories mentioned above indeed coincide. Therefore, the $bc$ ghosts indeed appear in the cohomology on-shell. For interacting systems, $D_{AB}$ is no longer zero on-shell, instead, it equals the hyper-K\"ahler moment map describing the D-term potential for hypermultiplets, and proper identification of observables in the cohomology involves the BRST reduction procedure as pointed out in~\cite{Beem:2013sza}. The final answer is again the same for either $\cQ^H$, $\rQ_i$ or the simultaneous cohomology.

Both free and interacting systems, however, share the same property: in order to obtain the correct contribution of vector multiplets to the Q-cohomology of local operators, we need to go on-shell in the vector multiplet, \emph{i.e.}, integrate out auxiliary fields. 

To be more precise, part of it may appear off-shell if we consider the cohomology of a single supercharge, which is what we do in the ordinary localization anyways. Note how $u^{A}(\bar z) \bar\lambda_{A +}(z,\bar{z})$ survives in the off-shell $\rQ_1$ cohomology, $u^{A}(\bar z) \lambda_{A +}(z,\bar{z})$ belongs to the off-shell $\rQ_2$ cohomology, and their difference belongs to the off-shell $\cQ^H$ cohomology. This brings about another subtle point: the identification of the vector multiplet contribution as a small $bc$ system is not unique and depends on the supercharge with respect to which we compute the cohomology. In fact, the cohomology is computed by the spectral sequence, where the zeroth page -- the zeroth order in $g_{\rm YM}$ -- identifies the gauge and matter contributions as $\beta-\gamma$ and small $bc$ ghosts, while the exact answer is interpreted as the BRST cohomology of the zeroth page. This seems to be a universal phenomenon in gauge theories, which also takes place in two-dimensional $\cN=(0,2)$ models~\cite{Dedushenko:2017osi} (see also~\cite{DelZotto:2018tcj,Eager:2019zrc} for more applications in two-dimensional theories with $\cN=(0,4)$ and $\cN=(0,2)$ supersymmetry).

What we point out here is that the choice of the supercharge -- whether it be one of the $\rQ_i$'s or $\cQ^H$ -- determines how we identify the small $bc$ ghosts in terms of gaugini at the zeroth page. The ghost that is found in the off-shell cohomology is always $\partial c$. Namely, for the $\rQ_2$ cohomology, we identify $[u^{A}(\bar z) \lambda_{A +}(z,\bar{z})]=\partial c$, -- which was the choice made in~\cite{Beem:2013sza} -- for the $\rQ_1$ cohomology, it is $[u^{A}(\bar z) \bar\lambda_{A +}(z,\bar{z})]=\partial c$, while for the $\cQ^H$-cohomology we claim $[u^{A}(\bar z) \lambda_{A +}(z,\bar{z})-u^{A}(\bar z) \bar\lambda_{A +}(z,\bar{z})]=\partial c$. Since these combinations are in the \emph{exact} cohomology of the four-dimensional theory, they should survive the BRST reduction. The latter plays well with the fact that in the ordinary $bc$ system, one has
\begin{equation}
\{Q_B,\partial c^A\} \ \propto \ f^A{}_{BC} c^B \partial c^C \,,
\end{equation}
where $f^A{}_{BC}$ are gauge structure constants, so that in the small $bc$ system, where $c$ without derivatives is thrown away, we simply have $\{Q_B,\partial c\}=0$.

Let us now return to the localization computation. Going on-shell is completely harmless from the vantage point of the abstract OPE approach taken in~\cite{Beem:2013sza}, but poses a fundamental challenge for the localization arguments. Indeed, in order for the localization to work, the supercharge $\cQ^H$ (with respect to which we localize) must be realized off-shell, which in the standard approach only allows to compute correlators of observables that are $\cQ^H$-closed off-shell. However, it was observed in~\cite{Pan:2019bor} that in some cases it is possible to modify the supersymmetric localization principle to include observables that are supersymmetric on-shell. To do so, one has to supplement the localizing deformation $\exp(-t \{\cQ^H, V\})$ (with $t\to\infty$) by the appropriate scaling of observables that are only $\cQ^H$-closed on-shell. Namely, each insertion of $u^{A}\lambda_{A +}$ and $u^{A}\bar\lambda_{A +}$ should come with an extra factor of $\sqrt{t}$. This small modification makes the path-integral again independent of the parameter $t$, and thus allows for taking the localization limit $t\to\infty$. In this limit, after integrating out the bosonic fields, fermions are effectively controlled by the quadratic action. Thus, to compute correlators, one simply needs to know the Green's function and perform Wick contractions.

The latter point implies that if we can find a quadratic two-dimensional action that both reproduces the correct partition function and the expected Green's function, we have the complete answer. It turns out that the action in~\eqref{bc_PI} has precisely this property; later we will compute its Green's function, and it will match the Green's function obtained in~\cite{Pan:2019bor} from the four-dimensional computation.

\subsection{Hypermultiplets and the gauged $\beta-\gamma$ system}\label{sec:betagamma}

Let us now turn towards the localization of hypermultiplets. We expect to reproduce the $\beta-\gamma$ action from the four-dimensional hypermultiplet action evaluated on the localization locus. This type of analysis (but for a different supercharge) has first appeared in~\cite{Pestun:2009nn} for theories on $S^4$. In a context closer to the current paper, a similar computation in three-dimensional $\cN=4$ theories on $S^3$ was performed in~\cite{Dedushenko:2016jxl}. Subsequently, it was repeated in~\cite{Pan:2017zie} for free hypermultiplets on $S^4$, where it was found that they indeed localize to the $\beta-\gamma$ system on $S^2\subset S^4$.

We expect that the resulting two-dimensional action on $S^{1}_{\varphi} \times S^{1}_{y}$ will be written in terms of fields in the $\cQ^H$-cohomology of the hypermultiplet. With a certain convenient normalization, they are given by
\begin{align}
\label{symp_bos}
\sZ_I(\varphi,y) \ = \sqrt{\ell}\ e^{-\ii\zeta\frac{y}{2\ell}+\ii\alpha \frac{y}{2\ell}+\frac{y}{2\ell} -\ii\frac{\varphi}{2}}\left(q_{1I} + \ii q_{2I} e^{\ii\zeta \frac{y}{\ell}-\ii\alpha \frac{y}{\ell}-\frac{y}{\ell}+\ii\varphi} \right)\Big|_{\theta=\pi/2} \,,
\end{align}
inserted anywhere on the torus located at $\theta=\pi/2$.

In the following, we again keep $\alpha = 2\pi N/\beta$ and $\zeta=2\pi M/\beta$ with integral $N$ and $M$, such that the supercharge $\cQ^{H}$ is conserved. Furthermore, we close our supercharge $\cQ^H$ off-shell: the hypermultiplet action~\eqref{hmaction} preserves off-shell supersymmetry $\cQ^H$, given that the equations for the auxiliary spinors $\check{\xi}_{A}$ and $\bar{\check{\xi}}_{A}$ in~\eqref{auxspinors} are satisfied; the solutions are given in~\eqref{checkxifix1} and~\eqref{checkxifix2}. It is straightforward to check the off-shell closure of $\cQ^H$, \emph{i.e.} that the hypermultiplet supersymmetry transformations~\eqref{Qhm1} square into a sum of bosonic symmetries off-shell~\cite{Hama:2012bg},
\begin{align}
\label{susy1hyp}
 Q ^2 q_A  & \ = \  \ii v^mD_mq_A+\ii\Phi q_A+wq_A +\Theta_{AB}q^B \,, \\
 Q ^2\psi  & \ = \ \ii v^mD_m\psi+\ii\Phi\psi+\frac32w\psi-\Theta\psi +\frac \ii 4\sigma^{kl}\psi D_{k}v_{l}\,,\\
 Q ^2\bar\psi & \ = \ \ii v^mD_m\bar\psi+\ii\Phi\bar\psi+\frac32w\bar\psi+\Theta\psi +\frac \ii 4\bar\sigma^{kl}\bar\psi D_{k}v_{l}\,,\\
 \label{susy2hyp}
 Q ^2 F_A & \ = \ \ii v^mD_m F_A+\ii\Phi F_A+2wF_A +\check{\Theta}_{AB}F^B \,,
\end{align}
where for $Q=\cQ^H$, we explicitly find that
\beaa\label{vetc}
 v_m \dd x^m  \ = \ & 4 \ii \cos ^2\theta \left(\ell \dd \tau +\frac{\zeta}{\ell} \dd y\right)  \,, \quad & w  \ = \ &0\,,   \\
\Phi \ = \ & 4 \ii \cos\theta e^{-\ii \tau - \ii \zeta \frac{y}{\ell}-\ii \alpha \frac{y}{\ell }} \left(\phi  \, e^{2 \ii \alpha  \frac{y}{\ell }}-\bar\phi \, e^{2 \ii \tau +2 \ii \zeta \frac{ y}{\ell}}\right)\,,
\quad & \Theta  \ = \ &\frac{2 \ii}{\ell }\,,  \\
 \check\Theta_{AB} \ = \ & -\frac{4\alpha}{\ell}\left(\begin{array}{cc}
 \ii e^{\ii \zeta \frac{ y}{\ell }-\ii \alpha \frac{  y}{\ell }-\frac{y}{\ell }+\ii \varphi } \sin\theta & 1 \\
 1 &  - \ii e^{\ii \alpha \frac{  y}{\ell }-\ii \zeta \frac{  y}{\ell }+\frac{y}{\ell }-\ii \varphi } \sin \theta \\
\end{array}\right) \,,
\quad &  \Theta_{AB}  \ = \ & 0 \,.
\eeaa

To localize the hypermultiplets, we use the canonical localizing deformation $\cQ^H V$, where
\begin{equation}
V \ = \ \sum_{\alpha, I} \psi_{\alpha I} (\cQ^H\psi_{\alpha I})^* + \sum_{\dot{\alpha}, I}\bar\psi^{\dot\alpha}{}_I (\cQ^H\bar\psi^{\dot\alpha}{}_I)^* \,.
\end{equation}
It has no fermionic zero modes (which is expected because the $\cQ^H$-cohomology of a hypermultiplet is purely bosonic). The bosonic zero modes -- \emph{i.e.} the localization locus -- are determined by the BPS equations $\cQ^H\psi=\cQ^H\bar\psi=0$, which explicitly read
\begin{align}\label{BPS11}
0  & \ = \ 
2 \sigma^{\mu} \bar \xi_{A} \cD_{\mu} q^{A}{}_{I}
+ \sigma^{\mu} \cD_{\mu} \bar \xi_{A} q^{A}{}_{I} 
- 4 \ii \xi_{A} \bar \phi q^{A}{}_{I} 
+2 \check{\xi}_{A} F^{A}{}_{I} \,,\\
\label{BPS21}
0 & \ = \ 
2 \bar\sigma^{\mu} \xi_{A} \cD_{\mu} q^{A}{}_{I} 
+ \bar\sigma^{\mu} \cD_{\mu} \xi_{A} q^{A}{}_{I} 
- 4 \ii \bar\xi_{A} \phi q^{A}{}_{I} 
+2\bar{\check{\xi}}_{A} F^{A}{}_{I} \,,
\end{align}
where the covariant derivatives act on the hypermultiplet scalars as follows
\be\label{covdq}
\cD_{\mu} q_{AI} \ = \ \partial_{\mu} q_{AI} - \ii (A_{\mu})_{I}{}^{J} q_{AJ} + \ii V^{B}{}_{A} q_{BI} \,,
\ee
with $V^{A}{}_{B}$ the $SU(2)_{R}$ background gauge field, and $(A_{\mu})_{I}{}^{J}$ a dynamical gauge field in the vector multiplet (of course one can also couple to background vector multiplet, which then induces a ``flavor-fugacity" and/or mass for the hypermultiplet).

For our purposes, it is enough to localize hypermultiplets in a background of localized vector multiplets. From the previous subsection we know that all the vector multiplet fields vanish in such a background, except for the component $A_y$ which takes on constant values parametrizing flat connections on $S^3\times S^1_y$. At this point of the computation, there is no distinction between the gauge and flavor background holonomies, but of course at the end we will integrate over the former but not over the latter. So, with the normalization as in~\eqref{flat_conn}, we have
\begin{equation}
A  \ = \  \frac{a}{\beta\ell}\dd y\,,\quad \text{with} \quad a\in \mathfrak{t} = \mathrm{Lie} \, \mathbb{T}\,.
\end{equation}
Furthermore, due to $\alpha$ being non-zero, there is a non-vanishing R-symmetry holonomy, given by
\begin{equation}
V^B{}_A  \ = \  \frac{\alpha}{2\ell}\dd y \, (\tau_3)^B{}_A  \,, \quad \text{with} \quad \alpha \ = \ \frac{2\pi N}{\beta} \,.
\end{equation}

Note that only the $\cD_y$ component of the covariant derivative in~\eqref{covdq} acting on $q_{AI}$ involves such holonomies, while the other derivatives can be (and are) replaced by the usual partial derivative $\partial_\mu$ in what follows. We can solve the BPS-equations for the auxiliary fields and upon substituting back into the equations, we obtain
\be\label{tauinvq}
\partial_{\tau} q_{AI} \ = \ 0  \,,
\ee
which effectively compactifies fields in the $\tau$ direction: $\tau$-independent configurations are described by fields on $D^2 \times S^1_y$, where $D^2$ is parametrized by $(\theta, \varphi)$.

Now, to proceed, we impose the reality conditions on $F$ and $q$ as given in~\eqref{hyp_real}. With these conditions, the remaining BPS equations become
\begin{align}\label{BPScoordsfugs01}
\left(\frac1{\cos\theta}\partial_\theta - \frac{\ii}{\sin\theta}\partial_\varphi \right)\left(e^{+\frac{y}{2\ell}} q_{1I} \right) &  \ = \ 
\ii e^{\ii\varphi +\frac{\ii (\zeta - \alpha)  y}{\ell }}(\ell\cD_y-\zeta\partial_\varphi) \left(e^{-\frac{y}{2\ell}}q_{2I} \right) \,,\\
\label{BPScoordsfugs02}
\left(\frac1{\cos\theta}\partial_\theta + \frac{\ii}{\sin\theta}\partial_\varphi \right)\left(e^{-\frac{y}{2\ell}} q_{2I} \right) &  \ = \ 
\ii e^{-\ii\varphi -\frac{\ii (\zeta-\alpha)  y}{\ell }}(\ell\cD_y -\zeta\partial_\varphi) \left(e^{\frac{y}{2\ell}}q_{1I} \right) \,.
\end{align}
Additionally, we get expressions for the auxiliary fields $F_{AI}$ in terms of $q_{AI}$ given as solutions of the partial differential equations~\eqref{BPScoordsfugs01} and~\eqref{BPScoordsfugs02},
\be
F_{AI}  \ = \  \frac{\ii}{2 \ell } \left(2 \tan\theta\, \partial_{\theta}+1\right)  q_{AI} \,.
\ee

To simplify~\eqref{BPScoordsfugs01} and~\eqref{BPScoordsfugs02}, it is useful to introduce complex coordinates $(v,\bar v)$ on the $(\theta,\varphi)$-disk, which undergo a twist as we move along the $y$ direction
\begin{equation}
\label{v_coord}
v \ = \ \sin\theta\,e^{\ii\left(\varphi+\zeta \frac{y}{\ell}\right)} \,.
\end{equation}
In terms of this coordinate, the BPS equations take the more suggestive form:
\begin{align}\label{BPScoordsfugs1}
\partial_v\left(e^{+\frac{y}{2\ell}} q_{1I} \right) &  \ = \ 
\frac{\ii}{2} e^{-\frac{\ii \alpha  y}{\ell }}\ell\cD_y \left(e^{-\frac{y}{2\ell}}q_{2I} \right) \,,\\
\label{BPScoordsfugs2}
\partial_{\bar v}\left(e^{-\frac{y}{2\ell}} q_{2I} \right) &  \ = \ 
\frac{\ii}{2} e^{\frac{\ii \alpha  y}{\ell }}\ell\cD_y \left(e^{\frac{y}{2\ell}}q_{1I} \right) \,.
\end{align}
To solve these equations, let us expand the fields $q_{AI}$ in terms of Fourier modes of $y$ (taking into account the twisted sector periodicity)
\begin{align}
\label{BPS_yModes1}
q_{1I} & \ = \  \sum_{n\in\Z}a^+_{I,n}(v,\bar{v}) e^{\frac{2\pi\ii n y}{\beta\ell}-\frac{y}{2\ell}}\,,\\
\label{BPS_yModes2}
\ii\, q_{2I} & \ = \  \sum_{n\in\Z}a^-_{I,n}(v,\bar{v}) e^{\frac{2\pi\ii n y}{\beta\ell}+\frac{y}{2\ell}}\,.
\end{align}
This separates the variables $v, \bar{v}$ from $y$, and including the background gauge holonomy $a$ and the R-symmetry holonomy $\alpha=2\pi N/\beta$, we obtain the following equations for the modes
\begin{align}
\label{BPS_modes1}
\partial_v a_{I, n}^+ &  \ = \ 
+\frac{\ii}{2} \left(\frac{2\pi (n+N)}{\beta} - \frac{a}{\beta}\right)a^-_{I, n+N} \,,\\
\label{BPS_modes2}
\partial_{\bar v}a_{I,n+N}^- &  \ = \ 
-\frac{\ii}{2} \left(\frac{2\pi n}{\beta} - \frac{a}{\beta}\right)a^+_{I, n} \,.
\end{align}
For ease of notation, let us define
\be
\sigma_{n} \ \equiv \ \frac{2\pi n}{\beta} - \frac{a}{\beta}\,.
\ee
Solutions to these equations that are regular on the disk $D^{2}$ take the following form
\begin{align}
a^+_{I,n}(v,\bar{v}) & \ = \  \sum_{m\in\Z} \ii\, x_I^{m,n}\sqrt{\frac{\sigma_{n+N}}{\sigma_{n}}} I_m(\sqrt{\sigma_n\sigma_{n+N}}\sin\theta)e^{\ii m \left(\varphi+\zeta \frac{y}{\ell}\right)}\,,\\
a^-_{I,n+N}(v,\bar{v}) & \ = \  \sum_{m\in\Z} x_I^{m+1,n} I_m(\sqrt{\sigma_n\sigma_{n+N}}\sin\theta)e^{\ii m \left(\varphi+\zeta \frac{y}{\ell}\right)}\,,
\end{align}
where $I_m$ are modified Bessel functions, and $x_I^{m,n}$ are complex constants. Inserting them back into the Fourier series, we find the following most general smooth solution on $D^2\times S^1_y$ in the form of a double Fourier series on $S^1_\varphi\times S^1_y$
\begin{align}
\label{BPS_sol}
q_{1I} & \ = \  \sum_{n,m\in\Z}\ii\,  x_I^{m,n} \sqrt{\frac{\sigma_{n+N}}{\sigma_{n}}} I_m(\sqrt{\sigma_n\sigma_{n+N}}\sin\theta)e^{\ii m \left(\varphi+\zeta \frac{y}{\ell}\right)+\frac{2\pi\ii n y}{\beta\ell}-\frac{y}{2\ell}}\,,\\
\ii\, q_{2I} & \ = \ \sum_{n,m\in\Z}x_I^{m+1,n-N} I_m(\sqrt{\sigma_{n-N}\sigma_n}\sin\theta)e^{\ii m \left(\varphi+\zeta \frac{y}{\ell}\right)+\frac{2\pi\ii n y}{\beta\ell}+\frac{y}{2\ell}}\,.
\end{align}
We can plug this into the definition of the two-dimensional fields~\eqref{symp_bos} to find that at the boundary $\theta=\pi/2$ of $D^2 \times S^1_y$ we obtain
\begin{align}
\label{ZI_formula}
\sZ_I(\varphi, y)  \ = \  \sqrt{\ell} e^{-\frac{\ii(\zeta-\alpha) y}{ 2\ell}-\frac{\ii}{2}\varphi} \sum_{n,m\in\Z} & x_I^{m,n}\left(I_{m-1}(\sqrt{\sigma_n\sigma_{n+N}}) + \ii\sqrt{\frac{\sigma_{n+N}}{\sigma_n}} I_{m}(\sqrt{\sigma_n\sigma_{n+N}})\right)\cr
&\quad\times e^{\ii m\left(\varphi + \zeta \frac{y}{\ell}\right) + \frac{2\pi\ii n y}{\beta\ell}}\,.
\end{align}
This expression immediately implies that all the modes $x_I^{m,n}$ are encoded in the two-dimensional fields $\sZ_I$ living on $S^1_\varphi\times S^1_y$. In other words, regular solutions to the BPS equations -- \emph{i.e.} the localization locus -- are indeed parametrized by fields on the torus (symplectic bosons, as we will see soon). The above equation also shows that the symplectic bosons $\sZ_I$ are naturally anti-periodic in the $\varphi$ direction, while the periodicity in the $y$ direction is determined by $\zeta-\alpha=2\pi(M-N)/\beta$: they are periodic for $M-N \equiv 0\ {\rm mod}\ 2$ and anti-periodic for $M-N \equiv 1\ {\rm mod}\ 2$.

Now, let us turn to the hypermultiplet action~\eqref{hmaction}, and evaluate it on the localization locus. At the saddle point we have the following bosonic action
\be
\cS_{\rm mat} \Big|_{\rm BPS}   \ = \ \int_{S^3 \times S^1} \left[ 
\frac{1}{2} \cD_{\mu} q^{A \, I} \cD^{\mu} q_{A \, I} 
+ \frac{1}{2 \ell^{2}}  q^{A \, I}q_{A\, I} 
 - \frac{1}{2} F^{{A} \, I} F_{{A}\, I}  \right] \vol_{S^3 \times S^1} \,,
\ee
where we integrate over the whole spacetime $S^3 \times S^1$, with $\vol_{S^3 \times S^1} = \ell^{3} \sin \theta \cos \theta\,\dd\theta\dd\varphi\dd\tau\dd y$. Now, using $\tau$-invariance and the expression for $F_{AI}$, we can rewrite this as
\be
\cS_{\rm mat} \Big|_{\rm BPS} \ = \ 
- \left[ \int_{0}^{2\pi}\ell\dd \tau \right]\int_{S^{1}_y\times D^{2}} 
\left( \nabla_{a} \Phi^{a} \right) \vol_{S^{1}_y\times D^{2}}  \,,
\ee
where we defined
\bea
\Phi^{a} & = & \frac{1}{\sqrt{S \bar S}}\epsilon^{abcd} \left( \xi^{A} \sigma_{b} \bar\xi_{A} \right)\left( \xi_{B} \sigma_{c} \bar \xi_{C} \right) q^{B}{}_{I}\cD_{d}q^{C I}
- 2\ii \sqrt{\frac{S}{\bar S}}\left( \xi^{\prime}{}_{A}\sigma^{a} \bar\xi_{B} \right) q^{A}{}_{I}q^{B I} \,,
\eea
and $S^{1}_y\times D^{2}$ is the space parametrized by $(y,\theta,\varphi)$, with $\vol_{S^{1}_y\times D^{2}} = \ell^{2} \sin \theta\, \dd y\dd\theta\dd\varphi$. Furthermore, we defined the quantities $S$ and $\bar S$ as follows
\be
S \ = \ \xi^{A} \xi_{A} \ = \ -2 e^{\ii \tau} \cos \theta \,,\qquad \bar S \ = \ \bar{\xi}^{A}\bar \xi_{A} \ = \ -2 e^{-\ii \tau} \cos \theta \,.
\ee
Hence, we apply Stokes' theorem and find that the now three-dimensional action further reduces to a two-dimensional action on the boundary torus $\partial (S^{1}_y\times D^{2}) = S^{1}_{y}\times S^{1}_{\varphi}$,
\be
\cS_{\rm mat} \Big|_{\rm BPS}  \ = \ 
- 2\pi \ell^{2} \int_{S_{y}^{1}\times S_{\varphi}^{1}} \Phi \cdot \dd  S\,,
\ee
where $\dd S$ is along the normal (\emph{i.e.} along the frame $e_1 = \frac1{\ell}\frac{\partial}{\partial \theta}$) of the surface $S_{y}^{1}\times S_{\varphi}^{1}$ . We now define the elementary fields $\sZ_I$ in the $\cQ^{H}$-cohomology as in~\eqref{symp_bos}. With this definition we can write the remaining action on torus as
\beaa\label{symp_action1}
\cS_{\rm mat} \Big|_{\rm BPS}  \ &= \  
4\pi \ii \int_{S^{1}_{y}\times S^{1}_{\varphi}} \dd\varphi\dd y\,
\varepsilon^{IJ}\sZ_I \nabla_{\bar w} \sZ_J\cr
&= \ 4\pi \ii\ell \int_{S^{1}_{y}\times S^{1}_{\varphi}} \dd^2w\,
\varepsilon^{IJ}\sZ_I \nabla_{\bar w} \sZ_J \,,
\eeaa
where
\be
\nabla_{\bar w} \ = \ \frac{\ii}{2}\left(1 + \ii \zeta \right) \partial_{\varphi } + \frac{\ell}{2} \cD_y \ = \ \partial_{\bar w} -\ii\frac{a}{2\beta}\, ,
\ee
and we used the complex coordinates $(w, \bar w)$ on the torus previously introduced in~\eqref{w_with_zeta}:
\begin{equation}
w  \ = \  \frac{y}{\ell} + \ii \left(\varphi + \zeta \frac{y}{\ell}\right)\,,\qquad \bar{w} \ = \ \frac{y}{\ell} - \ii \left(\varphi + \zeta \frac{y}{\ell}\right)\,.
\end{equation}
Notice that the induced two-dimensional metric on $S^{1}_{y}\times S^{1}_{\varphi}$ is given by
\be
\dd s^{2}_{2} \ = \ \ell ^2 \left[ \dd \phi^2+ \frac{2\zeta}{\ell} \dd \phi \dd y +\frac{1}{\ell ^2} \left(1+\zeta^2 \right)\dd y^{2} \right] \ = \ \ell^2 \dd w\dd\bar{w} \,,
\ee
where $\zeta$ can be identified as the complex structure parameter of the torus. 

If we replace $S^1_y$ by a line, $w, \bar{w}$ turn into coordinates on the cylinder and -- as previously mentioned -- we can conformally map it to the plane via (assume $\zeta=0$)
\begin{equation}
z \ = \  e^w\,,\qquad \bar{z} \ = \ e^{\bar w} \,,
\end{equation}
so the fields transform as
\begin{equation}
\sZ_I \ = \ \sqrt{z}\hat{\sZ}_I \,,
\end{equation}
and the action (assuming $a=0$) becomes $\propto \int \dd^2z\, \varepsilon^{IJ}\hat{\sZ}_I\partial_{\bar z}\hat{\sZ}_J$, \emph{i.e.} the symplectic boson on the plane. In this paper, however, we are mostly interested in the case when the two-dimensional theory lives on the torus.

The symplectic boson theory is of course a special case of the $\beta-\gamma$ system, where both fields are of conformal dimension $1/2$. Separating $\sZ_I$ into $\gamma$ and $\beta$ requires a choice of Lagrangian splitting of the representation space of half-hypermultiplets. In general, there is no preferred one. However, in the case of \emph{full} hypermultiplets, the representation of half-hypermultiplets is of the form $\rR\cong\cR \oplus\bar\cR$, where $\sZ_1\in\cR$ and $\sZ_2\in\bar\cR$. In that case, we can simply choose
\begin{equation}
\gamma \ = \ \sZ_1\,,\qquad \beta \ = \ \sZ_2\,,
\end{equation}
and the action becomes $8\pi\ii\ell\int\dd^2w\,\beta\nabla_{\bar w}\gamma$, \emph{i.e.} the standard action of the $\beta-\gamma$ system. However, because general superconformal field theories may also contain half-hypermultiplets, we will keep the action in the more general form of equation~\eqref{symp_action1}.

\subsubsection{One-Loop Determinant}

To complete the localization, we now need to determine the one-loop determinant $Z_{\rm mat}$ for small fluctuations around the BPS locus. In general, the one-loop determinant might depend on all parameters of the localization locus, which in our case are the holonomy $u$ and the symplectic bosons $\sZ_I(w, \bar{w})$. As it is easy to see in our problem, the quadratic action for small fluctuations of the hypermultiplet fields around the LL actually does not depend on $\sZ_I(w, \bar{w})$. Unlike the vector multiplet case, the supersymmetry variation $\cQ^H\psi$ is linear in the hypermultiplet fields, and hence the canonical localizing term $|\cQ^H\psi|^2+\dots$ is purely quadratic in the hypermultiplet fields, implying that the one-loop determinant can only depend on $u$.

To find $Z_{\rm mat}(u)$, we are going to utilize the fact that the four-dimensional hypermultiplet action is already quadratic after localizing the vector multiplet part. Therefore, we can compute the four-dimensional hypermultiplet path integral exactly, and equate it with the localization computation result for the hypermultiplet. In other words, we have the following relation:
\begin{align}\label{equality2d4d}
Z_{\rm mat}(u) \int \cD\sZ_I\, e^{-4\pi \ii\ell \int_{S^{1}_{y}\times S^{1}_{\varphi}} \dd^2w\,
	\varepsilon^{IJ}\sZ_I \nabla_{\bar w} \sZ_J}  \ = \  \int \cD\cH\, e^{-\cS_{\rm mat}[\cH,u]}\,.
\end{align}
So to find $Z_{\rm mat}(u)$, we simply have to compute the quadratic path integrals on the two sides of the equality and take their ratio.

The right hand side of~\eqref{equality2d4d} describes a free hypermultiplet on $S^3\times S^1_y$ coupled to the background $G$-holonomy $u$ and a possible $(R+r)$ holonomy $\alpha=2\pi N/\beta$. For $N=0\mod 2$, this is nothing but the flavored Schur index of a free hypermultiplet. Again referring to Appendix~\ref{App:N1intoN2} for a review of localization results in four dimensions, it is given by,
\begin{align}
&N=0\mod 2:\cr
&\eqref{equality2d4d} \ = \  q^{\frac{\dim\rR}{48} - \frac14 \sum_{w\in\rR}\langle w,a\rangle^2} \prod_{w\in\rR} \frac1{\left[\left( \sqrt{q} u^w; q \right)\left( \sqrt{q} u^{-w}; q \right)\right]^{1/2}}\,,\quad q \ = \ e^{-\beta(1+\ii\zeta)} \,,
\end{align}
where $\rR$ is the total representation of half-hypermultiplets (including the full hypermultiplets). With $N=1\mod 2$, we have an extra insertion of $(-1)^{2(R+r)}$, which does not affect the Casimir energy, but slightly modifies the Schur index of a hypermultiplet. Namely, the ``single-letters'' have $(-1)^{2(R+r)}=-1$, and so their contributions become $-\sqrt{q}$ instead of the usual $\sqrt{q}$. Thus, the answer reads
\begin{align}
&N=1\mod 2:\cr
&\eqref{equality2d4d} \ = \  q^{\frac{\dim\rR}{48} - \frac14 \sum_{w\in\rR}\langle w,a\rangle^2} \prod_{w\in\rR} \frac1{\left[\left( -\sqrt{q} u^w; q \right)\left( -\sqrt{q} u^{-w}; q \right)\right]^{1/2}}\,,
\end{align}
with $q=e^{-\beta(1+\ii\zeta)} $.

On the other hand, the two-dimensional symplectic boson path integral on $S^1_\varphi\times S^1_y$ -- anti-periodic in $\varphi$ and either periodic or anti-periodic in $y$ -- can be easily evaluated,
\begin{align}
&\text{Periodic in $y$:}\cr
&\int \cD\sZ_I\, e^{-4\pi \ii\ell \int_{S^{1}_{y}\times S^{1}_{\varphi}} \dd^2w\,
\varepsilon^{IJ}\sZ_I \nabla_{\bar w} \sZ_J} \ = \  q^{\frac{\dim\rR}{48}} \prod_{w\in\rR} \frac1{\left[\left( \sqrt{q} u^w; q \right)\left( \sqrt{q} u^{-w}; q \right)\right]^{1/2}}\,,\\
&\text{Anti-periodic in $y$:}\cr
&\int \cD\sZ_I\, e^{-4\pi \ii\ell \int_{S^{1}_{y}\times S^{1}_{\varphi}} \dd^2w\,
\varepsilon^{IJ}\sZ_I \nabla_{\bar w} \sZ_J} \ = \  q^{\frac{\dim\rR}{48}} \prod_{w\in\rR} \frac1{\left[\left( -\sqrt{q} u^w; q \right)\left( -\sqrt{q} u^{-w}; q \right)\right]^{1/2}}\,.
\end{align}
Matching this to the two expressions for Schur indices given above, implies that $Z_{\rm mat}(u)$ only receives contributions from the difference of Casimir energies in four and two dimensions,
\be
Z_{\rm mat}(u) \ = \ q^{- \frac14 \sum_{w\in\rR}\langle w,a\rangle^2} \,.
\ee
For a superconformal field theory (without flavor fugacities), this part of the Casimir term cancels against the similar one in the vector multiplet path integral. Namely, the $a$-dependent piece in~\eqref{Cas_vm} (which also did not cancel there) combines with what we have found here into
\begin{equation}
\sum_{\alpha\in\Delta^+} \langle\alpha,a\rangle^2 - \frac14 \sum_{w\in\rR}\langle w,a\rangle^2 \ = \ \frac12 \left(C(\mathbf{adj}) - \frac12 C(\rR)\right)\,\Tr a^2\,,
\end{equation}
where $C(\dots)$ denotes the index of the corresponding representation. The expression in parentheses is precisely the coefficient of the beta-function (for each simple gauge factor $G_i$) that vanishes in the superconformal field theory. One can also check that for each $G_i$,
\beaa
2C(\mathbf{adj}) - C(\rR) \ = \  & 2h^\vee(G_i) - \sum_{x}C(\rR_{i,x})\prod_{j\neq i}\dim\rR_{j,x} - \sum_{y} 2C(\cR_{i,y})\prod_{j\neq i}\dim\cR_{j,y} \ = \ 0 \,,
\eeaa
according to~\eqref{noanom}.

So in the end we find that in the absence of flavor fugacities, the one-loop determinants do not produce any additional factors, and the four-dimensional theory localizes precisely to the two-dimensional theory of gauged symplectic bosons
\begin{align}
\label{result_4d2d}
&\int \cD\cV\, \cD\cH e^{-\cS[\cV, \cH]} \ = \ \cr
&\hspace{.35 in} \frac{1}{|\cW|}\int_{\mathbb T} \frac{\dd u}{2\pi \ii u}\Delta_1(u) \int \cD\sZ_I \cD b\cD c\, \delta\left(\int b(\varphi,y)\dd\varphi\right) \delta\left(\int c(\varphi,y)\dd\varphi\right) e^{-\cS_{\rm 2d}} \,,
\end{align}
where $\cV$, $\cH$ denote the four-dimensional vector and hypermultiplets, and with the two-dimensional action given by
\begin{align}
\label{action2d}
\cS_{\rm 2d} \ = \ & \Tr\int_{S^1_y\times S^1_\varphi}\dd y\dd\varphi \, b\left(\bar\partial_w - \ii \frac{a}{2\beta}\right) c + 4\pi \ii \int_{S^{1}_{y}\times S^{1}_{\varphi}} \dd y\dd\varphi\,
\varepsilon^{IJ}\sZ_I \nabla_{\bar w} \sZ_J\,,
\end{align}
with
\begin{align}
\nabla_{\bar w}  \ = \  \partial_{\bar w} - \ii\frac{a}{2\beta}\,,
\end{align}
and where one integrates $u=e^{\ii a}$ over the maximal torus $\mathbb{T}\subset G$. Furthermore, changing periodicity of symplectic bosons along $S^1_y$ corresponds to the insertion of $(-1)^{2(R+r)}$ in the Schur index.

If we turn on flavor fugacities, \emph{i.e.} treat some components of $a$ as vevs $a_f$ of the background vector multiplets, then $Z_{\rm mat}(u)$ does not fully cancel against the vector multiplet contribution. In this case, there is a leftover factor producing a mismatch between the left and right hand sides of~\eqref{result_4d2d} given by the ``flavor Casimir term'',
\begin{equation}
\label{flav_Cas}
Z^{\rm f}_{\rm Cas}(q, a_f)  \ = \  q^{-\frac14 \sum_{w_f\in\mathbf{R}_f}\langle w_f, a_f\rangle^2}\,,
\end{equation}
where the summation is over weights of the flavor group representation.

The correlation functions of general BRST-closed observables (built from the $\sZ_I$ and $bc$ ghosts) can now be computed using the two-dimensional Gaussian path integral~\eqref{result_4d2d}-\eqref{action2d}, which is the final output of our localization computation in the absence of surface defects. They capture the full chiral algebra of the four-dimensional $\cN=2$ superconformal field theory.

Let us now also briefly discuss the Casimir energy in the presence of the deformations in terms of the integers $M$ and $N$. The way we have introduced those deformations as special values of the more general fugacities $\zeta$ and $\alpha$ (see equation~\eqref{alphazetaconds}), one would naively expect that the corresponding Casimir energy is simply obtained by evaluating the expressions of the $\cN=2$ Casimir energy in terms of $\zeta$ and $\alpha$ in equations~\eqref{eqn:Casimirfull} at those special values. However, recalling the definition of the (scheme-independent) supersymmetric Casimir energy as~\cite{Assel:2015nca}
\begin{equation}\label{eqn:defCas}
E_{\rm susy} \ = \ - \lim_{\beta\to \infty} \frac{\dd }{\dd \beta} Z_{S^{3} \times S^{1}_{\beta}} \,,
\end{equation}
we observe that $M$ and $N$ cannot contribute to it, as the special values of $\zeta$ and $\alpha$ in equation~\eqref{alphazetaconds} are explicitly $\beta$-dependent. 

Alternatively, we may apply a result of~\cite{Martelli:2015kuk}, which relates the supersymmetric Casimir energy,\footnote{Notice that, given a particular ``supersymmetric" regularization scheme, it was shown in~\cite{Assel:2015nca}, that the two definitions~\eqref{eqn:defCas} and~\eqref{eqn:defCas1} are in fact equivalent.}
\begin{equation}\label{eqn:defCas1}
E_{\rm susy} \ = \ \langle H_{\rm susy} \rangle\,,
\end{equation}
where $H_{\rm susy}$ is the supersymmetric Hamiltonian in the one-dimensional quantum mechanics, for $\cN=1$ theories on $S^{3} \times S^{1}$, to a special limit of the ``single letters" of the index, where one explicitly employs the ``supersymmetric" regularization scheme. However, as discussed above, the Schur index and thus its single-letters are unaffected by the inclusion of $M$ and $N$, and thus neither is the Casimir energy.

\subsubsection{Inclusion of the R-symmetry defect}
Let us now see how the analysis changes in the presence of a surface defect that creates the $(-1)^{2(R+r)}$ monodromy. Recall that the defect is located at $\theta=0$, stretching in the $\tau$ and $y$ directions, making all the hypermultiplet fields anti-periodic in the $\varphi$ coordinate. All the analysis above up to (and including) the equations~\eqref{BPS_modes1}-\eqref{BPS_modes2} holds without changes. The next step -- the Fourier expansion of $a^{\pm}_{I,n}$ as a function of $\varphi$ -- is modified since now we have to expand in half-integral modes. As a result, the general solution, if we require $q_{11}\sim \theta^{1/2}$ and $q_{21}\sim \theta^{-1/2}$ at $\theta=0$, is
\begin{align}
\label{ap1}
a^+_{I,n}(v,\bar{v}) \ = \ & \sum_{m\in\Z}\ii\, x_I^{m,n}\sqrt{\frac{\sigma_{n+N}}{\sigma_n}}I_{|m+\frac12|}(\sqrt{\sigma_n\sigma_{n+N}}\sin\theta) e^{\ii\left(m+\frac12\right)\left(\varphi+\zeta\frac{y}{\ell}\right)}\,,\\
\label{am1}
a^-_{I,n+N}(v,\bar{v})  \ = \ &  x_I^{0,n}I_{-\frac12}(\sqrt{\sigma_n\sigma_{n+N}}\sin\theta)e^{-\frac{\ii}{2}\left(\varphi+\zeta\frac{y}{\ell}\right)}\cr 
&+\sum_{\substack{m\in\Z\\ m\neq 0}} x_I^{m,n}I_{|m-\frac12|}(\sqrt{\sigma_n\sigma_{n+N}}\sin\theta)e^{\ii\left(m-\frac12\right)\left(\varphi+\zeta\frac{y}{\ell}\right)}\,.
\end{align}
Similarly, if we require $q_{11}\sim \theta^{-1/2}$ and $q_{21}\sim \theta^{1/2}$ at $\theta=0$, we find the following solution,
\begin{align}
\label{ap2}
a^+_{I,n}(v,\bar{v})  \ = \ & \ii\, x_I^{0,n}\sqrt{\frac{\sigma_{n+N}}{\sigma_n}}I_{-\frac12}(\sqrt{\sigma_n\sigma_{n+N}}\sin\theta) e^{\frac{\ii}{2}\left(\varphi+\zeta\frac{y}{\ell}\right)}\cr 
&+ \sum_{\substack{m\in\Z\\ m\neq 0}}\ii\, x_I^{m,n}\sqrt{\frac{\sigma_{n+N}}{\sigma_n}}I_{|m+\frac12|}(\sqrt{\sigma_n\sigma_{n+N}}\sin\theta) e^{\ii\left(m+\frac12\right)\left(\varphi+\zeta\frac{y}{\ell}\right)}\,,\\
\label{am2}
a^-_{I,n+N}(v,\bar{v})  \ = \ & \sum_{m\in\Z} x_I^{m,n}I_{|m-\frac12|}(\sqrt{\sigma_n\sigma_{n+N}}\sin\theta)e^{\ii\left(m-\frac12\right)\left(\varphi+\zeta\frac{y}{\ell}\right)}\,.
\end{align}
Using equations~\eqref{BPS_yModes1}-\eqref{BPS_yModes2} and the definition of $\sZ_I$, we find that $\sZ_I$ are now periodic in $\varphi$ and take the form
\begin{equation}
\sZ_I \ = \ e^{\frac{\ii\alpha y}{2\ell}}\sum_{m,n\in\Z} x_I^{m,n}(\dots) e^{\ii m\left(\varphi+\zeta\frac{y}{\ell}\right) + \frac{2\pi\ii ny}{\beta\ell}}\,,
\end{equation}
while previously they were anti-periodic in $\varphi$. Here $(\dots)$ is the combination of modified Bessel functions similar to that appearing in~\eqref{ZI_formula}; it depends on which of the two cases,~\eqref{ap1}-\eqref{am1} or~\eqref{ap2}-\eqref{am2}, we consider, and its precise form is not important. The four-dimensional action reduces to the same two-dimensional action~\eqref{action2d} as before, but now with $\sZ_I$ periodic in $\varphi$.

The one-loop determinant for hypermultiplets $Z_{\rm mat}(u)$ might of course be modified by the defect.\footnote{Recall that the vector multiplets are not affected, as the defect is defined to be invisible for them.} We are going to determine this indirectly now. Similar to the previous case, the path integral answer factorizes into an ``index'' part that counts operators and a ``Casimir'' part encoding the vacuum properties:
\begin{equation}
 Z_{\rm mat}(u)\int\cD\sZ_I e^{-4\pi\ii\ell\int _{S^1_\varphi\times S^1_y} \dd^2w\, \varepsilon^{IJ}\sZ_I\nabla_{\bar w}\sZ_J}  \ = \  \int\cD\cH e^{-\cS_{\rm mat}[\cH,u]} \ = \  Z_{\rm Cas}(q,u) \cI(q,u) \,,
\end{equation}
where in our notation, we drop the dependence on $q=e^{-\beta(1+\ii\zeta)}$ on the left.

The index $\cI$ still counts Schur operators, simply because $\cQ^H$ is conserved. However, the state-operator correspondence is slightly modified in the presence of a defect: the $E$ quantum number of a Schur \emph{state} receives a $\pm\frac12$ shift compared to the $E$ quantum number of the corresponding Schur \emph{operator} (this is related to the shift of angular momentum in the presence of a defect, as mentioned in~\cite{Cordova:2017mhb}). One way to understand this is by going back to flat space. Namely, we again replace $S^3\times S^1_y$ with $S^3\times \R$ and Weyl transform the latter cylinder to the $\R^4=\C_{z_1} \times \C_{z_2}$, with the chiral algebra plane being $\C_{z_1}\times \{0\}$ and the defect supported at $\{0\}\times \C_{z_2}$. Usually, to construct states in the radial quantization Hilbert space $\cH_{S^3}$, we bring operators to the origin of $\R^4$ and observe what states they create in $\cH_{S^3}$:
\begin{equation}
\cO \ \rightarrow \ |\cO\rangle  \ = \  \cO(0)|0\rangle\,, \quad \text{ where $|0\rangle$ is the no-defect vacuum}.
\end{equation}
Because our defect is defined to be invisible for vector multiplets, this procedure is not modified for the Schur operators built out of gaugini. However, the free hypermultiplet Schur operators,\footnote{Recall that it is enough to consider a free hypermultiplet coupled to a background flat connection for our purposes of computing the one-loop determinant.} which are $q_{11}$, $q_{12}$, and their $D_{+\dot{+}}$ derivatives, do feel the presence of a defect. Namely, for the two versions of the defect, they behave as (suppose we move them towards the origin along the $w=0$ plane)
\begin{align}
\text{Version I: } \qquad q_{11}(z_1,\bar z_1)|0\rangle_D & \ \sim \ \sqrt{r} |q_{11}\rangle_D + O(r^{3/2})\,,\\ 
 q_{12}(z_1,\bar z_1)|0\rangle_D & \ \sim \ \frac1{\sqrt{r}} |q_{12}\rangle_D + O(r^{1/2})\,,\\
\text{Version II: }\qquad q_{11}(z_1,\bar z_1)|0\rangle_D & \ \sim \ \frac1{\sqrt{r}} |q_{11}\rangle_D + O(r^{1/2})\,,\\ 
 q_{12}(z_1,\bar z_1)|0\rangle_D & \ \sim \ \sqrt{r} |q_{12}\rangle_D + O(r^{3/2})\,,
\end{align}
where $|0\rangle_D$ is the defect vacuum, and $r=|z_1|$ denotes the distance to the origin. Here $|q_{11}\rangle_D$ and $|q_{12}\rangle_D$ denote the corresponding Schur states in the radial quantization Hilbert space with the defect. These are the states counted by the Schur index. Explicit factors of $r^{\pm 1/2}$ show that conformal dimensions (eigenvalues of $E$) of such states are shifted compared to those of the operators by $\pm\frac12$. The R-charges are of course not modified, so this shift is straightforwardly incorporated into the Schur index ${\rm Tr} (-1)^F q^{E-R} u^f$ of a free hypermultiplet coupled to the background flat connection. Namely, every single-letter contribution should be additionally multiplied by $q^{1/2}$ or $q^{-1/2}$, reflecting the respective shift of $E$. As before, we consider the cases $N=0 \mod 2$ and $N=1\mod 2$, and find,
\begin{align}
\text{The case of $N=0\mod2$:}\cr
\quad\text{Version I:}\qquad
\cI(q,u)& \ = \ \prod_{w\in\rR} \frac1{\left[\left( q u^w; q \right)\left( u^{-w}; q \right)\right]^{1/2}}\,,\\
\quad\text{Version II:}\qquad
\cI(q,u)& \ = \ \prod_{w\in\rR} \frac1{\left[\left( u^w; q \right)\left( q u^{-w}; q \right)\right]^{1/2}}\,,\\
\text{The case of $N=1\mod2$:}\cr
\quad\text{Version I:}\qquad
\cI(q,u)& \ = \ \prod_{w\in\rR} \frac1{\left[\left( -q u^w; q \right)\left( -u^{-w}; q \right)\right]^{1/2}}\,,\\
\quad\text{Version II:}\qquad
\cI(q,u)& \ = \ \prod_{w\in\rR} \frac1{\left[\left( -u^w; q \right)\left( -q u^{-w}; q \right)\right]^{1/2}}\,.
\end{align}
In fact,\footnote{Because $\mathbf{R}$ is pseudoreal, even a stronger statement is true: for each $w\in\mathbf{R}$, there is $-w\in\mathbf{R}$.} $\sum_{w\in\mathbf{R}}w=0$, and one can easily check that in both cases, the ``Version I'' and ``Version II'' answers coincide, so there is no need to distinguish them in the following. We can also compute the two-dimensional path integral for symplectic bosons periodic in $\varphi$ and either periodic or anti-periodic in $y$,\footnote{Notice how the two-torus path integral does not distinguish the two Ramond modules. In order for the path integral answer to match the characters of these modules, one needs to assign appropriate opposite charges to the two Ramond vacua. In our case this does not matter, again due to $\sum_{w\in\mathbf{R}}w=0$.}
\begin{align}
&\text{Periodic in $y$:}\cr
&\int \cD\sZ_I\, e^{-4\pi \ii\ell \int_{S^{1}_{y}\times S^{1}_{\varphi}} \dd^2w\,
\varepsilon^{IJ}\sZ_I \nabla_{\bar w} \sZ_J} \ = \  q^{-\frac{\dim\rR}{24}} \prod_{w\in\rR} \frac{\left[u^{w/2}-u^{-w/2}\right]^{-1/2}}{\left[\left( q u^w; q \right)\left( q u^{-w}; q \right)\right]^{1/2}}\,,\\
&\text{Anti-periodic in $y$:}\cr
&\int \cD\sZ_I\, e^{-4\pi \ii\ell \int_{S^{1}_{y}\times S^{1}_{\varphi}} \dd^2w\,
\varepsilon^{IJ}\sZ_I \nabla_{\bar w} \sZ_J} \ = \  q^{-\frac{\dim\rR}{24}} \prod_{w\in\rR} \frac{\left[u^{w/2}+u^{-w/2}\right]^{-1/2}}{\left[\left( -q u^w; q \right)\left( -q u^{-w}; q \right)\right]^{1/2}}\,.
\end{align}
We find that the former matches the defect Schur index for $N=0\mod2$ if we include the appropriate four-dimensional Casimir energy factor, while the latter matches the $N=1\mod2$ case of the defect Schur index under the same assumption. We claim that the four-dimensional Casimir factor takes the following form in all cases:
\begin{align}
Z_{\rm Cas}(q,u)& \ = \ q^{-\frac{\dim\rR}{24}-\frac14\sum_{w\in\mathbf{R}}\langle w,a\rangle^2}\,,
\end{align}
where the additional term $\frac14\sum_{w\in\mathbf{R}}\langle w,a\rangle^2$ was included by hand to cancel similar term in the vector multiplet Casimir energy (so that the total two-dimensional answer matches the four-dimensional answer precisely). In particular, this implies that the matter one-loop determinant takes the same form as in the absence of defects,
\begin{equation}
Z_{\rm mat}(u) \ = \ q^{-\frac14\sum_{w\in\mathbf{R}}\langle w,a\rangle^2}\,,
\end{equation}
which heuristically makes sense because half of the hypermultiplet scalars receive a shift of $+\frac12$, and half are shifted by $-\frac12$, making it plausible for the total contribution to stay the same. It would be instructive to verify this by a direct evaluation of the one-loop determinant in the presence of a defect.

To summarize, we claim here that the precise equality of four-dimensional and two-dimensional path integrals~\eqref{result_4d2d} continues to hold in the presence of monodromy defects, with the only modification that the symplectic bosons become periodic in $\varphi$. Notice, however, that generally both sides in~\eqref{result_4d2d} might be infinite: in the two-dimensional language this is caused by the zero modes present in the Ramond sector of the $\beta-\gamma$ system. For this reason, we must turn on generic flavor fugacities, which render both sides in~\eqref{result_4d2d} finite. We expect the equality in~\eqref{result_4d2d} to remain true if we include the appropriate ``flavor Casimir term'' given in~\eqref{flav_Cas}.

\subsection{Deformations}

Let us now turn towards a short survey of the possible ``standard" deformations of our localization construction, and whether they preserve the two-dimensional chiral algebra construction (\emph{i.e.} whether or not $\cQ^{H}$ is preserved). We have already argued that deformations away from the Schur fugacities (\emph{i.e.} $\alpha, \zeta \neq 0$) are only allowed through the addition of $\zeta$ and $\alpha$ as in~\eqref{alphazetaconds}. In this section, we briefly explore the addition of other types of deformations including masses, Fayet-Iliopoulos (FI) parameters and holonomies.

We are exclusively dealing with superconformal field theories, but one could ask whether our construction admits a non-conformal  mass deformation. To answer this, note that the conservation of $\cQ^H$ requires that the $U(1)_r$ R-symmetry is preserved. On the other hand, supersymmetric twisted masses are introduced by giving a vev to a scalar $\phi$ in the background vector multiplet. This scalar has a $U(1)_r$-charge $r=1$, and thus its vev explicitly breaks the $U(1)_r$ symmetry. Therefore mass deformations are not available.

The inclusion of FI terms is also not possible, for two separate reasons, both related to the $U(1)_r$ symmetry. One trivial reason is that they can only be introduced for abelian factors of the gauge group, and there are no interacting four-dimensional $\cN=2$ superconformal field theories with abelian factors in the gauge group (any such theory would break $U(1)_r$ and conformal invariance quantum mechanically, as one can see from~\eqref{noanom}). But even classically, the standard FI term explicitly breaks $U(1)_r$. Indeed, supersymmetry allows the addition of~\cite{Hosomichi:2016flq}
\be
\cL_{\rm FI} \ = \ \zeta \left\{ \omega^{AB} D_{AB} - M \left( \phi + \bar \phi \right)  \right\}
\ee
to $\cN=2$ theories in conformal supergravity (with $T_{\m\n}=0$), if there exists an $SU(2)_{R}$-triplet background field $\omega^{AB}$, which satisfies equations
\beaa\label{mFIconds}
\omega^{AB} \xi_{B} \ = \ & \frac{1}{2} \sigma^{\mu} \cD_{\mu} \bar \xi^{A}\,, \qquad \omega^{AB} \bar \xi_{B} \ = \ \frac{1}{2} \bar \sigma^{\mu} \cD_{\mu} \xi^{A} \,.
\eeaa
We see that the term $M \left( \phi + \bar \phi \right)$ in $\cL_{\rm FI}$ explicitly breaks $U(1)_r$.

There exists a non-standard FI term on $S^3\times S^1$, which preserves $U(1)_r$ classically. In fact, this term is a dimensional uplift of a $\cQ^H$-preserving FI term in three-dimensional $\cN=4$ theories~\cite{Dedushenko:2016jxl}. It is given by the following insertion in the path integral
\begin{equation}
\exp[S_{\rm F.I.}] \ = \ \exp\left[ \frac{\ii K}{2\pi^2\ell^2}\int_{S^3\times S^1}\vol_{S^3} \dd y\left( D_{12} + \frac{2}{\ell}A_y \right) \right] \,,
\end{equation}
where $D_{12}$ and $A_y$ are fields in the abelian vector multiplet. One can check that classically, such an insertion preserves all supersymmetries. Notice that large gauge transformations require that the FI parameter is integer, $K\in\Z$. Such FI terms were previously considered in~\cite{Romelsberger:2005eg,Aharony:2013dha,Closset:2013vra}, and we mention them for completeness. Of course, we already know that quantum mechanically, there are no interacting theories with abelian gauge factors that preserve $U(1)_r$, so this type of FI terms is not available either.

The class of deformations that \emph{do} preserve all the symmetries required by our construction are flavor holonomies, briefly mentioned in the previous subsection. They are given by constant abelian vevs for $A_y$ in the background vector multiplets gauging flavor symmetries of the system. They correspond to introducing flavor fugacities in the Schur index, and the possibility of such a refinement was of course already mentioned in~\cite{Beem:2013sza}. Note also that such a deformation is a dimensional uplift of the $\cQ^H$-invariant mass in three-dimensional $\cN=4$ theories~\cite{Dedushenko:2016jxl}. The only effect flavor holonomies have on the two-dimensional action~\eqref{action2d} is that they modify the covariant derivative acting on the symplectic bosons in an obvious way, \emph{i.e.}
\begin{equation}
\nabla_{\bar w} \ \to \ \nabla_{\bar w} - \ii \frac{a_f}{2\beta}\,,
\end{equation}
where $u_f=e^{\ii a_f}$ is the flavor holonomy around $S^1_y$. When flavor holonomies are required to render the Schur index finite, we will allow $a_f$ to be complex (\emph{i.e.} include both the holonomy and the ``chemical potential''). We will give another interpretation of this deformation from the VOA point of view further below.

\section{The two-dimensional theory}
\label{sec:2dthy}

The main result of the previous section is the two-dimensional theory~\eqref{action2d}, which we derived through a localization computation with respect to the supercharge $\cQ^H$ on $S^3\times S^1_y$. In this section, we are going to elucidate various of its properties. We start by describing its integration cycle, then describe its propagators of the elementary fields, and finally discuss the effect of adding flavor holonomies. 

\subsection{The integration cycle}

Note that the reality conditions~\eqref{hyp_real} of the four-dimensional theory induce reality properties of the symplectic bosons, \emph{i.e.}, they determine the integration cycle for the theory~\eqref{action2d}. We now briefly discuss this cycle.

First, we consider the case without monodromy defect, that is when $\sZ_I$ is anti-periodic in $\varphi$. Using the decomposition~\eqref{BPS_sol} of general solutions to the BPS equations and the reality conditions of the hypermultiplets~\eqref{hyp_real}, we can find the reality conditions for the modes $x_I^{m,n}$ in~\eqref{BPS_sol}. For $N=0$, they simplify, and take the following form
\begin{equation}
(x_I^{m,n})^* \ = \ (-1)^m\varepsilon^{IJ}x_J^{-m+1,-n} \,.
\end{equation}

For convenience, let us slightly redefine the modes in~\eqref{BPS_sol}, so that the expression for $\sZ_I$ reads
\begin{equation}
\label{Fourier_c}
\sZ_I(\varphi,y)  \ = \  e^{-\frac{\ii (\zeta-\alpha)y}{2\ell}-\frac{\ii}{2}\varphi} \sum_{n,m\in\Z} c_I^{m,n} e^{\ii m\left(\varphi + \zeta\frac{y}{\ell}\right) + \frac{2\pi\ii ny}{\beta\ell}} \,.
\end{equation}
Then, for $N=0$, the corresponding reality conditions for the $c_I^{m,n}$ modes become
\begin{equation}
(c_I^{n,m})^* \ = \ \ii\, \varepsilon^{IJ}c_J^{-m+1,-n}\frac{I_{m-1}(\sigma_n)-\ii I_m(\sigma_n)}{I_{m-1}(\sigma_n)+\ii I_m(\sigma_n)} \,.
\end{equation}
Note that $\frac{I_{m-1}(\sigma_n)-\ii I_m(\sigma_n)}{I_{m-1}(\sigma_n)+\ii I_m(\sigma_n)}$ is a phase which is closer to $1$ for $m\geq1$ and closer to $-1$ for $m\leq 0$ -- in other words, it is close to $\sgn\left(m-\frac12\right)$ (especially for large $|m|$). Because the above equation defines a middle-dimensional integration cycle in the space of complexified fields, one expects the path integral to be invariant under small deformations of the cycle, as long as the convergence is preserved along the deformations. In particular, we expect that the above cycle can be deformed to a simpler one given by
\begin{equation}
(c_I^{n,m})^* \ = \ \ii\, \sgn\left(m-\frac12\right)\varepsilon^{IJ}c_J^{-m+1,-n}\,.
\end{equation}
We will now verify that this is indeed true, and further find that for general $N\in\Z$, the analogous simplified integration cycle still has a concise expression given by
\begin{equation}
\label{c_real}
(c_I^{m,n})^*  \ = \  \ii\, \sgn\left(m-\frac12\right)\varepsilon^{IJ}c_J^{-m+1,-n-N} \,.
\end{equation}
To check this, we evaluate the two-dimensional symplectic boson action in terms of the modes $c_I^{m,n}$ in~\eqref{Fourier_c}
\beaa
\label{action_modes}
&4\pi\ii\int\dd y\dd\varphi\,\varepsilon^{IJ}\sZ_I\nabla_{\bar w}\sZ_J \ = \ \\
&\hspace{.2 in} \ = \ 4\pi^2\ii\beta\ell\sum_{m,n\in\Z} \varepsilon^{IJ}c_J^{-m+1,-n-N}\left(m-\frac12 -\frac{2\pi\ii(n+N/2)}{\beta}+\ii\frac{a}{\beta}\right)c_I^{m,n} \\
&\hspace{.2 in} \ = \ 4\pi^2\beta\ell\sum_{m,n\in\Z}(c_I^{m,n})^* \sgn\left(m-\frac12\right)\left(m-\frac12 -\frac{2\pi\ii(n+N/2)}{\beta}+\ii\frac{a}{\beta}\right)c_I^{m,n}\,,
\eeaa
where going from the second to the third line we used~\eqref{c_real}. Thus, we see that the real part of this action is positive definite along the integration cycle~\eqref{c_real}, ensuring convergence of the path integral, while the imaginary part is sign indefinite. Using standard Morse theory arguments~\cite{Witten:2010zr,Witten:2010cx} (see~\cite{Dedushenko:2016jxl} for an application close to the current context), one can check that the integration cycle~\eqref{c_real} (up to convergence-preserving deformations) is the only allowed one in this theory, as long as we keep $a$ purely real.\footnote{If we include background flavor holonomies $a_f$, they can be complex, as we mentioned previously, however their imaginary part should be small enough to not alter the positive definiteness of~\eqref{action_modes}.} We do not consider alternative integration cycles for $a$ until Section~\ref{sec:taste}, where they are expected to be relevant for the study of modularity and general surface defects in four dimensions (which reduce to modules of the two-dimensional chiral algebra).

As we explained previously, in the presence of the monodromy creating defect, the symplectic bosons become periodic in $\varphi$. The Fourier expansion thus takes the form
\begin{equation}
\label{Fourier_peri}
\sZ_I(\varphi,y)  \ = \  e^{\frac{\ii \alpha y}{2\ell}} \sum_{n,m\in\Z} s_I^{m,n} e^{\ii m\left(\varphi + \zeta\frac{y}{\ell}\right) + \frac{2\pi\ii ny}{\beta\ell}} \,,
\end{equation}
and we can evaluate the two-dimensional action in terms of modes $s_I^{m,n}$ as
\begin{align}
\label{action_perio1}
&4\pi\ii\int\dd y\dd\varphi\,\varepsilon^{IJ}\sZ_I\nabla_{\bar w}\sZ_J \cr
&\hspace{.3 in} \ = \ 4\pi^2\ii\beta\ell \sum_{m,n}\varepsilon^{IJ}s_J^{-m,-n-N}\left( m-\frac{2\pi\ii}{\beta}\left(n+\frac{N}{2}\right) + \ii\frac{a}{\beta} \right)s_I^{m,n} \,.
\end{align}
We see that the integration cycle should be
\begin{equation}
(s_I^{m,n})^* \ = \ \ii\sgn(m)\varepsilon^{IJ}s_J^{-m,-n-N},
\end{equation}
where we choose to extend $\sgn(0)=+1$. The action then becomes
\begin{equation}
\label{action_perio2}
4\pi^2\beta\ell\sum_{m,n} (s_I^{m,n})^* \sgn(m)\left( m-\frac{2\pi\ii}{\beta}\left(n+\frac{N}{2}\right) + \ii\frac{a}{\beta} \right)s_I^{m,n}\,.
\end{equation}
Recall that the full theory~\eqref{result_4d2d} involves integration over the holonomy variable $a$, and the presence of the zero mode $m=0$ in~\eqref{action_perio2} can cause this integration to diverge, which is typical for characters in the Ramond sector. As we have mentioned previously, this divergence is cured by turning on flavor holonomies, $a \to a+ a_f$, and furthermore, we should keep them complex. Equation~\eqref{action_perio2} manifests that positive definiteness of the action requires $\Im(a_f)$ to be small enough to not alter the convergence for $m\neq 0$ modes. Importantly, for convergence of the integral over $m=0$ modes, with the choice $\sgn(0)=+1$, we should keep the imaginary part of $a_f$ negative,
\begin{equation}
\Im(a_f) \ < \ 0\,.
\end{equation}

\subsection{Green's functions}

In this subsection we compute the torus propagators of the elementary two-dimensional fields. For symplectic bosons, we do it in all four spin structures described above. The equations in this subsection are reminiscent of those appearing in standard treatments of genus-one superstring amplitudes. Nevertheless, we provide a detailed derivation here, both for completeness, and because the ghost systems we deal with are slightly different from the ones appearing in superstrings, and sometimes require an extra amount of care.

\subsubsection{Fermionic propagator}

We start by expanding the small $bc$ ghosts in Fourier modes,
\begin{align}
b \ = \ \sum_{\substack{m,n\in\Z\\ m\neq 0}}b_{m,n} e^{\ii m\left(\varphi + \zeta\frac{y}{\ell} \right) + \frac{2\pi\ii ny}{\beta\ell}}\,,\qquad
c \ = \ \sum_{\substack{m,n\in\Z\\ m\neq 0}}c_{m,n} e^{\ii m\left(\varphi + \zeta\frac{y}{\ell} \right) + \frac{2\pi\ii ny}{\beta\ell}}\,,
\end{align}
where the $m=0$ modes are absent due to the delta-functions in~\eqref{bc_PI}. The action then takes the form
\begin{equation}
\cS_{\rm 2d}^{(bc)}  \ = \  -\pi\beta\ell\,\Tr \sum_{\substack{m,n\in\Z\\ m\neq 0}} b_{-m,-n} \left(m-\frac{2\pi\ii n}{\beta} + \ii\frac{a}{\beta} \right)c_{m,n}\,,
\end{equation} 
and the propagator is given by
\begin{align}
\langle c^A(w_1) b^B(w_2) \rangle & \ = \  P^{AB}(w_1-w_2, \bar{w}_1-\bar{w}_2)\,,\\
P(w,\bar{w}) & \ = \  -\frac{1}{\pi\beta\ell} \sum_{\substack{m,n\in\Z\\ m\neq 0}}\frac{e^{\ii m\left(\varphi + \zeta\frac{y}{\ell} \right) + \frac{2\pi\ii ny}{\beta\ell}}}{m-\frac{2\pi\ii n}{\beta} + \ii\frac{a}{\beta}}
\ \equiv \ \cG(w,\bar{w})+g(w,\bar{w})\,,
\end{align}
with
\begin{align}
\cG(w,\bar{w}) & \ = \  -\frac{1}{\pi\beta\ell} \sum_{m,n\in\Z}\frac{e^{\ii m\left(\varphi + \zeta\frac{y}{\ell} \right) + \frac{2\pi\ii ny}{\beta\ell}}}{m-\frac{2\pi\ii n}{\beta} + \ii\frac{a}{\beta}}\,,\\
g(w,\bar{w}) & \ = \  -\frac1{\pi\beta\ell}\sum_{n\in\Z} \frac{e^{\frac{2\pi\ii n y}{\beta\ell}}}{\frac{2\pi\ii n}{\beta} - \ii\frac{a}{\beta}}\,,
\end{align}
and where $A,B$ are the adjoint indices, and $P$ is a $|G|\times |G|$ matrix determined by the diagonal matrix $a\in\mathfrak{g}$. We have introduced an auxiliary Green's function $\cG$ that includes zero modes (and thus describes the \emph{full} $bc$ ghost system), and the correction term $g(w,\bar{w})$ that removes $m=0$ modes. This correction term is essentially a propagator of the one-dimensional analog of the $bc$ system.

The function $\cG(w,\bar{w})$ satisfies an equation
\begin{equation}
\nabla_{\bar w}\cG(w,\bar{w})  \ = \  \delta\left(\varphi\right) \delta(y)\,,
\end{equation}
and is required to be doubly-periodic on the torus, namely, it obeys (\emph{c.f.}~\eqref{w_with_zeta})
\begin{align}
\cG\left(w+2\pi\ii\tau,\bar{w}-2\pi\ii\bar\tau\right)& \ = \ \cG(w+2\pi\ii, \bar{w}-2\pi\ii) \ = \ \cG(w,\bar{w})\,,
\end{align}
where
\begin{align}
 \tau & \ = \ \frac{\ii\beta}{2\pi}(1+\ii\zeta) \,.
\end{align}
We can also define the function $\cG_0(w,\bar{w})$,
\begin{equation}
\cG(w,\bar{w})  \ = \  e^{\ii\frac{a}{2\beta}(w+\bar{w})}\cG_0(w,\bar{w})\,,
\end{equation}
which satisfies a simpler equation, but has a twisted periodicity along $\Re(w)$,
\begin{align}
\label{eqnG0}
\partial_{\bar w}\cG_0(w,\bar{w}) & \ = \  \delta\left(\varphi\right) \delta(y)\,,\\
\cG_0(w+2\pi\ii\tau,\bar{w}-2\pi\ii\bar\tau)& \ = \ e^{\ii a} \cG_0(w,\bar{w})\,.
\end{align}
The equation implies that the function is holomorphic away from $w=0$, and has a simple pole at $w=0$ with residue $(\pi\ell)^{-1}$. This would be impossible for $a=0$, since the function would have to be elliptic, and such functions have poles of total order at least two. However, it is possible at generic $a$. The standard way to construct functions of a given periodicity on elliptic curves is by taking ratios of Jacobi theta-functions. The basic Jacobi theta-function is defined as follows
\begin{equation}
\vartheta(z; \tau) \ = \ \sum_{n\in\Z} \exp(\ii\pi n^2\tau + 2\pi\ii nz) \,.
\end{equation}
In the present case, the relevant function is
\begin{equation}
\theta_1(z;\tau) \ = \ -\exp\left(\frac14\ii\pi\tau + \ii\pi(z+1/2) \right)\vartheta\left(z+\frac12 \tau + \frac12; \tau\right)\,.
\end{equation}
It has a simple zero at $z=0$ and satisfies 
\begin{equation}
\theta_1(z+n+m\tau; \tau) \ = \ e^{\ii\pi(n-m)-\ii\pi m^2\tau - 2\ii\pi mz}\theta_1(z;\tau)\,.
\end{equation} 
One can easily check that the following function,
\begin{equation}
\frac{\theta_1\left(\frac{w}{2\pi\ii}-\frac{a}{2\pi}; \tau\right)}{\theta_1\left(\frac{w}{2\pi\ii}; \tau\right)}\,,
\end{equation}
has a simple pole at $w=0$ and satisfies the periodicity as required for $\cG_0$. If we further normalize it to have a residue $(\pi\ell)^{-1}$ at $w=0$, as required by~\eqref{eqnG0}, we find
\begin{equation}
\label{G0sol}
\cG_0(w)  \ = \  \frac1{2\pi^2\ii\ell}\cdot\frac{\theta_1'\left(0; \tau\right)}{\theta_1\left(-\frac{a}{2\pi}; \tau\right)} \frac{\theta_1\left(\frac{w}{2\pi\ii}-\frac{a}{2\pi}; \tau\right)}{\theta_1\left(\frac{w}{2\pi\ii}; \tau\right)}\,,
\end{equation}
where $\theta_1'$ denotes the derivative in the first argument of $\theta_1$. This is actually a unique solution for $\cG_0$. Any other solution, according to~\eqref{eqnG0}, would differ from~\eqref{G0sol} by a holomorphic function $f(w)$ satisfying $f(w+2\pi\ii)=f(w)$ and $f(w+2\pi\ii\tau)=e^{\ii a} f(w)$. Such a function extends to a bounded holomorphic function on $\C$, which thus must be a constant. Additionally, $f(w+2\pi\ii\tau)=e^{\ii a} f(w)$ (together with $a$ being generic) implies that it ought to vanish.

We next turn to the correction term $g(w,\bar{w})$. Summing its Fourier series, we find
\begin{equation}
g(w,\bar{w}) \ = \  \frac{1}{\pi\ell}\cdot\frac{e^{\ii a \frac{y}{\beta\ell}}}{e^{\ii a}-1}\,, \quad \text{for} \quad 0<y<\beta\ell\,.
\end{equation}
Notice that this function is discontinuous on the whole circle $S^1_y$, namely it has a jump at $y=0\mod \beta\ell\,\Z$. However, as explained in the paragraph after equations~\eqref{schrod_modes}, we must keep operator insertions separated in the $y$ direction, \emph{i.e.} work in the interval $0<y<\beta\ell$, where $g(w,\bar{w})$ is smooth. Collecting all the pieces, we find the propagator
\begin{equation}
P(w,\bar{w}) \ = \ \frac1{2\pi^2\ii\ell}e^{\ii\frac{a}{2\beta}(w+\bar{w})} \frac{\theta_1'\left(0; \tau\right)}{\theta_1\left(-\frac{a}{2\pi}; \tau\right)}  \frac{\theta_1\left(\frac{w}{2\pi\ii}-\frac{a}{2\pi}; \tau\right)}{\theta_1\left(\frac{w}{2\pi\ii}; \tau\right)} + \frac{1}{\pi\ell} \frac{e^{\ii \frac{a}{2\beta}(w+\bar{w})}}{e^{\ii a}-1}\,,
\end{equation}
for $0<\Re(w)<\beta$. In the small $bc$ ghost system, all vertex operators are constructed from $b$, $\nabla_w c$, and their derivatives, so it is more useful to know $\langle \nabla_w c(w,\bar{w})\, b(0)\rangle = \nabla_w P(w,\bar{w})$. We notice that $\nabla_w e^{\ii \frac{a}{2\beta}(w+\bar{w})}=0$, which implies that
\begin{equation}
\langle \nabla_w c(w,\bar{w})\, b(0)\rangle  \ = \  \frac1{2\pi^2\ii\ell}\cdot e^{\ii\frac{a}{2\beta}(w+\bar{w})} \frac{\theta_1'\left(0; \tau\right)}{\theta_1\left(-\frac{a}{2\pi}; \tau\right)} \partial_w \frac{\theta_1\left(\frac{w}{2\pi\ii}-\frac{a}{2\pi}; \tau\right)}{\theta_1\left(\frac{w}{2\pi\ii}; \tau\right)}\,.
\end{equation}
In fact, this is the expression that must be compared against the four-dimensional correlator of twisted-translated operators $\langle\lambda(w,\bar{w})\tilde\lambda(0)\rangle$ computed in~\cite{Pan:2019bor}. To do this, we define a new variable $z=-\ii w$, and note that
\begin{equation}
\theta_1'(0,\tau) \ = \ 2\pi \eta(\tau)^3.
\end{equation}
Then, we obtain
\begin{equation}
\langle \nabla_z c(z,\bar{z})\, b(0)\rangle  \ = \  \frac1{\pi\ii\ell} \eta(\tau)^3 \cdot e^{-\ii\frac{a}{2\pi}\cdot\frac{z-\bar{z}}{\tau-\bar{\tau}}} \partial_z \frac{\theta_1\left(-\frac{a}{2\pi}+\frac{z}{2\pi}; \tau\right)}{\theta_1\left(\frac{z}{2\pi}; \tau\right)\theta_1\left(-\frac{a}{2\pi}; \tau\right)}\,.
\end{equation}
Upon rescaling fields by numerical factors, we see complete agreement with (4.24) of~\cite{Pan:2019bor}, which implies that the $bc$ ghost action in~\eqref{bc_PI} is the correct one, namely it fully reproduces all the necessary correlation functions in the VOA. Finally, we should also note that on components of the $bc$ fields valued in the Cartan subalgebra, $a$ acts by zero, and one can check that the above formula still applies as it has a well-defined $a\to0$ limit.

\subsubsection{Bosonic propagators}

We now repeat this procedure for the symplectic bosons. In the absence of monodromy defects, the Fourier expansion of the bosonic two-dimensional action is given in~\eqref{action_modes}, while the presence of a defect modifies it to~\eqref{action_perio1}. Throughout this section, $a\in\mathfrak{g}$ is assumed to be a diagonal matrix acting in the matter representation $\mathbf{R}$. To avoid clutter, we still write it as $a$, with the understanding that on a weight $\rho\in\mathbf{R}$ it acts as a multiplication by $\rho(a)$. The two-point function in all cases is
\beaa
\langle \sZ_I(w_1) \sZ_J(w_2) \rangle& \ = \ \frac{\ii}{4\pi\ell}\varepsilon_{IJ} B(w_1-w_2, \bar{w}_1-\bar{w}_2)\,,
\eeaa
\\
where the function $B(w,\bar{w})$ has a double Fourier series representation
\begin{align}
\text{Without defect:}\quad B(w,\bar{w})& \ = \ \frac1{\pi\beta}\sum_{m,n\in\Z} \frac{e^{\ii\left(m-\frac12\right)\left(\varphi+\zeta \frac{y}{\ell}\right) + 2\pi\ii \left(n+\frac{N}2\right)\frac{y}{\beta\ell}}}{m-\frac12 - \frac{2\pi\ii}{\beta}\left(n+\frac{N}{2}\right) + \ii \frac{a}{\beta}} \,,\\
\text{With defect:}\quad B(w,\bar{w})& \ = \ \frac1{\pi\beta}\sum_{m,n\in\Z} \frac{e^{\ii m\left(\varphi+\zeta \frac{y}{\ell}\right) + 2\pi\ii \left(n+\frac{N}2\right)\frac{y}{\beta\ell}}}{m - \frac{2\pi\ii}{\beta}\left(n+\frac{N}{2}\right) + \ii \frac{a}{\beta}} \,.
\end{align}
Of course, the function $B(w,\bar{w})$ is the Green's function of $\nabla_{\bar w}$, \emph{i.e.}
\begin{align}
\label{green_eqn1}
\text{Without defect:} \quad \nabla_{\bar w}B(w,\bar w) & \ = \ - e^{-\frac{\ii}{2}\varphi + \ii\pi (N-M) \frac{y}{\beta\ell}}\,\ell\, \delta\left(\varphi\right) \delta\left(y\right) \,,\\
\label{green_eqn2}
\text{With defect:} \quad \nabla_{\bar w}B(w,\bar w) & \ = \ - e^{ \ii\pi N \frac{y}{\beta\ell}}\,\ell\, \delta\left(\varphi\right) \delta\left(y\right) \,.
\end{align}
The delta functions on the right hand side are periodic, and the exponentials in front of them are necessary to match the periodicity of the left hand side. The periodicity of $B$ in $\varphi$ is determined by the presence of a defect, while its periodicity in $y$ is determined by $N$ and $M$.

As before, the form of the covariant derivative $\nabla_{\bar w}=\partial_{\bar w}-\frac{\ii a}{2\beta}$ suggests defining $B_0$,
\begin{equation}
B(w,\bar{w}) \ = \ e^{\ii\frac{a}{2\beta}(\bar{w}+w)}B_0(w) \,.
\end{equation}
Equations~\eqref{green_eqn1}-\eqref{green_eqn2} then reduce to
\begin{align}
\text{Without defect:}\quad \partial_{\bar w}B_0(w) \ &= \ - e^{-\frac{\ii}{2}\varphi + \ii\pi (N-M) \frac{y}{\beta\ell}}\,\ell\, \delta\left(\varphi\right) \delta\left(y\right) \,,\\
\text{With defect:}\quad \partial_{\bar w}B_0(w) \ &= \ - e^{ \ii\pi N \frac{y}{\beta\ell}}\,\ell\, \delta\left(\varphi\right) \delta\left(y\right) \,,
\end{align}
implying that $B_0$ is always meromorphic, with a simple pole at $w=0$ with residue $\left( -\pi^{-1} \right)$. It satisfies twisted periodicity along $y$ due to the explicit factor of $e^{\ii\frac{a}{\beta\ell}(w+\bar{w})}$, and acquires an additional sign, either $(-1)^{N-M}$ or $(-1)^N$, upon going around $S^1_y$.

To summarize, the periodicity properties of $B_0$ are given as follows
\begin{align}
\label{perioB01}
\text{Without defect:}\quad B_0(w+2\pi\ii)  \ = \  & -B_0(w)\,,\quad B_0(w+2\pi\ii\tau)  =  (-1)^{N-M} e^{\ii a}B_0(w)\,,\\
\label{perioB02}
\text{With defect:}\quad B_0(w+2\pi\ii)   \ = \  & B_0(w)\,,\quad \quad \, B_0(w+2\pi\ii\tau)   =  (-1)^N e^{\ii a}B_0(w)\,,
\end{align}
where
\begin{equation}
\label{eqn:tau}
\tau \ = \ \frac{\ii\beta}{2\pi}(1+\ii\zeta)\,.
\end{equation}
Again, we can easily write a ratio of theta-functions satisfying these properties and having a simple pole at $w=0$ with residue $\left( -\pi^{-1} \right)$. To this end, we recall the notation $(\nu_1, \nu_2)$ for the spin structure on $S^1_\varphi\times S^1_y$; symplectic bosons pick up a phase $e^{\ii\pi\nu_1}$ upon going around $S^1_\varphi$, and a phase $e^{\ii\pi\nu_2}$ around $S^1_y$. Therefore, we have\footnote{Thus, the R-R sector corresponds to $(0,0)$, R-NS to $(0,1)$, NS-R is $(1,0)$ and NS-NS to $(1,1)$. In the literature, one can sometimes find the opposite notation $(\alpha,\beta)=(1-\nu_1, 1-\nu_2)$.}
\begin{align}
\text{Without defect:}\quad \nu_1& \ = \ 1,\quad \nu_2 \ = \  N-M \mod 2 \,,\\
\text{With defect:}\quad \nu_1& \ = \ 0,\quad \nu_2 \ = \ N\mod 2\,.
\end{align}
Using these notations, the answer is
\begin{equation}
\label{general_B0}
B_0(w) \ = \ \frac{\ii}{2\pi^2}\frac{\theta_1'\left(0; \tau\right)}{\vartheta_{1-\nu_1, 1-\nu_2}\left(-\frac{a}{2\pi}; \tau\right)} \times \frac{\vartheta_{1-\nu_1, 1-\nu_2}\left(\frac{w}{2\pi\ii}-\frac{a}{2\pi}; \tau\right)}{\theta_1\left(\frac{w}{2\pi\ii}; \tau\right)}\,,
\end{equation}
written in terms of the standard half-period theta functions,
\begin{equation}
\vartheta_{n,m}(z;\tau)  \ = \  \exp\left( \frac{n}{4} \ii\pi\tau + n\ii\pi \left(z + \frac{m}{2}\right) \right) \vartheta\left(z + \frac{n}{2}\tau + \frac{m}{2}; \tau\right)\,.
\end{equation}
Utilizing the basic property $\vartheta(z+n+m\tau; \tau)=\exp(-\ii\pi m^2\tau - 2\pi\ii mz)\vartheta(z; \tau)$, one can check that~\eqref{general_B0} satisfies the correct periodicity in all four cases~\eqref{perioB01}-\eqref{perioB02}. We prove the uniqueness of solution~\eqref{general_B0} as before. Because it has the correct residue at the (only) pole $w=0$, any other solution for $B_0$ would differ by a holomorphic function $f(w)$, obeying $f(w+2\pi\ii n +2\pi\ii\tau m)=(-1)^{n\nu_1+m\nu_2}e^{\ii a}f(w)$. Such a function extends to a bounded holomorphic function on $\C$, and so is constant. Again, for generic $a$, this constant can only be zero.

Thus, we obtain the propagator of symplectic bosons for arbitrary spin structure $(\nu_1,\nu_2)$,
\beaa
\label{generalB}
B^{(\nu_1,\nu_2)}(w,\bar{w}) \ &= \ e^{\ii\frac{a}{2\beta}(\bar{w}+w)}\frac{\ii}{2\pi^2}\cdot\frac{\theta_1'\left(0; \tau\right)}{\vartheta_{1-\nu_1, 1-\nu_2}\left(-\frac{a}{2\pi}; \tau\right)} \cdot \frac{\vartheta_{1-\nu_1, 1-\nu_2}\left(\frac{w}{2\pi\ii}-\frac{a}{2\pi}; \tau\right)}{\theta_1\left(\frac{w}{2\pi\ii}; \tau\right)}\cr
&= \ e^{\frac{a}{2\pi}\frac{\bar{w}+w}{\bar\tau-\tau}}\frac{\ii}{\pi}\eta\left(\tau\right)^3 \cdot \frac{\vartheta_{1-\nu_1, 1-\nu_2}\left(\frac{w}{2\pi\ii}-\frac{a}{2\pi}; \tau\right)}{\vartheta_{1-\nu_1, 1-\nu_2}\left(-\frac{a}{2\pi}; \tau\right)\theta_1\left(\frac{w}{2\pi\ii}; \tau\right)}\,.
\eeaa
It is worth noting that $(\nu_1,\nu_2)=(1,0)$ is the spin structure considered in~\cite{Pan:2019bor}, and~\eqref{generalB} agrees with their answer in this case.

\subsection{Flavor deformations and screenings}

The expressions for the two-point functions derived in the previous subsection can also serve as the two-point functions when we switch on background holonomies for flavor symmetries. This corresponds to treating some components in $a$ as vevs for flavor symmetries, \emph{i.e.} we now replace $a$ by $a+a_f\in\mathfrak{g}\oplus \mathfrak{f}$, where $\mathfrak{f}$ is the flavor symmetry Lie algebra. Both $a$ and $a_f$ are valued in the Cartan.

Operators which are charged under the Cartan $\mathfrak{t}_f$ of $\mathfrak{f}$ acquire non-holomorphic dependence when $a_f\neq 0$. Indeed, as follows from~\eqref{generalB}, a correlator of two such operators, \emph{e.g.} $\langle \cO_\rho (w,\bar{w}) \cO_{-\rho}(0,0)\rangle$, is proportional to $e^{\frac{\rho(a_f)}{2\pi}\frac{\bar{w}+w}{\bar\tau - \tau}}$, where  $\rho\in \mathfrak{t}_f^*$ is a flavor weight.\footnote{This non-holomorphy has a direct counterpart in the three-dimensional case discussed in~\cite{Dedushenko:2016jxl}, where observables charged under global symmetries failed to be topological in the presence of twisted masses.} We might still choose to focus on the holomorphic sector of a theory with generic holonomies $a_f\neq 0$ turned on; it is formed by $\mathfrak{t}_f$-invariant Schur operators, and provides a consistent truncation $V^{\mathfrak t_f}$ of the full VOA $V$.

In fact, such a truncation is familiar in the study of vertex operator algebras. Before describing it, first recall that when a four-dimensional superconformal field theory has flavor symmetry $\mathfrak{f}$ with four-dimensional central charge $k_{\rm 4d}$, the corresponding VOA $V$ contains an affine Kac-Moody subalgebra,
\begin{equation}
\widehat{\mathfrak{f}}_{k_{\rm 2d}} \ \subset \ V\,,
\end{equation}
whose two-dimensional level is $k_{\rm 2d}=-\frac{k_{\rm 4d}}{2}$~\cite{Beem:2013sza}. This $\widehat{\mathfrak{f}}_{k_{\rm 2d}}$ is generated by currents corresponding to the four-dimensional moment map operators. Suppose the currents $J_i(w)$, $i=1\dots {\rm rk}\,\mathfrak{f}$ correspond to the Cartan elements of $\mathfrak{f}$. We define their respective charges,
\begin{equation}
S_i  \ = \  \oint \dd w\, J_i(w)\,,
\end{equation}
and the $\mathfrak{t}_f$-invariant subsector of the VOA $V$ can be determined as their common kernel,
\begin{equation}
V^{\mathfrak t_f}  \ = \  \bigcap_{i=1}^{{\rm rk}\,\mathfrak{f}} \ker S_i\,,
\end{equation}
\emph{i.e.} $V^{\mathfrak{t}_f}$ is formed by the vertex operators commuting with all $S_i$'s. Note that $V^{\mathfrak t_f}$ obviously contains the Virasoro element of $V$. This is an example of screening; such $S_i$ are known as \emph{screening charges}, and $J_i(w)$ are called \emph{screening currents} (see~\cite{Feigin:2017edu} for a general construction).

Therefore, we conclude that turning on flavor holonomies corresponds, at the level of the chiral algebra, to screening by Cartan elements of the flavor symmetry. Screening is one of the standard operations for extracting vertex subalgebras. Another standard operation -- the BRST reduction (which should be viewed as the VOA analog of quotient) -- was previously identified as the two-dimensional avatar of four-dimensional gauging~\cite{Beem:2013sza}. While BRST reduction involves severe constraints on the matter content reflecting superconformal invariance in four dimensions and nilpotency of the BRST charge in two dimensions, no such constraints are necessary for screenings. Therefore, by turning on flavor holonomies one, at least formally, significantly enlarges the class of VOAs appearing in four-dimensional superconformal field theories. Of course this enlargement is somewhat formal, because we simply consider various sub-VOAs of the larger VOA that exists when all flavor holonomies are zero. Nevertheless, this observation might be important for understanding various properties (in particular modularity) of the flavored Schur index.

Note also that although the non-holomorphic factors cancel in $V^{\mathfrak t_f}$, the holonomies $a_f$ still appear in the arguments of various theta-functions in~\eqref{generalB}. Thus, the corresponding torus correlation functions may still depend on $a_f$. In other words, in general, the $a_f$ parametrize a family of torus correlation functions for $V^{\mathfrak t_f}$.

\paragraph{Example of a free hypermultiplet.} 
To illustrate the above ideas, let us consider the simplest example of a single free four-dimensional hypermultiplet. We turn on generic holonomy $a_f\in\R/2\pi\Z$ for its $U(1)_F\subset SU(2)_F$ flavor symmetry. The corresponding VOA is, as is well-known, a single symplectic boson (or a weight-$\frac12$ $\beta-\gamma$ system), and $U(1)_F$ corresponds to its ghost number symmetry. Namely, the full VOA $V$ (placed on $\C$) is generated by
\begin{equation}
\beta(z)\,, \ \ \gamma(z)\,, \quad \text{with} \quad \beta(z)\gamma(0) \ \sim \ \frac1z\,,
\end{equation}
and the screening is defined by the ghost number current,
\begin{equation}
J(z)  \ = \  :\beta(z)\gamma(z):\,.
\end{equation}
The sub-VOA $V^{\mathfrak{u}(1)_F}$ annihilated by $S=\oint \dd w\, :\beta(w)\gamma(w):$ is formed by words built from $\beta(w)$, $\gamma(w)$ and their derivatives, such that each word has equal number of $\beta$'s and $\gamma$'s, \emph{e.g.} $J$ itself. It is quite easy to give a simpler description for this sub-VOA in the present case using the bosonization isomorphism. Recall that for the $\beta-\gamma$ system, it is given by
\begin{equation}
\beta \ \cong \ e^{-\phi+\chi}\partial\chi\,, \qquad \gamma \ \cong \ e^{\phi-\chi}\,,
\end{equation}
where $\chi$ and $\phi$ are chiral bosons satisfying
\begin{align}
\phi(z)\phi(0)\ \sim \ -\ln z\,,\qquad \chi(z)\chi(0) \ \sim \ \ln z\,,
\end{align}
and furthermore, for the weight-$\frac12$ $\beta-\gamma$ system, the stress energy tensor becomes
\begin{equation}
T \ = \ -\frac12 \partial\phi \partial\phi +\frac12\partial\chi \partial\chi + \frac12 \partial^2\chi\,,
\end{equation}
with the same central charge $c=-1$. One can then easily see that the $V^{\mathfrak{u}(1)_F}$ subalgebra can be identified as generated by $\partial\phi$, $\partial\chi$, and their further derivatives, while the fields $\phi$ and $\chi$ themselves, as well as their exponentials, are not allowed in $V^{\mathfrak{u}(1)_F}$. That is, as a vector space,
\begin{equation}
V^{\mathfrak{u}(1)_F} \ \cong \ \C\left[\left\{\partial^n\phi, \partial^n\chi| n\geq 1\right\}\right]\,.
\end{equation}
In the $\beta-\gamma$ language, the flavor holonomy deformation corresponds to the insertion of 
\begin{equation}
e^{-\mu\int\dd^2w\, \beta(w)\gamma(w)}
\end{equation}
in correlators. Since $J=\beta\gamma$ corresponds to $\partial\chi$ in bosonization, we conclude that in the dual picture, the correlators are deformed according to
\begin{equation}
\langle\dots e^{-\mu \int\dd^2 w\, \partial\chi} \rangle \,.
\end{equation}
 This of course does not affect the flat space correlators.

\section{A taste of modularity}\label{sec:taste}

We have defined four versions of the Schur index, or rather the $S^3\times S^1_y$ partition function. Those versions differ by a $(-1)^{2(R+r)}$ refinement and by a presence or absence of the monodromy defect. In the two-dimensional language, the four versions correspond to the four spin structures on the torus labeled by $(\nu_1, \nu_2)$. Denoting the corresponding partition function as $Z^{(\nu_1, \nu_2)}(\tau, a_f)$, and rewriting our previous expressions in terms of theta-functions, we have
\begin{align}
\label{Z_integral}
 Z^{(\nu_1,\nu_2)}(\tau, a_f) \ & = \ 
 \frac{Z^{\rm f}_{\rm Cas}(q,a_f)}{|\cW|}\int_{\mathbb T}\frac{\dd u}{2\pi\ii u} \eta(\tau)^{3r_G -\dim(G) + \frac12 \dim(\mathbf{R})}\cr
& \times \prod_{\alpha\in\Delta\setminus\{0\}}\theta_1\left(\frac{\langle\alpha,a\rangle}{2\pi};\tau\right) \prod_{w\in\mathbf{R}}\vartheta_{1-\nu_1, 1-\nu_2}\left(\frac{\langle w,a\rangle}{2\pi} + \frac{\langle w_f,a_f\rangle}{2\pi};\tau\right)^{-1/2}\,,
\end{align}
where $\tau$ is the complex structure on the torus, as given in~\eqref{eqn:tau}, and we included flavor holonomies $a_f$. For completeness, we also included the flavor Casimir term $Z^{\rm f}_{\rm Cas}(q,a_f)=q^{-\frac14 \sum_{w_f\in\mathbf{R}_f}\langle w_f, a_f\rangle^2}$ from~\eqref{flav_Cas} in front of the integral. While $Z^{(1,0)}$ and $Z^{(1,1)}$ have finite $a_f\to 0$ limits, the other two quantities, $Z^{(0,0)}$ and $Z^{(0,1)}$, diverge as $a_f\to 0$, so we keep $a_f$ generic and complex. 

Note that $Z^{(1,0)}(\tau, a_f)$ corresponds to the standard Schur index, while $Z^{(1,1)}(\tau, a_f)$ corresponds to the ``modified Schur index" as introduced by Razamat~\cite{Razamat:2012uv}, and these two are related by the $T$-modular transformation, namely
\begin{equation}
\label{T1}
Z^{(1,0)}(\tau \pm 1, a_f)  \ = \  e^{\pm\frac{\ii\pi}{12}\left(2\dim(G) + \frac12\dim(\mathbf{R})\right)} Z^{(1,1)}(\tau, a_f)\,.
\end{equation}
This relation has a clear geometric interpretation; In the spin structure $(1,0)$, a spinor is anti-periodic along $S^1_\varphi$ (call it the $A$ cycle) and periodic along $S^1_y$ (call it the $B$ cycle). Then, upon a Dehn twist along $A$ -- which is what the $T$-transformation is -- the $B$ cycle is ``replaced'' by $A+B$. The spinor is clearly anti-periodic both along $A$ and $A+B$, which is why we obtain $Z^{(1,1)}(\tau, a_f)$ in this way. The four-dimensional lift of this operation is a similar large diffeomorphism on $S^3\times S^1_y$, \emph{i.e.} we cut along $S^3$ at a point ${\rm pt}\in S^1_y$, twist the boundary by $e^{4\pi \ii j_1}$, that is by $\Delta\varphi=\Delta\tau=2\pi$, and then glue it back. In our four-dimensional equations this corresponds to a shift $M \mapsto M+1$. From this geometric picture, it is obvious that the other two partition functions transform into themselves, and indeed we explicitly find that
\begin{align}
\label{T2}
Z^{(0,0)}(\tau + 1, a_f) & \ = \  e^{\frac{\ii\pi}{6}\left(\dim(G) - \frac12\dim(\mathbf{R})\right)} Z^{(0,0)}(\tau, a_f)\,,\\
\label{T3}
Z^{(0,1)}(\tau + 1, a_f) & \ = \  e^{\frac{\ii\pi}{6}\left(\dim(G) - \frac12\dim(\mathbf{R})\right)} Z^{(0,1)}(\tau, a_f)\,.
\end{align}
The phase factors appearing in~\eqref{T1}-\eqref{T3} must be interpreted in terms of global gravitational anomalies, but we leave an in-depth analysis of this for future work.

Note that the way the various spin structures transform amongst themselves is of course well-known from the study of superstring one-loop amplitudes. The novel aspect here is how these properties arise from the Schur index, and have a clear four-dimensional interpretation. 

Similarly, we might ask how the $S$-modular transformation acts on $Z^{(\nu_1,\nu_2)}(\tau, a_f)$. For that, let us first slightly modify the integral formula~\eqref{Z_integral}. There, we integrate over the maximal torus of the gauge group, which can be thought of as
\begin{equation}
\mathbb{T} \ \cong \ \mathfrak{t} / 2\pi\Lambda_{\rm ch}^\vee\,,
\end{equation}
with $\mathfrak{t}$ the Cartan subalgebra, and $\Lambda_{\rm ch}^\vee$ the cocharacter lattice of the gauge group G. We may complexify $\mathbb{T}$ by using the same complex structure as on the spacetime torus, \emph{i.e.}
\begin{equation}
\mathbb{T}_\C \ \cong \ \mathfrak{t}_\C / (2\pi\Lambda_{\rm ch}^\vee  +2\pi\tau\Lambda_{\rm ch}^\vee) \ \cong \ \mathbb{T} \oplus \tau\mathbb{T}\,,
\end{equation}
and consider a family of middle-dimensional cycles in $\mathbb{T}_\C$ parametrized by two integers $(m,n)$,
\begin{equation}
\mathbb{T}_{(m,n)} \ = \  m\mathbb{T} \oplus n\tau \mathbb{T}\,.
\end{equation}
This notation means that $\mathbb{T}_{(m,n)}$ wraps the original real cycle $\mathbb{T}$ exactly $m$ times, and $n$ times the ``imaginary'' cycle $\tau\mathbb{T}$. This is of course a higher-dimensional analog of an $mA + nB$ cycle on a two-torus.

In equation~\eqref{Z_integral}, $u=e^{\ii a}$, and we integrate $a$ over the real cycle $\mathbb{T}_{(1,0)}$. We may consider an integral over a more general cycle, and define the following quantity\footnote{For now, we take this as a definition of some function $Z_{(m,n)}^{(\nu_1,\nu_2)}(\tau,a_f)$, whose physical meaning is unclear. However, we expect them to be related to the partition function with the inclusion of more general defects, and leave a detailed analysis for the future.} 
\begin{align}
\label{Z_mn_nunu}
Z_{(m,n)}^{(\nu_1,\nu_2)}(\tau, a_f) \ & = \  \frac{Z^{\rm f}_{\rm Cas}(q,a_f)}{|\cW|}\int_{\mathbb T_{(m,n)}\subset \mathfrak{t}_\C / (2\pi\Lambda_{\rm ch}^\vee  +2\pi\tau\Lambda_{\rm ch}^\vee)}\frac{\dd a}{2\pi} \eta(\tau)^{3r_G -\dim(G) + \frac12 \dim(\mathbf{R})}\times \cr
&\times\prod_{\alpha\in\Delta\setminus\{0\}}\theta_1\left(\frac{\langle\alpha,a\rangle}{2\pi};\tau\right) \prod_{w\in\mathbf{R}}\vartheta_{1-\nu_1, 1-\nu_2}\left(\frac{\langle w,a\rangle}{2\pi} + \frac{\langle w_f,a_f\rangle}{2\pi};\tau\right)^{-1/2}\,,
\end{align}
such that the original integral~\eqref{Z_integral} is $Z^{(\nu_1, \nu_2)}_{(1,0)}(\tau, a_f)$. A straightforward computation shows:
\begin{equation}
\label{S_with_af}
Z_{(m,n)}^{(\nu_1, \nu_2)}\left(-\frac1{\tau}, \frac{a_f}{\tau}\right) \ = \  e^{-\frac{\ii\pi}{2}\left(\dim(G) - \frac12\dim(\mathbf{R})\delta_{\nu_1,0}\delta_{\nu_2,0}\right)}  Z^{(\nu_2, \nu_1)}_{(-n,m)}(\tau, a_f) e^{\frac{\ii\pi}{2}\left(\frac1{\tau^3} + \frac1{2\pi^2\tau} +\tau \right)\sum_{w_f}\langle w_f,a_f\rangle^2}\,.
\end{equation}
Again, we postpone explanation of the overall phase for future work. We see that the $S$-transformation both flips the spin structure, $(\nu_1, \nu_2)\mapsto (\nu_2, \nu_1)$ (as expected), and changes the integration cycle, $(m,n)\mapsto (-n,m)$. In particular, the original cycle $(1,0)$ goes to $(0,1)$, which is \emph{not} equivalent to the $(1,0)$-cycle.

We can also check how the more general object $Z_{(m,n)}^{(\nu_1,\nu_2)}(\tau, a_f)$ transforms under the $T$-modular transformation, generalizing the case of $(m,n)=(1,0)$ described at the beginning of this section. We explicitly find that
\begin{equation}
\label{T_with_af}
Z_{(m,n)}^{(\nu_1,\nu_2)}(\tau+1, a_f)  \ = \  e^{\frac{\ii\pi}{6}\left(\dim(G) - \frac12\dim(\mathbf{R}) + \frac34 \dim(R)\delta_{\nu_1,1}\right)} Z_{(m+n,n)}^{(\nu_1, \nu_2+\nu_1)}(\tau, a_f) \,,
\end{equation}
where $\nu_1+\nu_2$ should be taken as a remainder modulo $2$, \emph{i.e.} $1+1$ is replaced by $0$.

To summarize, we have found that the functions  $Z_{(m,n)}^{(\nu_1,\nu_2)}(\tau, a_f)$ transform in an interesting way under the modular group. It permutes their discrete labels $(m,n)$ according to the $SL(2,\Z)$ action on the lattice $\Z^2$. As for the $\Z_2\times \Z_2$ labels $(\nu_1, \nu_2)$, the $(0,0)\in\Z_2\times \Z_2$ is fixed, while the other three, $(0,1), (1,0), (1,1)$, are permuted by $SL(2,\Z)$.

The transformation property~\eqref{S_with_af} involves a cumbersome $a_f$-dependent exponential factor, which we would like to remove by taking the $a_f \to 0$ limit. This limit is also interesting due to the relation with~\cite{Beem:2017ooy}. However, as we know, this limit might be singular. More precisely, it is singular for those integration contours $\mathbb{T}_{(m,n)}$ that pass through poles of the integrand of~\eqref{Z_mn_nunu} as $a_f\to 0$. It turns out that this property is preserved by the modular group, and the following holds:

\textbf{An observation. } If $Z_{(m,n)}^{(\nu_1,\nu_2)}(\tau, a_f)$ has a finite $a_f\to 0$ limit, then so do its images under the $SL(2,\Z)$ action described above.

This simple observation implies that for a given $(\nu_1, \nu_2)$, there is a set $\Sigma^{(\nu_1,\nu_2)}$ of integration cycles $\mathbb{T}_{(m,n)}$ that admit a well-defined $a_f\to 0$ limit. Let us denote the corresponding limits as
\begin{equation}
Z_{(m,n)}^{\nu_1,\nu_2}(\tau)  \ = \  \lim_{a_f\to 0} Z_{(m,n)}^{(\nu_1,\nu_2)}(\tau, a_f)\,,
\end{equation}
whose modular transformations follow trivially from~\eqref{S_with_af} and~\eqref{T_with_af}. Notice that for $Z^{(0,0)}_{(m,n)}(\tau, a_f)$, there does not exist a tuple of integers, $(m,n)$, which renders the limit $a_f\to 0$ regular. This is because there is a pole at $a=0$ in the respective integrand. Nevertheless, for the other three spin structures, there are many such integers $(m,n)$ with finite answer in the $a_f\to 0$ limit. Thus, we find a triplet of $\Z^2$-labeled functions,
\begin{equation}
\left(Z^{(0,1)}_{(m,n)}(\tau),Z^{(1,0)}_{(m,n)}(\tau),Z^{(1,1)}_{(m,n)}(\tau)\right)\,,
\end{equation}
where each function has its own set $\Sigma^{(\nu_1,\nu_2)}$ of allowed $(m,n)\in\Sigma^{(\nu_1,\nu_2)}$ labels. The modular group permutes them, and thus, up to phases, they all transform as components of a weight-$0$ modular vector. In particular, the standard Schur index $Z^{(1,0)}_{(1,0)}(\tau)$ belongs to this family, and so do its $T$-transform $Z^{(1,1)}_{(1,0)}(\tau)$ and its $S$-transform $Z^{(0,1)}_{(0,1)}(\tau)$.

What we described so far looks like an infinite-dimensional projective representation of $SL(2,\Z)$. In concrete physical theories, it is expected to truncate. Namely, not all functions $Z_{(m,n)}^{(\nu_1,\nu_2)}(\tau)$ will be independent, and a (possibly finite) set of independent functions will transform as some modular vector~\cite{Beem:2017ooy}.

Finally, physically, we expect that the quantities $Z_{(m,n)}^{(\nu_1,\nu_2)}(\tau, a_f)$, defined as integrals over the non-standard cycles $\mathbb{T}_{(m,n)}$, correspond to partition functions in the presence of more general defects, \emph{i.e.} characters of some non-trivial modules of the VOA. Those functions that are singular in the limit $a_f\to 0$ correspond to modules that have infinite degeneracies at each energy level. Such modules are not pathological, and in fact we have seen above that Ramond modules of the underlying symplectic boson exhibit such a behavior, the quantities $Z^{(0,0)}_{(1,0)}(\tau, a_f)$ and $Z^{(0,1)}_{(1,0)}(\tau, a_f)$ being our basic examples of those.

Let us now briefly review a couple of simple examples.

\subsection{Examples}

\subsubsection{$SU(2)$ with $N_f=4$}

An $SU(2)$ gauge theory with four hypermultiplets is the simplest interacting Lagrangian superconformal field theory. Its Schur index is given by
\begin{align}
Z^{(1,0)}_{(1,0)}(\tau)  & \ = \
-\frac{1}{4\pi}\int_0^{2\pi}\dd a\, \eta(\tau)^8
\frac{\prod_{\alpha\in\{\pm1\}}\theta_1\left(\frac{\langle\alpha,a\rangle}{2\pi};\tau\right)}{\left[\prod_{w\in\{\pm\frac12\}}\vartheta_{0, 1}\left(\frac{\langle w,a\rangle}{2\pi};\tau\right) \right]^4}\cr
&  \ = \ q^{\frac{14}{24}}\left( 1 + 28q + 329 q^2 + 2632 q^3 + 16380 q^4 + O(q^5) \right) \,.
\end{align}
Up to a constant phase, the $T$-transform is expressed by the same formula with $\vartheta_{0,1}$ replaced by $\vartheta_{0,0}$, and it gives the same result, \emph{i.e.} we find
\begin{equation}
Z^{(1,1)}_{(1,0)}(\tau) \ = \ Z^{(1,0)}_{(1,0)}(\tau) \,.
\end{equation}
This is not surprising because the corresponding VOA is believed to be $\mathfrak{so}(8)_{-2}$, which does not have any half-integer graded operators. Therefore, changing the spin structure from $(1,0)$ to $(1,1)$ should not affect the answer.

We can also take a look at the $S$-transform of the Schur index that, up to an overall phase factor, is given by
\begin{align}
Z^{(0,1)}_{(0,1)}(\tau) \ = \ &
-\frac{1}{4\pi}\int_0^{2\pi\tau}\dd a\, \eta(\tau)^8
\frac{\prod_{\alpha\in\{\pm1\}}\theta_1\left(\frac{\langle\alpha,a\rangle}{2\pi};\tau\right)}{\left[\prod_{w\in\{\pm\frac12\}}\vartheta_{1, 0}\left(\frac{\langle w,a\rangle}{2\pi};\tau\right) \right]^4}\cr
 \ = \ & \frac{q^{\frac{14}{24}}}{120\pi}\Bigg( \frac1{q} + 250 + 60\log(q) + (4625 + 1680 \log(q))q \cr&+ (44250 + 19740 \log(q)) q^2 + (305750 + 157920 \log(q)) q^3+\dots \Bigg)\,.
\end{align}
The logarithmic pieces appearing in this expression might look surprising, however such a behavior was already announced in~\cite{Beem:2017ooy}. Additionally, the leading order term $\frac1{q}$ also agrees with their results. We leave the interpretation of this phenomenon for future work.

\subsubsection{$\cN=4$ super Yang-Mills with $G=SU(2)$}

Here, we consider another illustrative example, the $\cN=4$ SYM with gauge group $SU(2)$. The Schur index is computed by
\begin{align}
Z^{(1,0)}_{(1,0)}(\tau) \ = \ &
-\frac{1}{4\pi}\int_0^{2\pi}\dd a\, \eta(\tau)^3
\frac{\prod_{\alpha\in\{\pm1\}}\theta_1\left(\frac{\langle\alpha,a\rangle}{2\pi};\tau\right)}{\prod_{w\in\{-1,0,+1\}}\vartheta_{0, 1}\left(\frac{\langle w,a\rangle}{2\pi};\tau\right)}\cr
\ = \ &q^{\frac{9}{24}}\left( 1 + 3q + 4 q^{3/2} + 9 q^2 + 12 q^{5/2} + 22 q^3 + 36 q^{7/2} + 60 q^4 + 88 q^{9/2} + O(q^5) \right)\,.\cr
\end{align}
In this case, since there are half-integral operators in the spectrum, the $T$-transformed Schur index has alternating signs, \emph{i.e.}
\begin{align}
Z^{(1,1)}_{(1,0)}(\tau) \ = \ q^{\frac9{24}}\left( 1 + 3q - 4 q^{3/2} + 9 q^2 - 12 q^{5/2} + 22 q^3 - \R36 q^{7/2} + 60 q^4 - 88 q^{9/2} + O(q^5) \right)\,.\cr
\end{align}
Similarly, we can also compute the $S$-transformed Schur index, and find
\begin{align}
Z^{(0,1)}_{(0,1)}(\tau) \ = \ &
-\frac{1}{4\pi}\int_0^{2\pi\tau}\dd a\, \eta(\tau)^3
\frac{\prod_{\alpha\in\{\pm1\}}\theta_1\left(\frac{\langle\alpha,a\rangle}{2\pi};\tau\right)}{\prod_{w\in\{-1,0,+1\}}\vartheta_{1, 0}\left(\frac{\langle w,a\rangle}{2\pi};\tau\right)}\cr
\ = \ &\frac{q^{\frac9{24}}}{8\pi}\times q^{-3/8}\Big( 2 + \log(q) +(24 + 12 \log(q)) q + (24 + 36 \log(q)) q^2 \cr &+ (192 + 
128 \log(q)) q^3 + (152 + 324 \log(q)) q^4 + O(q^5) \Big).\cr
\end{align}
Again, this is in agreement with~\cite{Beem:2017ooy}, and an explanation of the logarithmic behavior calls for further investigation.

\section{Relation to the $\Omega$-background $\R_{\epsilon}^{2} \oplus \R^{2}$}\label{sec:omega}

We now switch gears and discuss another application of our $S^{3}\times S^{1}$ background. In particular, it has long been argued~\cite{OldIdea} (see also~\cite{Costello:2018fnz,Costello:2018zrm}), that the two-dimensional chiral algebra should be obtained from putting a four-dimensional superconformal field theory on an $\Omega$-deformed background~\cite{Nekrasov:2002qd,Nekrasov:2003rj}, whilst imposing a holomorphic-topological (``Kapustin") twist~\cite{Kapustin:2006hi}.\footnote{In fact, in the very recent papers~\cite{Oh:2019bgz,Jeong:2019pzg}, this was confirmed explicitly. Here, we take a different approach, but we claim to obtain the relevant four-dimensional background on flat space $\R^{2}_{\epsilon}\oplus \R^{2}$.} Intuitively, it is clear that this ought to be related to some ``local" version of the $S^{3}\times S^{1}$ background treated in our paper, and by expanding the latter, we will explicitly see the corresponding (flat) $\Omega$-background of geometry $\R^{2}_{\epsilon}\oplus \R^{2}$ in the following. The four-dimensional theory of course localizes to the tip of the cigar/origin of $\R^{2}_{\epsilon}$, and the theory thus effectively reduces to the two-dimensional chiral algebra, in the same way as discussed in the previous sections for $S^{3}\times S^{1}$. For simplicity, we start by setting
\begin{equation}
\alpha  \ = \  \zeta  \ = \  0 \,,
\end{equation}
so $S^3 \times S^1$ has the standard product metric, and no background R-symmetry holonomies are present.

The Killing vector arising from the Killing spinors in~\eqref{KillingSpins} (with $\alpha = \zeta = 0$), is given by
\be
K^{\mu} \partial_\mu \ = \ -\frac{1}{2}\xi \sigma^{\mu} \bar\xi \partial_{\mu} \ = \ -\frac{\ii}{\ell} \partial_{\tau} \,.
\ee
This Killing vector naturally defines a $U(1)$-action with respect to which we can define the $\Omega$-background. In this vein, we now define
\be\label{eps1ell}
\epsilon \ = \ \frac{1}{\ell} \,,
\ee
and introduce the new (complex) coordinates
\be
x \ = \  \rho e^{\ii\tau} \,, \qquad \bar x \ = \ \rho e^{-\ii \tau} \,,
\ee
where we conveniently defined 
\be\label{rhocoord}
\rho  \ = \ \frac{1}{\epsilon} \left( \frac{\pi}{2}-\theta \right) \,,
\ee
which is valued in $\rho \in \big[0, \frac{\pi}{2\epsilon} \big]$.\footnote{In the following we will take the $\epsilon \to 0$ expansion, and thus $\rho$ turns into the decompactified radial coordinate on $\R_{\epsilon}^{2}$.} The coordinates $x$ and $\bar x$ parametrize directions orthogonal to the chiral algebra plane, which we recall is located at $\theta=\pi/2$, and spanned by the coordinates $(y,\varphi)$. We can then rewrite the Killing vector $K$ in the more suggestive form
\be
K^{\mu} \partial_{\mu} \ = \ \frac{\ii \epsilon}{2}  \left( x \partial_{x} - \bar x \partial_{\bar x} \right) \,.
\ee
We then recall that our supercharge $\cQ^{H}$ squares to
\be
\left( \cQ^{H} \right)^{2} \ \equiv \ \{\rQ_1,\rQ_2\}  \ = \ \ii \cL_{K} + r \,,
\ee
where $\cL_{K}$ is the Lie derivative with respect to the vector field $K$, and $r$ is the $U(1)_{r}$-symmetry generator. This equation is inherited from the $S^3\times S^1$ background.

As we want to end up with the non-compact space $\R^{2}_{\epsilon} \oplus \R^{2}$, -- or alternatively with the local version of $S^{3}\times S^{1}$ around $\theta = \frac{\pi}{2}$ ($\rho = 0$) -- we ought to expand our metric in the large $\ell \to \infty$ -- or small $\epsilon\to 0$ -- limit. To decompactify the $\varphi$ direction, we rescale 
\be\label{varphihat}
\varphi \ \to \ \hat \varphi  \ = \  \ell \varphi \,,
\ee
and our $S^{3}\times S^{1}$-metric, written in terms of $\rho$ in~\eqref{rhocoord}, $\epsilon$ in~\eqref{eps1ell}, and $\hat \varphi$ in~\eqref{varphihat}, reads
\be
\dd s^{2} \ = \ \left[ \dd y^2+ \dd\hat\varphi^2 \right]+ \left[ \dd \rho^2+ \rho^{2} \dd \tau^2  \right] + o ( \epsilon^{2} )\,,
\ee
corresponding to the flat space $\R_{\epsilon}^{2}\oplus \R^{2}$.

Now, given our explicit expansion in terms of $\epsilon$, we can write down the corresponding (flat space) Killing spinors (we refer to Appendix~\ref{app:omega} for details), and proceed analogously to the $S^{3}\times S^{1}$ analysis performed in the earlier parts of this paper. We can explicitly check that the supersymmetry algebra still closes appropriately, solve the BPS equations (which imply the gradient flow equations mentioned in~\cite{Oh:2019bgz} and~\cite{Jeong:2019pzg}), show that the hypermultiplet action reduces to the symplectic boson action in two (flat) dimensions, the Yang-Mills action remains $\cQ^{H}$-exact, $\emph{et cetera}$.

Notice that the supercharge we obtain in the $\epsilon\to0$ expansion is simply the flat space supercharge $\cQ^1_- - \tilde\cQ_{2\dot-} + \epsilon(\cS_1^- + \tilde\cS^{2\dot{-}})$, and the invariant action does not contain $\epsilon$ -- it is the usual superconformal action in flat space. This corresponds to the statement in~\cite{Oh:2019bgz} that for a superconformal field theory in flat space, the relevant $\Omega$-deformation does not change the action, but rather simply instructs to study the cohomology of the corresponding ``$\cQ+\cS$'' supercharge.

\section{Discussion and open questions}\label{sec:disc}

In this paper, we discussed a variety of aspects of four-dimensional superconformal field theories on the $S^{3}\times S^{1}$ background, and their relation to the two-dimensional chiral algebra on the torus $S^{1}_{\varphi}\times S^{1}_{y}$. The main tool for our study was supersymmetric localization. We started by analyzing the relevant (rigid) supersymmetric background, and studied the preserved subalgebras of the four-dimensional superconformal algebra. By doing so, we classified the fugacities one may turn on, whilst still preserving the two-dimensional chiral algebra structure. Indeed, we found that, in addition to the ``Schur" fugacity, we may turn on quantized (non-vanishing) fugacities $\zeta$ and $\alpha$. Although, we argued that they do not affect the class of operators we count, they affect the complex structure and spin structure of the chiral algebra torus $S^{1}_{\varphi}\times S^{1}_{y}$. This naturally lead us to the discussion of the relevance of the torus spin structure and its relation to the $S^{3}\times S^{1}$ partition function. In particular, a thorough treatment of the latter requires the introduction of what we call the ``canonical" R-symmetry interface, which might end on a surface defect responsible for the change in spin structure along the $\varphi$ direction. We were able to localize (in some detail) the four-dimensional theory (with the inclusion of the fugacities $\zeta$ and $\alpha$ as well as the R-symmetry defect), to arrive at the relevant two-dimensional chiral algebra quantities. Indeed, by doing so, we explained the appearance of the different spin structures and characters in two-dimensions from a four-dimensional perspective, which is intimately related to modularity properties of the four-dimensional superconformal index. We further discussed various detailed properties of the two-dimensional chiral algebras arising from this construction, and made remarks upon an expansion that allows to derive the flat $\Omega$-background underlying the chiral algebra.

Let us now briefly remark upon some open questions and future directions.

One obvious, but interesting future direction involves understanding the modular properties of the Schur index for Lagrangian theories using the technology introduced in Section~\ref{sec:taste}; in particular, this involves understanding the relation of functions $Z^{(\nu_1,\nu_2)}_{(m,n)}(\tau, a_f)$ to surface defects, as well as their place in the story of~\cite{Beem:2017ooy}, as well as related puzzles.

With the localization of the chiral algebra in hand, it is also natural to ask whether there are extensions to other geometries. For instance the lens index~\cite{Benini:2011nc,Alday:2013rs,Razamat:2013jxa,Razamat:2013opa} would be a natural place to start. Results in~\cite{Fluder:2017oxm} suggest that some data of the chiral algebra is stored in it. Similarly, the extensions discussed in~\cite{Cecotti:2015lab} seem to be related to some slightly modified geometry, which allows for a study using localization. By turning on background fields (and thus implementing an appropriate twist), it should be straightforward to extend the rigid $\Omega$-background in Section~\ref{sec:omega} to a curved space of the form $\cC \times \Sigma$, where $\cC$ and $\Sigma$ are two Riemann surfaces (this was very recently done in~\cite{Oh:2019bgz,Jeong:2019pzg}). This leads to generalizations of Schur indices/characters to higher-genus Riemann surfaces $\cC$, with the natural questions arising almost immediately: how do these quantities count Schur operators, what can they tell us about the VOA structure, and of course how they behave under the mapping class group action of $\cC$.

Next, the inclusion of various types of (non-local) defects, whose relation to the VOA was studied from a variety of perspectives in~\cite{Cordova:2016uwk,Cordova:2017ohl,Cordova:2017mhb,Pan:2017zie,Nishinaka:2018zwq}, deserves a study from the point of view of localization on $S^{3}\times S^{1}$. In this paper, we have introduced the novel type of R-symmetry surface defects, and studied their simplest version for Lagrangian theories in relation to the chiral algebra. Extending this analysis to more general defects of this type, as well as other defects is a next logical step. Similarly, it might be interesting to study similar defects in the full superconformal index. 

At this point, it is widely accepted that one may define the Schur index for non-conformal theories (see \emph{e.g.}~\cite{Cordova:2015nma,AreYouEverGonnaPublishThis}). Similarly, there are constructions in  three dimensions (see \emph{e.g.}~\cite{Gaiotto:2016wcv}), which suggest that one may relax the condition of theory being conformal in deriving the VOA. These results are very suggestive that there is a way to move away from ``conformality'' of the four-dimensional theory, and it might be feasible to find a simple modification of our construction reproducing the VOA in such more general setting. In fact, in~\cite{Oh:2019bgz} the authors proposed a possible way to achieve this by introducing anomalous surface defects and adding a certain rotational symmetry breaking term in the action, though this proposal calls for further clarification.

It is of course intriguing to study similar constructions (\emph{i.e.} preserving some interesting subalgebra) in other dimensions, such as~\cite{Beem:2014kka,Chester:2014mea,Beem:2016cbd,Dedushenko:2016jxl,Mezei:2018url}. Even in the three-dimensional $\cN=4$ case, which has been scrutinized for Lagrangian theories in~\cite{Dedushenko:2016jxl,Dedushenko:2017avn,Dedushenko:2018icp}, there is no known result on extended objects and modules of the corresponding associative algebra yet. Further, there is a known connection, on the one hand, between the three-sphere partition function and the quantized Coulomb and Higgs branches~\cite{Dedushenko:2016jxl,Dedushenko:2017avn,Dedushenko:2018icp}. On the other hand, there is a similar story relating three-dimensional $\cN=4$ theories on a space of the form $\R\times \R^{2}_{\epsilon}$, where the latter factor is the $\Omega$-background~\cite{Bullimore:2015lsa,Beem:2016cbd,Bullimore:2016nji,Bullimore:2016hdc,Dimofte:2018abu}. Roughly speaking, the latter construction corresponds to a ``local" version of the former, akin to the logic in Section~\ref{sec:omega}. Indeed, here, we treat the (``uplifted") version in four dimensions, and thus expect a three-dimensional analysis to be straightforward. In this spirit, we remark that the role of short star-products related to the distinguished basis in the associative algebra~\cite{Beem:2016cbd} has never been discussed from the $\Omega$-deformation viewpoint, though the results of~\cite{Dedushenko:2018icp} suggest that the defining properties of such products largely constrain the algebra.

Finally, it would be very interesting to understand whether there exists a ``four-dimensional mirror'' version of the VOA construction. In three dimensions this was obviously the case; the two constructions producing special ``short'' quantizations of the Higgs and Coulomb branches~\cite{Dedushenko:2016jxl,Dedushenko:2017avn,Dedushenko:2018icp} had connections to the mathematical construction of symplectic duality~\cite{Bullimore:2016nji}. It is also understood that the VOA in four dimensions is closely related to the Higgs branch of the theory~\cite{Beem:2017ooy,Beem:2019tfp}, although there is intriguing evidence that it retains data about the Coulomb branch as well~\cite{Fredrickson:2017yka,Dedushenko:2018bpp}. Understanding whether there exists some sort of ``four-dimensional mirror'' counterpart of the VOA naturally attached to the Coulomb sector would be a very exciting development.

\section*{Acknowledgements}

We thank Thomas~T.~Dumitrescu, Sergei~Gukov, Shu-Heng~Shao, Nikita~Sopenko, Lev~Spodyneiko, and Yifan~Wang for discussions and correspondence. The work of MD was supported by the Walter Burke Institute for Theoretical Physics and the U.S. Department of Energy, Office of Science, Office of High Energy Physics, under Award No de-sc0011632, as well as the Sherman Fairchild Foundation. MF is partially supported by the JSPS Grant-In-Aid for Scientific Research Wakate(A) 17H04837, the WPI Initiative, MEXT, Japan at IPMU, the University of Tokyo, the David and Ellen Lee Postdoctoral Scholarship, and the U.S. Department of Energy, Office of Science, Office of High Energy Physics, under Award Number de-sc0011632.

\appendix
\section{Notations and conventions}\label{sec:Notation}

In this Appendix, we introduce the relevant notations and conventions. We follow the conventions of~\cite{Hosomichi:2016flq}, but recall them here for the convenience of the reader. We denote by $\alpha, \beta, \ldots = 1,2$ and $\dot \alpha, \dot \beta, \ldots = 1,2$ chiral and anti-chiral spinors $\psi_{\alpha}$ and $\bar \psi^{\dot \alpha}$, respectively. Those indices are raised and lowered by multiplying with the totally anti-symmetric tensor $\varepsilon$, in the NW-SE and SW-NE convention (\emph{i.e.} upper left indices contract with lower right indices, \emph{etc.})
\be
\psi^{\alpha} \ = \ \varepsilon^{\alpha \beta} \psi_{\beta} \,, \qquad \bar \psi_{\dot\alpha} \ = \ \varepsilon_{\dot \alpha \dot \beta} \bar\psi^{\dot \beta} \,,
\ee
where we take $\epsilon$ to be defined via $\varepsilon^{12} = - \varepsilon_{12} = \varepsilon^{\dot 1 \dot 2} = - \varepsilon_{\dot 1 \dot 2} = 1$. Furthermore, spinors are contracted as 
\be
\psi \chi \ = \ \psi^{\alpha} \chi_{\alpha} \,, \qquad \bar \psi \bar \chi \ = \ \bar\psi_{\dot \alpha} \bar \chi^{\dot \alpha} \,.
\ee
Additionally, the doublet-indices $A,B, \ldots = 1,2$ are raised and lowered with $\varepsilon^{AB}$ and $\varepsilon_{AB}$, with $\varepsilon^{12} = - \varepsilon_{12} = 1$, with $\xi^{A} = \varepsilon^{AB} \xi_{B}$, as well as $\xi_{A} = \varepsilon_{AB} \xi^{B}$. Finally, throughout the paper we identify the auxiliary indices (usually denoted by $\check{A}, \check{B}, \ldots$) with the $SU(2)_{R}$ doublet indices $A, B, \ldots$, and thus remove the checks.

The four-dimensional gamma matrices can be decomposed into two $2 \times 2$ blocks, for which we use the following choice\footnote{Throughout the text we denote by $\tau$ the Pauli matrices and by $\sigma$ the $2\times 2$ blocks inside the four-dimensional gamma matrices.} 
\be
\label{gammas}
(\sigma^{a})_{\alpha\dot\alpha}\ = \ \left(-\ii \vec{\sigma}, \mathbbm{1} \right) \,, \qquad (\bar \sigma^{a})^{\dot\alpha \alpha} = \left( \ii \vec{\sigma}, \mathbbm{1} \right) \,,
\ee
with $\tau^{a}$ the standard Pauli matrices
\be
\tau^{1} \ = \ \left( \begin{array}{cc} 0& 1\\ 1& 0 \end{array} \right) \,,\qquad
\tau^{2} \ = \ \left( \begin{array}{cc} 0& -\ii \\ \ii & 0 \end{array} \right) \,,\qquad
\tau^{3} \ = \ \left( \begin{array}{cc} 1 & 0\\ 0&-1 \end{array} \right) \,.
\ee
Furthermore, we define the tensors as follows
\bea
\left( \sigma_{ab}  \right)_{\alpha}{}^{\beta} \ = \ \frac{1}{2}\left( \sigma_{a} \bar \sigma_{b}- \sigma_{b} \bar \sigma_{a} \right) \,, 
\qquad 
\left( \bar \sigma_{ab} \right)^{\dot \alpha}{}_{\dot \beta} \ = \ \frac{1}{2} \left( \bar \sigma_{a}\sigma_{b} -  \bar \sigma_{b}\sigma_{a}\right) \,,
\eea
and correspondingly use the shorthand for the spinors $\xi_{A} \equiv \xi_{A \alpha}$ as well as $\bar \xi_{A} \equiv \bar \xi_{A}{}^{\dot \alpha}$.

For the group theory notations, we use $r_G$ to denote the rank of $G$, $\Delta_+$ to denote positive roots, and $\Delta$ to denote the whole root system, including $r_G$ copies of a zero root.

\section{The Yang-Mills action is $\cQ^{H}$-exact}\label{sec:YM_exact}

In this Appendix we show that the $\cN=2$ Yang-Mills Lagrangian as given in equation~\eqref{YM_act} can be written as a $\mathcal{Q}^{H}$-exact piece (plus some surface terms). In fact, it is even possible to turn on $\alpha = 2\pi N/\beta $ and $\zeta = 2 \pi M/\beta$, with $M, N \in \mathbb{Z}$. An involved yet explicit calculation shows\footnote{\label{tracenegl}For ease of notation we will neglect the trace over the gauge indices here.}
\beaa\label{QYM}
&\hspace{- .2 cm} \cL_{{\rm YM}} \, \widehat{\vol}_{S^3 \times S^1} \ = \ \\
&   \mathcal{Q}^{H} \Bigg[  
- \frac{2\ii}{\bar S} \bar\phi \left( (\xi')^{A \alpha} \lambda_{A \alpha}  \right)
- \left\{ \frac{4}{\ell S^{2}} \phi +\frac{\ii}{S} \left[ \phi, \bar \phi \right] \right\}\left(  \xi^{A\alpha} \lambda_{A\alpha}  \right)
+ \frac{1}{\bar S}\left(  \cD_{\mu} \bar \phi \right)  \left(  \left( \sigma^{\mu} \right)^{\alpha}{}_{ \dot \alpha} \bar\xi^{A\dot \alpha} \lambda_{A \alpha}  \right)\\
&
\hspace{1 cm}
+ \frac{1}{4 S } F_{\m\n} \left(  \lambda^{A \alpha}  \left( \sigma^{\mu\nu} \right)_{\alpha}{}^{\beta} \xi_{A\beta} \right)
+ \frac{1}{2S}\left( \xi^{A\alpha}  D_{AB}  \lambda^{B}{}_{\alpha}  \right)\\
&
\hspace{1 cm}
-\frac{2\ii }{S} \phi \left( ( \bar\xi' )^{A}{}_{\dot \alpha} \bar \lambda_{A}{}^{\dot \alpha} \right)
- \left\{ \frac{4}{\ell \bar S^{2}}\bar\phi - \frac{\ii}{\bar S} \left[ \phi, \bar \phi \right] \right\} \left( \bar \xi^{A}{}_{\dot \alpha} \bar \lambda_{A}{}^{\dot \alpha} \right)
+\frac{1}{S} \left( \cD_{\mu}\phi \right) \left( \left( \bar\sigma^{\mu} \right)_{\dot\alpha}{}^{\alpha} \xi^{A}{}_{\alpha} \bar \lambda_{A}{}^{\dot \alpha} \right) \\
&
\hspace{1 cm}
+\frac{1}{4\bar S} F_{\mu\nu}  \left(  \bar \lambda_{A}{}_{\dot \alpha} \left( \bar \sigma^{\mu\nu} \right)^{\dot \alpha}{}_{\dot \beta} \bar \xi^{A \dot \beta} \right)
+\frac{1}{2\bar S} \left(  \bar \xi^{A}{}_{\dot \alpha} D_{AB} \bar \lambda^{B \dot \alpha} \right)
\Bigg] \widehat{\vol}_{S^3 \times S^1} \\
&+ \left( \text{~surface~terms~} \right)\,,
\eeaa
where we used the definitions for $S$ and $\bar S$, which we recall here
\be
S \ = \ \xi^{A} \xi_{A} \ = \ -2 e^{\ii \tau} \cos \theta \,,\qquad \bar S \ = \ \bar{\xi}^{A}\bar \xi_{A} \ = \ -2 e^{-\ii \tau} \cos \theta \,,
\ee
and where
\begin{equation}
\widehat{\vol}_{S^3 \times S^1} \ = \ \ell^{3} \sin\theta \cos \theta \, \dd \theta \wedge \dd \varphi \wedge \dd \tau \wedge \dd y 
\end{equation}
is the four-dimensional volume form on $S^3 \times S^1$. 

The bosonic piece of the surface -- or total derivative -- terms in the last line of~\eqref{QYM} can be written as follows\footnote{Our conventions for the four-dimensional Hodge star operation is fixed as follows; for two $p$-forms $\alpha_1$ and $\alpha_2$ we have $\alpha_{1} \wedge * \alpha_{2} = \frac{1}{p!} (\alpha_1)_{m_1 \cdots m_p}(\alpha_2)^{m_1 \cdots m_p}$.}
\begin{align}\label{surftermsbos}
\left(\, \text{bosonic surface~terms} \,\right) \ = \ \dd \bigg[ \left(  \Xi + \bar \Xi \right) \wedge F +  \left( * \Pi  \right)\bigg]\,,
\end{align}
where $F$ is the two-form field strength (we again neglected the trace over gauge indices), and the one-forms $\Xi$ and $\bar \Xi$ are given by
\begin{align}
\Xi \ = \ & \frac{1}{S} v \phi \,,\qquad
\bar\Xi_{\mu} \ = \ \frac{1}{\bar S} v \bar \phi \,,
\end{align}
where we recall from~\eqref{vetc} that the one-form $v$ is given by
\begin{align}
 v \ = \ & 4 \ii \cos ^2\theta \left(\ell \dd \tau +\frac{\zeta}{\ell} \dd y\right) \,.
\end{align}
The third one-form $\Pi$ can be written as
\begin{equation}
\Pi \ = \ \frac{2}{\ell S^{2}} v \, \phi^{2} - \frac{2}{\ell \bar{S}^{2}} v \, \bar{\phi}^{2} + \frac{16\ii}{S \bar{S}} \left( \bar{S} v^{\prime}  - \frac{\ii}{4\ell} v \right)\, \phi \bar \phi\,,
\end{equation}
where we introduced an additional bilinear $v^{\prime}$ as follows
\begin{align}
v^{\prime} \ = \ -\frac{1}{2} \xi \sigma_{\mu} \bar{\xi}^{\prime} 
\ = \ \frac{1}{2} e^{\ii \tau +\ii \frac{\zeta-\alpha}{\ell}  y} \left( \cos \theta \,\dd \tau +\ii \sin\theta\,\dd\theta +\frac{\zeta}{\ell} \cos\theta\,\dd y\right) \,.
\end{align}

Now, the $F$-dependent pieces in~\eqref{surftermsbos} include the pieces $\Xi$ and $\bar \Xi$, which are vanishing at the boundary of $S^{3}\times S^{1}$ at $\theta = \pi/2$, and hence, they will not contribute to the action.\footnote{Note that the derivatives of $F$ are vanishing due to the Bianchi identity.} However, the piece dependent on $\phi \bar \phi$ in~\eqref{surftermsbos} is not vanishing at $\theta = \pi/2$. Nevertheless, we can remove it (together with the fermionic surface terms) by the addition of a $\cQ^{H}$-exact piece at the $\theta=\pi/2$ surface. In fact, upon removing the vanishing $F$, $\phi^{2}$, and $\bar{\phi}^{2}$ dependent pieces at the boundary we end up with the remnant terms (recall that the $\tau$ direction degenerates at $\theta=\pi/2$)
\begin{align}\label{surftermstot}
\left(\, \text{surface~terms} \,\right)\Big|_{\theta=\pi/2} \ = \ - 4 \ell^{2}  \left[ \phi \bar \phi + \frac{\ii \ell }{4}( \cQ^{H} \phi )(\cQ^{H}\bar \phi) \right]_{\theta=\pi/2}  \dd \varphi  \wedge \dd y \,.
\end{align}
We recall that 
\begin{align}
\left( \cQ^{H}  \right)^{2}\phi \Big|_{\theta=\pi/2}  \ = \ \frac{4\ii}{\ell}\phi \,, \qquad \left( \cQ^{H}  \right)^{2}\bar\phi \Big|_{\theta=\pi/2}  \ = \ -\frac{4\ii}{\ell}\bar\phi \,,
\end{align}
and thus~\eqref{surftermstot} is $\cQ^{H}$-exact with
\begin{align}
\left(\, \text{surface~terms} \,\right)\Big|_{\theta=\pi/2} \ = \ -\frac{\ii \ell ^3}{2}  \cQ^{H}\left[  (\cQ^{H}\phi ) \bar \phi -  \phi ( \cQ^{H} \bar \phi )  \right] \dd \varphi  \wedge \dd y \Big|_{\theta=\pi/2} \,.
\end{align}

\section{$\cN=1$ description of the $\cN=2$ background}
\label{App:N1intoN2}

In this Appendix we describe the embedding of the $\cN=1$ background on $S^3 \times S^1$ into our $\cN=2$ background outlined in the main text. This allows us to import $\cN=1$ localization results for one-loop determinants and Casimir energies of the $\cN=2$ vector and hypermultiplets. We further describe those localization results and take the necessary Schur limit.

\subsection{$\cN=1$ Supersymmetric background}

Let us first write down the corresponding $\cN=1$ background, to then proceed and leverage the localization results of~\cite{Assel:2014paa} for our $\cN=2$ background. We start by defining the new coordinates
\be
\phi_1 \ \equiv \ -\tau \,, \qquad \phi_2 \ \equiv \ \varphi \,,
\ee
which brings our metric in~\eqref{metric} into a more compatible form with other literature~\cite{Assel:2014paa}. Furthermore, we pick the following vierbeins,
\beaa\label{hate1}
\hat e^{1} &  \ = \  \dd y \,,\\
\hat e^{2} &  \ = \ 
 \ell  \left[  \cos ^2 \theta \dd \phi_1+\sin ^2\theta \dd \phi_2  \right]
- \zeta   \cos 2 \theta \dd y \,,\\
\hat e^{3} &  \ = \  
\frac{1}{2} \sin 2 \theta \sin (\phi_2+\phi_1) \left[2 \zeta   \dd y - \ell  (\dd\phi_1-\dd\phi_2) \right] 
-\ell \cos (\phi_1+\phi_2) \dd\theta \,,\\
\hat e^{4} & \ = \ 
\frac{1}{2} \sin 2 \theta \cos (\phi_2+\phi_1) \left[2 \zeta   \dd y - \ell  (\dd\phi_1-\dd\phi_2) \right] 
+\ell \sin (\phi_1+\phi_2) \dd\theta \,.
\eeaa
As we will see, this choice of orthonormal frame will lead to simplifications of some expressions down the line.\footnote{They were inspired by the choice in~\cite{Cabo-Bizet:2018ehj}.} Of course, we are dealing with a complex (Hermitian) manifold, and in particular, picking the complex coordinates
\beaa
\label{Ccoords1}
z_1 & \ = \ \ell \sin \theta e^{\frac{y}{\ell} - \ii \left( \phi_2 + \frac{\zeta}{\ell} y \right)}  \,,\\
z_2 &  \ = \  \ell \cos \theta e^{\frac{y}{\ell} + \ii \left( - \phi_1 + \frac{\zeta}{\ell} y \right)} \,,
\eeaa
we can write our metric as follows
\be\label{Cmet}
\dd s^{2} \ = \ h_{\alpha \bar\beta} \, \dd z^{\alpha}  \dd \bar z^{\bar \beta} \,, \qquad h_{\alpha \bar\beta} \ = \ e^{-\frac{2y}{\ell}} \left(\begin{array}{cc}1&0\\0&1 \end{array}\right) \,,
\ee
with $h_{\alpha\bar\beta}$ the complex metric.

In order to derive the correct rigid $\cN=1$ supersymmetric background, we take the rigid limit of new minimal supergravity and follow the analysis in~\cite{Festuccia:2011ws,Dumitrescu:2012ha}. The bosonic fields of new minimal supergravity multiplet consist of the metric as well as two auxiliary gauge fields $A_{\mu}$ and $\tilde A_{\mu}$, where the real part of the former is related to the $u(1)_{R}$ symmetry of the $\cN=1$ theory, and the latter ought to be divergence less (it arises as the Hodge dual of a three-form), \emph{i.e.} $\nabla^{\mu} \tilde A_{\mu}=0$. The conditions for supersymmetry are given by the existence of a nontrivial solution to the following Killing spinor equations\footnote{The ambiguous use of the letter $\zeta$ as the Killing spinors and the deformation of the metric~\eqref{metric} in the main text, is unfortunate. However, it should be clear from the context which one is referred to.}
\bea
\label{N1KSE1}
(\nabla_\mu - \ii A_\mu) \zeta + \ii \tilde A_{\mu} \zeta + \ii \tilde A^{\mu} \sigma_{\mu\nu} \zeta & = & 0 \,,\\
\label{N1KSE2}
(\nabla_\mu + \ii A_\mu) \bar\zeta - \ii \tilde A_{\mu} \bar\zeta - \ii \tilde A^{\mu} \bar\sigma_{\mu\nu} \bar\zeta & = & 0 \,.
\eea
Here, the covariant derivatives are defined as follows
\beaa
\nabla_\mu \zeta_{\alpha}  \ = \ & \partial_{\mu} \zeta_{\alpha} - \frac{1}{2} \omega_{\mu}{}^{ab} (\sigma_{ab})_{\alpha}{}^{\beta} \zeta_{\beta} \,,\\
\nabla_\mu \bar\zeta^{\dot\alpha}  \ = \ & \partial_{\mu} \bar\zeta^{\dot\alpha} - \frac{1}{2} \omega_{\mu}{}^{ab} (\bar\sigma_{ab})^{\dot\alpha}{}_{\dot\beta} \bar\zeta^{\dot\beta} \,.
\eeaa

It was shown that the existence of nontrivial solutions to equations~\eqref{N1KSE1} and~\eqref{N1KSE2} are equivalent to the existence of an almost complex structure of the four-dimensional manifold~\cite{Dumitrescu:2012ha}. Thus, there exists a fundamental two-form $J_{\mu\nu}$ of type $(1,1)$, which is expressed in terms of the Killing spinors
\bea
J_{\mu\nu} \ = \ \frac{\ii}{\left| \zeta \right|^{2}} \zeta^{\dagger} \sigma_{\mu\nu}\zeta \,,
\eea
where $\left( \zeta^{\dagger} \right)^{\alpha} = \left( \zeta^* \right)_{\alpha}$, with the latter being the complex conjugate. In fact, together with the complex two-form of type $(2,0)$, $\zeta \sigma_{\mu\nu} \zeta$, this generates a $U(2)$ structure. We can compute $J_{\mu\nu}$ from the metric in complex coordinates~\eqref{Cmet}, and obtain\footnote{Notice, that compared with~\cite{Dumitrescu:2012ha} we have slightly differing conventions, which explains the appearance of some alternate normalization in the following.}
\be
\frac{1}{2} J_{\mu\nu} \hat e^{\mu}\wedge \hat e^{\nu} \ = \ - \hat e^{1} \wedge \hat e^{2}+ \hat e^{3} \wedge \hat e^{4}\,,
\ee
where we use the frame in equation~\eqref{hate1}. Now, if we further define the function
\be
s \ = \ \left| \zeta \right|^{2} \ = \ \left( \zeta^{\dagger} \right)^{\alpha}  \zeta_{\alpha} \,,
\ee
we can express the background fields $V_{\mu}$ and $\tilde V_{\mu}$ in terms of this data. Namely,
\bea
\label{Atmu}
\tilde A_{\mu} & = & -\frac{1}{2} \nabla^{\rho} J_{\rho \mu} + u_\mu\,,\\
\label{Amu}
A_{\mu} & = & A_{\mu}^{\mathrm{c}} - \frac{1}{4} \left( \delta^{\nu}_{\mu} - \ii J_{\mu}{}^{\nu} \right) \nabla^{\rho}J_{\rho\nu} + \frac{3}{2} u_{\mu}\,,
\eea
where $u_{\mu}$ is an arbitrary holomorphic (\emph{i.e.} $J^{\mu}{}_{\nu} u^{\nu} = \ii u^{\mu}$) vector field satisfying $\nabla^{\mu} u_{\mu} = 0$. Furthermore, $A_{\mu}^{\mathrm{c}} $ is given by
\be
A_{\mu}^{\mathrm{c}} \ = \ \frac{1}{4} J_{\mu}{}^{\nu} \partial_{\nu} \log h - \frac{\ii}{2} \nabla_{\mu} \log s\,,
\ee
with $h= \det h_{\alpha\bar\beta}$. 

Now, for 
\be
 s \ = \ \frac{C e^{\kappa y}}{\sqrt{h}} \,,
\ee
with some arbitrary constants  $C$ and $\kappa$, we find that a convenient choice for $u_{\mu}$ in~\eqref{Amu} and~\eqref{Atmu} gives
\be
\tilde A_{\mu} \dd x^{\mu} \ = \ \frac{\ii}{\ell} \dd y \,,\qquad u_{\mu} \dd x^{\mu} \ = \ \frac{\ii}{\ell}\left( \hat e^{1}+\hat e^{2} \right) \,,
\ee
as well as
\be
A_{\mu} \dd x^{\mu} \ = \ - \frac{\ii \kappa}{2} \dd y \,,
\ee
and thus -- in accordance with our $\cN=2$ background -- both background fields are purely gauge holonomies for the (temporal) cycle. 

This choice of background fields allows us to go back and solve the Killing spinor equations. We find
\bea
\zeta \ = \ 
\sqrt{C} e^{y \left(\frac{\kappa }{2}+\frac{1}{\ell }\right)}  \left( \begin{array}{c} 1 \\ 0 \end{array} \right)\,,
\qquad
\bar \zeta \ = \ 
\sqrt{\tilde C} e^{-y \left( \frac{\kappa}{2} + \frac{1}{\ell} \right)}  \left( \begin{array}{c} 1 \\ 0 \end{array} \right) \,,
\eea
for some additional constant $\tilde C$, and where we used the choice of gamma matrices in~\eqref{gammas}.\footnote{Notice that although this choice of gamma matrices agrees with the ones in the main text, they do differ from the conventions in~\cite{Dumitrescu:2012ha}.} Now, in order to get appropriate boundary conditions of the Killing spinors upon going around the $y$-circle, \emph{i.e.}
\be
\zeta (y+\beta \ell) \ = \ e^{\frac{\beta  \kappa \ell }{2}+\beta} \zeta (y) \,,
\qquad
\bar\zeta (y+\beta \ell) \ = \ e^{- \frac{\beta  \kappa \ell }{2}-\beta} \bar\zeta (y) \,,
\ee
we fix\footnote{Notice, that the spinors could be fixed to be periodic by absorbing an appropriate factor into the background fields. However, with an eye towards embedding the $\cN=1$ background into the $\cN=2$ one, we have chosen this (unusual) periodicity, which turns out to match with the twisted-sector requirement of the $\cN=2$ spinors.}
\be\label{kappachoice}
\kappa \ = \ -\frac{5}{2 \ell }\,,
\ee
and thus the background one-form gauge field $A_{\mu}$ reduces to
\be
A_{\mu} \dd x^{\mu} \ = \ \frac{5 \ii}{4 \ell} \dd y \,.
\ee

With our Killing spinors $\zeta$ and $\bar \zeta$, we can compute the Killing vector
\be
K^{\mu} \partial_{\mu} \ = \ -\frac{1}{s^{2}} \zeta \sigma^{\mu} \bar \zeta \partial_{\mu} \ = \ 
\frac{1}{2} \left[
\left(\frac{1}{\ell}-\frac{\ii \zeta }{\ell}\right)\partial_{\phi_1}
+\left(\frac{1}{\ell}+\frac{\ii \zeta }{\ell}\right)\partial_{\phi_2}
-\ii\partial_{y }\right] \,.
\ee
Given the general expression for the Killing vector of Hopf surfaces~\cite{Assel:2014paa}
\be
K \ = \ \frac{1}{2} \left( b_{1} \partial_{\phi_1}+b_{2} \partial_{\phi_2} - \ii \partial_{y} \right) \,,
\ee
we can identify
\be
\ell b_{1} \ = \ \left( 1-\ii \zeta \right)\,,\qquad
\ell b_{2} \ = \ \left( 1+ \ii \zeta \right) \,.
\ee
Let us for completeness mention that one can also use $\bar \zeta$ to define the two-form $\bar J_{\mu\nu} \propto \bar \zeta^{\dagger} \bar \sigma_{\mu\nu} \bar \zeta$. Analogous to the case discussed above, there are formulas that relate the geometry to the background fields, and a careful analysis shows that one ends up with the equivalent background fields.

Now, in~\cite{Assel:2014paa} it was shown that the $\cN=1$ partition function of Hopf surfaces only depend on the complex structure parameters $p$ and $q$. Further, it was argued that these parameters are related to $b_1$ and $b_2$ appearing above as follows
\be\label{pqN1}
p \ = \ e^{-\beta \ell b_1} \,, \qquad q \ = \ e^{-\beta \ell b_2}\,,
\ee
where as before $\beta$ is the periodicity of $y$, \emph{i.e.} $y \sim y +\beta\ell$. Thus, we end up with exactly the same $p$ and $q$ as in the $\cN=2$ case we deal with in the main text and whose relation to our deformation parameters appears in equations~\eqref{pqtbetazetagamma1}~--~\eqref{pqtbetazetagamma3}. 

Alternatively, these complex structure parameters $p$ and $q$ are obtained by observing that in the complex coordinate of equation~\eqref{Ccoords1}, the identification upon going around the circle, $y \sim y +\beta\ell$, reads
\be
\left( z_1, z_2 \right) \ \sim \ \left( e^{- \beta \left( 1 - \ii \zeta \right)} z_1, e^{- \beta\left( 1+\ii \zeta \right)} z_2 \right) \,,
\ee
and thus one can identify $p$ and $q$ in agreement with~\eqref{pqN1}.

\subsection{$\cN=1$ Localization results}

Let us now cite the relevant $\cN=1$ localization results, as derived in~\cite{Assel:2014paa}. We reiterate that the partition function is only dependent on the complex structure of the four-dimensional manifold~\cite{Dumitrescu:2012ha}. In our particular case, this means that it will only depend on the parameters $p$ and $q$.

The $\cN=1$ vector multiplet, as computed from localization, see~\cite{Closset:2013sxa,Assel:2014paa} (we will use the latter results; for index-results, see~\cite{Romelsberger:2007ec} and~\cite{Rastelli:2016tbz} as well as references therein)
\be\label{N1vmS1S3}
Z_{\mathrm{vec}} ^{(\cN=1)} \ = \ e^{- \pi \beta \ell E^{(0)}_{\mathrm{v},\,\cN=1}} (p;p)_{\infty}^{r_G} (q;q)_{\infty}^{r_G} \Delta_{\mathrm{VdM}}^{-1} \prod_{\alpha \in \Delta_{+}} \theta\left( u^{\alpha}, p\right)\theta\left( u^{-\alpha},q\right)\,,
\ee
where the Casimir energy prefactor is given by
\be
E^{(0)}_{\mathrm{v},\,\cN=1} \ = \ 
\frac{1}{12} \left( b_1+b_2 -\frac{b_1+b_2}{b_1b_2} \right) \dim G
- \frac{b_1+b_2}{b_1b_2} \sum_{\alpha \in \Delta_+} \langle\alpha,a \rangle^{2} \,.
\ee
Furthermore, by $r_G$ we denote the rank of the gauge group $G$, by $\Delta_{\mathrm{VdM}}$ the Vandermonde determinant,
\be
\Delta_{\mathrm{VdM}} \ = \ \prod_{\alpha \in \Delta_{+}} \left( 1- u^{\alpha} \right)\left( 1- u^{-\alpha} \right)
\ee
of the given gauge group, by $\Delta_{+}$ the positive roots of $G$, and finally we defined 
\be
u^{\alpha} \ = \ e^{\ii \langle\alpha,a \rangle}\,,
\ee
where $u=e^{\ii a}$ is the holonomy for the gauge group $G$, and by $\langle \cdot , \cdot \rangle$ we denote the canonical pairing of the Lie algebra $\mathfrak{g} = \mathrm{Lie}\,G$ with its dual. In addition, $p$ and $q$ are given in terms of our parameters as in~\eqref{pqN1}. Lastly, the elliptic theta function is defined as follows
\be
\theta(z, p) \ = \ \left(z;p\right)_{\infty}  \left(p/z;p\right)_{\infty} \,,
\ee
with the Pochhammer symbol defined as
\be
\left(z;q\right)_{\infty} \ = \ \prod_{i\geq 0} \left(1-z q^{i}\right) \,.
\ee

Similarly, we can write down the partition function of an $\cN=1$ chiral multiplet in a representation $\cR$ of the gauge group $G$ (and/or flavor symmetry group) and of $\cN=1$ $U(1)_r$-charge $r_{\cN=1}$, 
\be\label{N1cmS1S3}
Z_{\mathrm{mat}} ^{(\cN=1)} \ = \ e^{-\pi \beta \ell E^{(0)}_{\mathrm{mat},\,\cN=1}} \prod_{w\in \cR} \Gamma\left( u^{w} (pq)^{\frac{r_{\cN=1}}{2}}; p,q \right) \,,
\ee
where we again use the shorthand $u^{w} = e^{\ii \langle w, a \rangle}$, take the product over weights of $\cR$, and with the Casimir energy given as follows
\beaa
E^{(0)}_{\mathrm{mat},\,\cN=1}  \ = \ & 
 \sum_{w \in \cR}\frac{1}{b_1b_2} \bigg\{- \frac{\ii }{3} \langle w, a \rangle^{3} 
 + \left(r_{\cN=1}-1\right) \frac{b_1+b_2}{2} \langle w, a \rangle^{2} \\
 &\hspace{2.1 cm}+ \frac{\ii}{12} \left[ 3\left(r_{\cN=1}-1\right)^{2}(b_1+b_2)^{2} -2 -b_1^{2} - b_2^{2} \right] \langle w, a \rangle\bigg\} \\
& +  \frac{b_1+b_2}{24 b_1 b_2}\left[ \left(r_{\cN=1}-1\right)^{3} \left(b_1+b_2\right)^{2}-\left(r_{\cN=1}-1\right) \left(b_1^{2}+b_2^{2} +2\right) \right]  \dim \cR \,,
\eeaa
where we sum over the weights of the representation $\cR$. Here, we write $p$ and $q$ in terms of the squashing parameters $b_1$ and $b_2$ as in~\eqref{pqN1}. Finally, the elliptic gamma function is defined as follows
\be
\Gamma(z, p, q) \ = \ \prod_{i,j\geq 0} \frac{1-p^{i+1} q^{i+1}/z}{1- p^{i} q^{i}z} \,.
\ee

\subsection{Embedding into $\cN=2$ background}

Now let us carefully embed the above $\cN=1$ background in new minimal supergravity into our $\cN=2$ background in the standard Weyl multiplet detailed in the main text. As we already mentioned, the complex structure parameters $p$ and $q$, which completely determine the $\cN=1$ partition function precisely coincide with the aptly named $\cN=2$ parameters, $p$ and $q$. These parameters can thus be interpreted geometrically. The remaining $\cN=2$ parameter, $t$, is related to the constant $\alpha$, which enters the background gauge fields $V$ and $\tilde V$ as in equations~\eqref{VN2} and~\eqref{tildeVN2}.

We start by recalling the $\cN=1$ actions for the chiral and vector multiplets and then move to the study of their embedding into the $\cN=2$ vector and hypermultiplet actions~\eqref{YM_act} and~\eqref{hmaction}. 

The $\cN=1$ vector multiplet, consisting of the components
\be\label{vmcomp}
\left( \cA_{\mu} ,~\lambda,~\bar\lambda,~D \right) \,,
\ee
with $\cA_{\mu}$ a gauge field, $\lambda$, $\bar \lambda$ two spinors of opposite chirality and $D$ an additional auxiliary field. The fields in~\eqref{vmcomp} have assigned $\cN=1$ R-charges $(0,1,-1,0)$. Then, the $\cN=1$ vector multiplet action reads,
\be
\cL^{(\cN=1)}_{\mathrm{vec}}  \ = \  
{\rm Tr} \left( \frac{1}{4} \cF^{\m\nu}\cF_{\mu\nu} -\frac{1}{2} D^{2}  + \frac{\ii}{2} \lambda \sigma^{\mu}D_{\mu}^{\mathrm{cs}} \bar \lambda + \frac{\ii}{2} \bar \lambda \bar \sigma^{\mu} D_{\mu}^{\mathrm{cs}} \lambda \right) \,,
\ee
with
\be
D_{\mu}^{\mathrm{cs}}   \ = \  \nabla_{\mu} - \ii \cA_{\mu} \cdot - \ii q_{R} \left( A_{\mu} - \frac{3}{2} \tilde A_{\mu} \right)\,,
\ee
where by $q_{R}$ we denote the $\cN=1$ $U(1)_{r}$ R-charge of the respective fields, and by $\cdot$ the action in the appropriate representation. 

Similarly, the $\cN=1$ chiral multiplet 
\be
\label{cmcomp}
\left(\phi,~\bar\phi,~\psi,~\bar \psi,~F,~\bar F\right) \,,
\ee
consists of two complex scalars $\phi$ and $\bar \phi$, spinors $\psi$ and $\bar \psi$, and auxiliary fields $F$ and $\bar F$. The $U(1)_{R}$ charges of~\eqref{cmcomp} are given by $\left(\pm r_{\cN=1},\pm (r_{\cN=1}-1),\pm (r_{\cN=1}-2)\right)$. Then, the $\cN=1$ chiral multiplet action reads
\bea
\cL^{(\cN=1)}_{\mathrm{mat}} & = & 
D_{\mu} \bar\phi D^{\mu} \phi 
+ \tilde A^{\mu}\left( \ii D_{\mu} \bar \phi \phi - \ii \bar \phi D_{\mu} \phi \right) 
+\frac{r_{\cN=1}}{4}\left( R + 6 \tilde A_{\mu} \tilde A^{\mu}\right) \bar \phi \phi
+ \bar\phi D \phi
-\bar F F \nonumber\\
&&
+\ii \bar \psi \bar \sigma^{\mu} D_{\mu} \psi
+\frac{1}{2} \tilde A^{\mu} \bar \psi \bar \sigma_{\mu} \psi
+\ii \sqrt{2} \left( \bar \phi \lambda \psi - \bar \psi \bar \lambda \phi \right) \,,
\eea
where the covariant derivatives contain the $\cN=1$ $u(1)_r$ background gauge field $A_{\mu}$ as well as the (dynamical) gauge field $\cA_{\mu}$ in the vector multiplet, $\emph{i.e.}$
\be
D_{\mu} \ = \ \nabla_{\mu} - \ii \cA_{\mu} \cdot - \ii q_R A_\mu \,,
\ee
with the notation understood as above.

To arrive at the required $\cN=2$ vector multiplet action, we start with the $\cN=1$ background introduced above, and take a vector multiplet in some representation of the gauge group $G$, together with a chiral multiplet in the adjoint of $G$ and of R-charge $r_{\cN=1}=\frac{2}{3}$, and should arrive at the required $\cN=2$ Lagrangian. Similarly, the $\cN=2$ (full) hypermultiplet action is obtained from two $\cN=1$ chiral multiplets in representations $\cR$, $\bar \cR$ respectively.\footnote{Recall that half-hypermultiplets in a representation $\rR=\cR \oplus \bar \cR$ are called full hypermultiplets.}

In the index-realization, to match to the enriched fugacity space of the $\cN=2$ superconformal index, we are required to turn on a distinguished combination of flavor fugacities in the $\cN=1$ index (see \emph{e.g.}~\cite{Gadde:2010en,Rastelli:2016tbz,Maruyoshi:2016aim}). In particular, we define the $\cN=1$ index (we neglect additional flavor fugacities for simplicity now)
\be
\cI_{\cN=1} \left( p,q,\xi \right)\ = \ \mathrm{Tr}_{\mathcal{H}_{S^{3}}} \left( -1 \right)^{F}p^{j_1+j_2+\frac{r_{\cN=1}}{2}}q^{j_2-j_1+\frac{r_{\cN=1}}{2}} \xi^{\cF} \,,
\ee
with $\mathrm{Tr}_{\mathcal{H}_{S^{3}}}$ being the trace over the $\cN=1$ Hilbert space in radial quantization, $j_i$ and $F$ defined as in the main text, and $r_{\cN=1}$ the corresponding $\cN=1$ R-charge. Furthermore, the additional fugacity $\xi$ is chosen to match with the $\cN=2$ fugacities $p$, $q$ and $t$,
\be
\xi  \ = \  \left( \frac{t}{(pq)^{2/3}} \right) \,, \label{xipqt}
\ee
and $\cF$ corresponds to the $\cF$-charge, which is given in terms of $\cN=2$, $SU(2)_{R}\times U(1)_{r}$ charges $R$ and $r$ as follows
\be
\cF \ = \ R+r\,.
\ee
In particular, the adjoint chiral multiplet inside the $\cN=2$ vector multiplet has charge $\cF = -1$, and similarly the chiral multiplets inside the $\cN=2$ hypermultiplet have $\cF = \frac{1}{2}$.

Now, in the localization-realization this is translated to introducing the additional background gauge field
\be
a_{\mathrm{bg}} \ = \  \cF a_0 \dd y \,,
\ee
with $\cF$ the $\cF$-charge as specified above and $a_0$ to be determined from a precise matching of the Lagrangians. Indeed, an explicit computation shows that the respective $\cN=1$ actions reproduce the $\cN=2$ vector and hypermultiplet Lagrangians in their respective backgrounds if and only if we precisely introduce the following background gauge field 
\be\label{bgn1gf}
a_0 \ = \ \frac{1}{3}+ \ii \beta  \alpha \,,
\ee
with $\beta=-1$ for the vector multiplet and $\beta=\frac{1}{2}$ for the hypermultiplet. As expected this precisely corresponds to the fugacity combination in~\eqref{xipqt}. 

Hence, our analysis shows that we can safely import the $\cN=1$ localization results~\eqref{N1vmS1S3} and~\eqref{N1cmS1S3} to reproduce the $\cN=2$, $S^3 \times S^1$ localization. Our discussion and the explicit embedding of the $\cN=1$ background into the $\cN=2$ background implies that\footnote{We stress again, that we are dealing with a full hypermultiplet in a representation $\cR$, which corresponds to half-hypermultiplets in $\rR=\cR \oplus \bar\cR$.}
\begin{align}
Z_{\mathrm{vec}}^{(\cN=2)}  &  \ = \  
e^{\beta \ell E_{\mathrm{vec}, \,\cN=2}^{(0)}}\left( p;p \right)^{r_G} \left( q;q \right)^{r_G} \Delta_{\mathrm{VdM}}^{-1}
\prod_{\alpha \in \Delta_{+}} \theta\left( u^{\alpha} ; p \right) \theta\left( u^{-\alpha} ; q \right)
\Gamma \left( u^{\pm\alpha} \xi^{-1} (pq)^{2/3} ; p ; q \right)\,,\\
Z_{\mathrm{mat}}^{(\cN=2)}  &  \ = \ 
e^{\beta \ell E_{\mathrm{mat}, \, \cN=2}^{(0)}} \prod_{w \in \cR} \Gamma \left( u^{\pm w} \xi^{1/2} (pq)^{2/3} ; p ; q \right) \,,
\end{align}
with additional (distinguished) background holonomy $\xi$ as in equation~\eqref{xipqt} due to the required addition of the background gauge field $a_{\mathrm{bg}}$ (as given in~\eqref{bgn1gf}), and where we fixed $r_{\cN=1}=\frac{2}{3}$. Finally, the same procedure gives the additional Casimir energy piece as follows (see also~\cite{Bobev:2015kza})
\beaa\label{eqn:Casimirfull}
E_{\mathrm{vec}, \,\cN=2}^{(0)} & = & \frac{1+\ii \alpha}{12} \left[12 \sum_{\delta \in \Delta_{+}} \langle \delta, a \rangle^{2} + \dim G \left(1+\zeta ^2-2 \alpha ^2-2 \ii \alpha \right)\right] \,,\\
E_{\mathrm{mat}, \,\cN=2}^{(0)} & = & \frac{1+\ii \alpha}{24} \left[-12 \sum_{w \in \cR} \langle w , a \rangle^2+\dim \cR \left(1 -2 \zeta ^2+\alpha ^2-2 \ii \alpha\right)\right] \,.
\eeaa
Of course, all these expressions agree with the known expressions for the $\cN=2$ superconformal index~\cite{Kinney:2005ej,Romelsberger:2005eg,Romelsberger:2007ec,Gadde:2009kb,Gadde:2010en,Gadde:2011uv,Gaiotto:2012xa,Bobev:2015kza}.

\subsection{Schur limit}

Let us now collect the relevant formulae of the $\cN=2$ partition function in the Schur limit ($q=t$ and $p$ arbitrary)~\cite{Gadde:2011ik,Gadde:2011uv}. By doing so, we arrive at the following well-known results\footnote{In the first expression for the vector multiplet, we take product over \emph{all} the roots of $G$, including the zero-roots, in the latter we rewrite it in terms of positive roots.}
\bea
Z_{\mathrm{vec}}^{(\cN=2)} \Big|_{q=t}  & = & q^{\frac{\dim G}{12}+\sum_{\alpha \in \Delta_{+}} \langle\alpha, u \rangle^{2} } \prod_{\alpha \in \Delta} \left( q u^{\alpha}; q \right)^2 \\
 & = & q^{\frac{\dim G}{12}+\sum_{\alpha \in \Delta_{+}} \langle\alpha, u \rangle^{2} }  \left(q;q\right)^{2 r_G} \prod_{\alpha \in \Delta_+} \left(q u^\alpha; q\right)^2\left(q u^{-\alpha}; q\right)^2\,,\label{ZvmSchur}\\
Z_{\mathrm{mat}}^{(\cN=2)} \Big|_{q=t}  & = & q^{\frac{\dim \cR}{24}  - \frac{1}{2} \sum_{w \in \cR} \langle w , a \rangle^2 } \prod_{w \in \cR} \frac{1}{\left( \sqrt{q} u^{w} ; q \right)\left( \sqrt{q} u^{- w} ; q \right)}\\
& = & q^{\frac{\dim \rR}{48}  - \frac{1}{4} \sum_{w \in \rR} \langle w , a \rangle^2 } \prod_{w \in \rR} \frac{1}{\left[ \left( \sqrt{q} u^{w} ; q \right)\left( \sqrt{q} u^{-w} ; q \right) \right]^{1/2}}
 \,,
\eea
with the notation understood as in the previous sections, and where in the last line, we have rewritten the matter contribution in terms of $\rR = \cR \oplus \bar\cR$, so that the latter formula is applicable to half-hypermultiplets as well.

\section{The R-symmetry monodromy defect}\label{app:defect}

Defect operators in a given theory are often defined either by coupling to the lower-dimensional quantum field theory (which upon integrating out the lower dimensional degrees of freedom leads to an alteration of fields in the higher dimensional theory), or by straight away modifying the space of fields one integrates over in the path integral. This then usually imposes some sort of conditions at the defect location. Our case of interest is going to be the surface defect of the latter kind. Even though the two types of constructions often happen to be related (see \emph{e.g.}~\cite{Gaiotto:2012xa,Gadde:2013dda,Hosomichi:2017dbc}), we will not attempt to find an equivalent definition in terms of a ``2d-4d coupled system''.

Here, we would like to define a defect (located at $\left\{ \theta=0 \right\} \subset S^{3}\times S^{1}$) whose main property is that it creates the monodromy $(-1)$ for the $U(1)$ symmetry generated by $2(R+r)$. In a Lagrangian theory all fields have $\frac12\Z$-valued $R+r$ charges, and those that have half-integral $R+r$ are required to become anti-periodic along the circle that links the defect. All the vector multiplet fields have integral $R+r$, so they remain periodic, and the minimal choice consistent with full supersymmetry is to define the defect to be completely ``invisible'' for the vector multiplets.

However, on the contrary, all the hypermultiplet fields have half-integral, \emph{i.e.} $\left( \Z + \frac12 \right)$-valued, $R+r$ charges, so they become anti-periodic around the defect. Since for the hypermultiplets, the supersymmetry variations are linear in the hypermultiplet fields, this seems naively consistent with all supersymmetry as well. The latter obviously cannot be true because the location of the defect breaks part of the isometries, and thus can \emph{at most} preserve half of the supersymmetry. This happens because simply stating that the fields are anti-periodic around the defect is incomplete. For example, for the scalar fields, it allows a behavior of the kind $q_{11}\sim {\rm const} \times e^{\ii\varphi/2}$ (for the defect sitting at $\theta=0$), and one can easily see that such configurations have infinite kinetic energy. Thus, in order to ensure that the action remains finite, one has to more carefully specify the behavior of the fields at the defect. This behavior, as well as possibly a boundary term in the supersymmetry variation, end up breaking part of supersymmetry.

Now let us consider full hypermultiplets (as opposed to half-hypermultiplets), and separate their component fields into the following four groups,
\begin{align}
A &\ =\ \{q_{11}, F_{11}, \psi_{21}, \bar\psi^{\dot1}{}_{1}\}\,,\qquad A'\ =\ \{q_{22}, F_{22}, \psi_{12}, \bar\psi^{\dot2}{}_{2}\}\,,\cr
B &\ =\ \{q_{21}, F_{21}, \psi_{11}, \bar\psi^{\dot2}{}_{1}  \}\,, \qquad B'\ =\ \{q_{12}, F_{12}, \psi_{22}, \bar\psi^{\dot1}{}_{2}\}\,,
\end{align}
where the index structure is as before, \emph{i.e.} $q_{AI}$, $F_{AI}$, $\psi_{\alpha I}$ and $\bar\psi^{\dot{\alpha}}{}_I$, with $A=1,2$ the $SU(2)_R$ index, $I=1,2$ the $SU(2)_F$ index, $\alpha=1,2$ the left and $\dot{\alpha}=1,2$ the right spinorial indices, while any additional flavor or gauge indices are suppressed. Notice, that complex conjugates of bosons in $A$ and $B$ belong to $A'$ and $B'$, respectively, which is only possible in the \emph{full} hypermultiplet case (as opposed to a half-hypermultiplet). Also, notice that we separate components of the same spinor into different groups, thus breaking the Lorentz symmetry, with the surviving generators $\sigma_3$ and $\bar\sigma_3$ corresponding precisely to the $U(1)_\ell\times U(1)_r$ isometry preserved by the defect sitting at $\theta=0$.

Also, we recall a convenient complex coordinate,
\begin{equation}
v \ = \ \sin\theta e^{\ii\varphi}\,,
\end{equation}
which in~\eqref{v_coord} was generalized to $\sin\theta e^{\ii\left(\varphi + \zeta\frac{y}{\ell}\right)}$ for $\zeta\neq 0$. We find two consistent versions of the defect characterized by the following asymptotic behavior of the fields at $\theta=0$,
\begin{align}
\label{def1}
\text{Version I: }\quad & A \text{ and } A'\ =\ \sqrt{v} \times (\text{smooth}) + \sqrt{\bar{v}}\times (\text{smooth})\,,\cr
& B\ =\ \frac1{\sqrt v} \times (\text{smooth})\,,\ \ B'\ =\ \frac1{\sqrt{\bar v}}\times (\text{smooth})\,,\\
\label{def2}
\text{Version II: }\quad & B \text{ and } B'\ =\ \sqrt{v} \times (\text{smooth}) + \sqrt{\bar{v}}\times (\text{smooth})\,,\cr
& A\ =\ \frac1{\sqrt{\bar v}} \times (\text{smooth})\,,\ \ A'\ =\ \frac1{\sqrt{v}} \times (\text{smooth})\,.
\end{align}
It is straightforward, but somewhat tedious, to check that for such a behavior, the kinetic energy ends up finite; it is obvious for fields that vanish as $\sqrt{\theta}$ near the defect, but slightly less obvious for those that behave as $\frac1{\sqrt v}$ or $\frac1{\sqrt{\bar v}}$. In fact, to ensure that the kinetic energy is finite, we are supposed to write the kinetic term for scalars $q_{AI}$ in the following form,
\begin{equation}
\label{finite_kin}
-\frac12 q^{AI}\cD^\mu \cD_\mu q_{AI} \,,
\end{equation}
which thus differs from a manifestly positive definite term $\frac12 \cD^\mu q^{AI}\cD_\mu q_{AI}$ by a surface term supported at the defect. To show that~\eqref{finite_kin} indeed leads to a finite kinetic energy, one should excise a tubular neighborhood of radius $\epsilon$ around the defect, and expand the integrated kinetic energy in powers of $\epsilon$, assuming the particular boundary behavior either in~\eqref{def1} or in~\eqref{def2}. We then find that the singular term in this expansion cancels due to the angular integration.

Having shown that the defect admits field configurations with finite action means that, at least naively, it is a well-defined object in the path integral. Next, we have to show that it preserves the necessary supersymmetry~\eqref{defect_SUSY}. The first step is to show that supersymmetry variations of the hypermultiplet fields are consistent, namely that the $\theta\to 0$ behavior of the variation $\delta X$ for some field $X$ matches the $\theta\to0$ behavior of $X$ itself. We have explicitly checked this by another tedious computation that this is indeed the case.\footnote{Notice, that on a technical level, our frame in~\eqref{fframe} is ill-suited for this, as it does not smoothly continue through $\theta=0$. Thus, we performed this calculation in an alternative frame, closely related to the one in~\eqref{Ccoords1}.}

Now, because for each preserved supersymmetry, $Q\cL = \nabla_\mu\Sigma^\mu$, we must determine the fate of the total derivative term in the presence of a defect. Namely, it might produce an extra term supported at the defect, and we must either show that it is vanishing, or cancel it by an additional boundary term. In our case, the defect is defined to be invisible for the vector multiplets, so we only have to worry about the hypermultiplet action,
\begin{equation}
Q\cL_{\rm mat}  \ = \  \nabla_\mu\Sigma^\mu\,.
\end{equation}
In the presence of the boundary conditions~\eqref{def1}-\eqref{def2}, the variation indeed produces a non-trivial surface term at the defect. This surface term can be canceled by the supersymmetry variation of the appropriate boundary correction term. Such boundary terms were encountered, \emph{e.g.} in the construction of vortex defects on $S^3$ in~\cite{Drukker:2012sr}. Since our defect is equivalent to the flavor vortex for a particular vorticity $(-1)\in U(1)_F\subset SU(2)_F$ (as mentioned in the main text), the results of~\cite{Drukker:2012sr}, uplifted to four-dimensions, are expected to apply.

We also observe that, quite importantly, solutions to the BPS equations in~\eqref{ap1}-\eqref{am1} are consistent with Version I of the defect, while solutions in~\eqref{ap2}-\eqref{am2} are consistent with the boundary conditions for the Version II. Finally, it must be obvious by now that the two versions of the defect correspond to the two Ramond modules of the symplectic boson.

\section{On the $\Omega$-background limit}\label{app:omega}

We provide some more details on the $\Omega$-background discussed in Section~\ref{sec:omega} as a particular expansion of the $S^{3}\times S^{1}$ background. As discussed there, we take the (flat) metric to be
\begin{equation}
\dd s^{2}_{\R_{\epsilon}^{2}\oplus \R^{2}} \ = \ \dd y^{2} + \dd \hat \varphi^{2} + \dd \rho^{2} + \rho^{2} \dd \tau^{2} \,,
\end{equation}
where $y,\hat \varphi \in \R$ span the chiral algebra plane, while $\rho \in [0, \infty)$ and $\tau \in \left[ 0 , 2\pi \right]$ are polar coordinates on the orthogonal plane. Next we introduce the (somewhat inconvenient) frame
\begin{equation}
e^{1} \ = \ - \dd \rho \,,\qquad e^{2} \ = \ \dd \hat\varphi \,,\qquad e^{3} \ = \ \rho \dd \tau \,,\qquad e^{4} \ = \ \dd y\,,
\end{equation}
which we find by employing our expansion for the $S^{3}\times S^{1}$-frame given in~\eqref{fframe} (with $\zeta = 0$). 

Our expansion then immediately gives the Killing spinors (now in flat space), given by\footnote{We use the shorthand matrix notation for $ \xi  = \left( \xi_{A \alpha} \right)$ and $\bar \xi = \left( \bar\xi_{A}{}^{\dot\alpha} \right)$.}
\begin{align}
\xi \ = \ &
\frac{e^{\frac{\ii \tau }{2}}}{\sqrt{2}}
\left(
\begin{array}{cc}
 -1+ \frac{y-\rho -\ii \hat \varphi}{2} \epsilon &-\ii+ \frac{\ii y+\ii \rho +\hat \varphi}{2} \epsilon\\
-\ii -\frac{\ii y-\ii\rho + \hat \varphi}{2} \epsilon & 1+ \frac{y+\rho -\ii \hat \varphi}{2}  \epsilon \\
\end{array}
\right) \,,\\
\bar\xi  \ = \ &
\frac{e^{-\frac{\ii \tau }{2}}}{\sqrt{2}}
\left(
\begin{array}{cc}
 1-\frac{ \epsilon  (y-\rho -\ii \hat \varphi )}{2} & -\ii + \frac{\ii y+\ii \rho +\hat \varphi}{2} \epsilon\\
 \ii +\frac{\ii y-\ii \rho +\hat \varphi}{2}  \epsilon & 1 + \frac{ y+\rho -\ii \hat \varphi }{2} \epsilon \\
\end{array}
\right) \,.
\end{align}
It is then straightforward to check that these satisfy the equations~\eqref{KSE1}--\eqref{KSE4} with 
\begin{align}
\xi' \ = \ &
\frac{e^{-\frac{\ii \tau }{2}}\epsilon}{2\sqrt{2}} \left(
\begin{array}{cc}
 -\ii  & -1 \\
 -1 & \ii \\
\end{array}
\right) \,,\qquad
\bar\xi'  \ = \ 
\frac{ e^{\frac{i \tau }{2}} \epsilon }{2 \sqrt{2}} 
\left(
\begin{array}{cc}
\ii & -1\\
 1 & \ii \\
\end{array}
\right) \,,
\end{align}
and all background fields turned off, \emph{i.e.} $(V_{\mu})^{B}{}_{A} =  \tilde{ V}_{\mu}= T_{\mu\nu}= \bar{T}_{\mu\nu} = M = 0$.
In fact, these are just the flat space Killing spinors for the supercharge $\cQ^1_- - \tilde\cQ_{2\dot-} + \epsilon(\cS_1^- + \tilde\cS^{2\dot{-}})$. The leading term in the action is just the standard superconformal action in flat space.

We can immediately reproduce our results for $S^{3}\times S^{1}$. In particular, the supersymmetry algebra closes as in equations~\eqref{susy1hyp}--\eqref{susy2hyp}, and similarly for the vector multiplet we have
\beaa
&Q^{2} A_{\mu} &&  = \  \ii v^{\nu} F_{\nu\mu} + D_{\mu} \Phi \,, \qquad
&&Q^{2}D_{AB} && = \  \ii v^{\nu} D_{\nu}D_{AB}+\ii [\Phi, D_{AB}] \,,\\
&Q^{2} \phi && = \  \ii v^{\nu} D_{\nu} \phi + \ii [\Phi, \phi]\,,\qquad
&&Q^{2} \lambda_A && = \ \ii v^{\nu} D_{\nu} \lambda_{A} + \ii [\Phi,\lambda_A] + \frac{\ii}{4} \sigma^{\mu \nu} \lambda_{A}D_{\mu} v_{\nu}\,,\\
&Q^{2}\bar \phi && = \  \ii v^{\nu} D_{\nu} \bar\phi + \ii [\Phi, \bar \phi] \,,\qquad
&&Q^{2} \bar\lambda_A && = \ \ii v^{\nu} D_{\nu}\bar \lambda_{A} + \ii [\Phi,\bar\lambda_A] + \frac{\ii}{4} \bar\sigma^{ \mu \nu} \bar \lambda_{A}D_{\mu} v_{\nu}\,,
\eeaa
with
\beaa\label{Omegavetc}
 v_m \dd x^m  \ = \ & 4 \ii \rho^2 \epsilon  \dd\tau \,, \quad & w  \ = \ &0\,,   \\
\Phi \ = \ & 4 \ii \rho \epsilon  \left(  e^{-\ii \tau } \phi -e^{ \ii \tau } \bar \phi \right)\,,
\quad & \Theta  \ = \ & 2 \ii \epsilon\,,  \\
 \check\Theta_{AB} \ = \ & 0 \,,
\quad &  \Theta_{AB}  \ = \ & 0 \,.
\eeaa

The hypermultiplet Lagrangian localizes to the flat space symplectic boson Lagrangian, where the (flat space) $\cQ^{H}$-cohomology is now spanned by the fields
\begin{align}
\sZ_I(\varphi,y) \ &= \  \frac{1}{2} \left(y \epsilon -\ii \epsilon  \hat\varphi +2\right) q_{1I}-\frac{1}{2} \left(y \epsilon -\ii \epsilon  \hat\varphi -2\right) \ii q_{2I}\cr
&=\ q_{+I} + \frac{\epsilon}{2}(y-\ii\hat\varphi)q_{-I},\ \text{where } q_{\pm I}=q_{1I}\pm\ii q_{2I}.
\end{align}
Furthermore, the $\cN=2$ Yang-Mills Lagrangian remains $\cQ^{H}$-exact.

\bibliographystyle{utphys}
\bibliography{refs}
\end{document}